\documentclass[draft]{agujournal2019}
\usepackage{url} 
\usepackage{lineno}
\usepackage[finalnew]{trackchanges} 
\usepackage{soul,color,colortbl}
\usepackage{longtable}

\draftfalse

\journalname{Space Weather}

\begin{document}

\title{Satellite orbital drag during magnetic storms}

\authors{D. M. Oliveira\affil{1,2}, E. Zesta\affil{2}}

\affiliation{1}{Goddard Planetary Heliophysics Institute, University of Maryland, Baltimore County, Baltimore, MD USA}
\affiliation{2}{Geospace Physics Laboratory, NASA Goddard Space Flight Center, Greenbelt, MD USA}

\correspondingauthor{Denny Oliveira}{denny.m.deoliveira@nasa.gov}

\begin{keypoints}
\item Comprehensive spatiotemporal study of satellite orbital drag effects during magnetic storms caused by coronal mass ejections
\item Stronger drag effects during early main phase of extreme storms occur at low latitudes due to Joule heating propagation from high latitudes
\item Densities during the strongest storms show large uncertainties that continue throughout the recovery phase 

\end{keypoints}

Manuscript published in {\it Space Weather} (2019), 17, https://doi.org/10.1029/2019SW002287  

\begin{abstract}

  We investigate satellite orbital drag effects at low-Earth orbit (LEO) associated with thermosphere heating during magnetic storms caused by coronal mass ejections. CHAllenge Mini-satellite Payload (CHAMP) and Gravity Recovery And Climate Experiment (GRACE) neutral density data are used to compute orbital drag. Storm-to-quiet density comparisons are performed with background densities obtained by the Jacchia-Bowman 2008 (JB2008) empirical model. Our storms are grouped in different categories regarding their intensities as indicated by minimum values of the SYM-H index. We then perform superposed epoch analyses with storm main phase onset as zero epoch time. In general, we find that orbital drag effects are larger for CHAMP (lower altitudes) in comparison to GRACE (higher altitudes). Results show that storm-time drag effects manifest first at high latitudes, but for extreme storms particularly observed by GRACE stronger orbital drag effects occur during early main phase at low/equatorial latitudes, probably due to \change{dayside prompt penetration electric fields}{heating propagation from high latitudes}. We find that storm-time orbital decay along the satellites' path generally increases with storm intensity, being stronger and faster for the most extreme events. For these events, orbital drag effects decrease faster probably due to elevated cooling effects caused by nitric oxide, which introduce modeled density uncertainties during storm recovery phase. Errors associated with total orbit decay introduced by JB2008 are generally the largest for the strongest storms, and increase during storm times, particular during recovery phases. We discuss the implication of these uncertainties for the prediction of collision between space objects at LEO during magnetic storms.

\end{abstract}

\section*{Plain Language Summary}

  In this work, we investigate the effects caused by atmospheric drag forces on satellites orbiting Earth in the upper atmosphere ($\sim$300-500 km altitude) during magnetic storms caused by solar perturbations. Atmospheric drag forces during storms have been recognized by the Federal Government as a natural hazard that could severely impact satellites and human assets that fly in the upper atmosphere by affecting their continuous tracking and re-entry processes, as well as significantly reducing their lifetimes. Examples are communications satellites and the International Space Station. We use atmospheric density data, a very important ingredient for orbital drag calculations, to investigate spatiotemporal patterns of their response. By assessing the performance of an empirical model when predicting orbital drag effects, we find that the model introduces uncertainties in atmospheric density computations particularly during the recovery of the most extreme storms. We then suggest physical mechanisms that should be incorporated by the model in order to reduce errors associated with atmospheric densities.

\section{Introduction}\label{introduction}

  The Sun is a very active and the closest star to the Earth. Due to the high variability of the solar magnetic field, the Sun presents a solar cycle whose period is around 22 years, but the alternation between maximum and minimum sunspot numbers (SSNs) is approximately 11 years  \cite{Eddy1976}. The Sun produces disturbances that propagate explosively away from it in the interplanetary space and eventually encounter Earth in their way. The most important disturbances with potential space weather impacts are the so called coronal mass ejections (CMEs). CMEs are more numerous when the Sun displays an active behavior linked to high SSN levels \cite{GOsling1991,Ramesh2010}. CMEs usually consist in three different parts: a leading edge with a shock, magnetic material or inner core, and a magnetic void \cite{Illing1985}. Usually CMEs are the cause of intense-to-extreme magnetic storms as a result of the occurrence of magnetic reconnection between the interplanetary magnetic field (IMF) and the geomagnetic field when IMF is directed southward and lasts for at least a few hours \cite{Gonzalez1987,Gonzalez1994}. In general, the magnetic field structure that follows CME leading edges lead the high-latitude thermosphere to higher heating levels than CME-shock compressions \cite{Lugaz2016,Kilpua2019}. \par

  The magnetosphere is strongly driven during magnetic storms. During active times, large amounts of energy and momentum are transported from the magnetosphere through field-aligned currents and injected into the high-latitude ionosphere-thermosphere system \cite{Fuller-Rowell1994,Liu2005a,Prolss2011,Emmert2015,Lu2016a,KalafatogluEyiguler2018}. This electromagnetic energy input increases the ion motion which in turn enhances the collision between ions and neutral particles in the ionosphere. As a result, neutral particles are heated and move towards higher regions in the upper atmosphere, or the thermosphere \cite{Forbes2007,Prolss2011,Emmert2015}. In addition, global wind surges characterized by large-scale gravity waves are generated at high latitudes and transport energy towards middle and low latitudes \cite{Fuller-Rowell1994}. Such waves are known as traveling atmospheric disturbances, or TADs \cite{Richmond1975}. TADs play an important role in causing global heating and energy distribution in the thermosphere \cite{Hunsucker1982,Hocke1996,Bruinsma2007}. For instance, \citeA{Sutton2009a} showed that a series of CME-driven magnetic storms was the cause of thermosphere heating 2 hours at high latitudes and 3.5-4.0 hours at equatorial latitudes after the CME impacts. Such energy propagation was associated with TADs. \citeA{Oliveira2017c} showed with a superposed epoch analysis (SEA) study of thermosphere heating due to magnetic storms caused by CMEs that energy and heating propagate via TADs from high latitudes to equatorial latitudes within the average time of 3 hours after storm main phase onset. As a result, LEO satellites that happen to fly in those regions find themselves in regions with enhanced neutral mass density levels and experience increased effects related to atmospheric air drag forces \cite{Chopra1961,King-Hele1987,Moe2005,Prolss2011,Prieto2014,Emmert2015,Zesta2016b}.  \par

  Perhaps the first clear connection between enhanced satellite orbital drag effects and magnetic activity was reported very early in space age by \citeA{Jacchia1959}. He studied ephemeris data from the Russian Sputnik 1958$\delta$1 spacecraft and discovered that the satellite acceleration increased significantly during a magnetic active period. Current data show that the main phase of that storm of 8-9 July 1958 ended at Dst (disturbance storm time index, used to quantify magnetic storm intensities) values smaller than $-$300 nT. The spacecraft decelerated because of its loss of gravitational potential energy and the consequent gain of kinetic energy, which led to a decrease in altitude \cite{Prolss2011,Prieto2014,Emmert2015,Zesta2016b}. \citeA{Jacchia1959} correctly attributed this effect to the upwelling of neutral particles from the lower to the upper atmosphere. This pioneer observation sparked the interest of many scientists in the following decades in producing empirical models to estimate densities in the thermosphere \cite{Jacchia1970,King-Hele1987,Mayr1990,Picone2002,Storz2002,Prolss2011,Bruinsma2015,Emmert2015,Yamazaki2015b,Weng2017,Bruinsma2018,Mehta2018,Sutton2018}. \citeA{He2018} have provided a review on the comparison between empirical thermosphere neutral mass density models commonly used in the past decades. \par

  The understanding and improvement of prediction capabilities of satellite orbital drag during storm times is of paramount importance in current space weather investigations and orbit-prediction data product conceptions. Correctly predicting and forecasting orbital drag can lead to satellite collision avoidance, reentry precision of decommissioned satellites, and increase of satellite life times \cite{Prolss2011,Emmert2015,Zesta2016b,He2018}. For example, the U.S. Department of Defense's Space Surveillance Network is a program whose mission is to identify, catalogue, and track artificial space objects/debris orbiting Earth. This network keeps track of more than 2,000 objects larger than 10 cm that could potentially collide with active LEO satellites which could in turn lead to their complete damage and losses. For instance, space debris produced by the event involving the total destruction of the Chinese meteorological satellite Fengyan 1C in 2007 immediately increased the collision risk of LEO satellites by approximately 10\% \cite{Pardini2009,Zesta2016b}. In addition, the unexpected collision between two communications satellites, the active American Iridium 33 satellite and the inactive Russian Cosmos 2251 satellite, increased the space debris by nearly 25\% \cite{Wang2010c}. The accuracy improvement of satellite orbital decay predictions during magnetic storms can play an important role as a tool in preventing the occurrence of the Kessler Syndrome, which was suggested by \citeA{Kessler1978}. The authors predicted with simulations that successive collisions of LEO spacecraft with other spacecraft and/or space debris would increase the debris population exponentially and subsequently lead to a cascade of high collision levels triggered by the augmented debris levels in space. This would lead to the ultimate scenario of some forbidden orbit regions to LEO satellites due to high risk of collisions there. Satellite orbital drag has been identified by the U.S. Federal Government as a natural hazard that could ``disrupt satellite, aircraft, and spacecraft operations; telecommunications; position, navigation, and timing services", as identified by the National Space Weather Strategy. The subsequent National Space Weather Action Plan further outlines goals and roadmaps in order to successively protect human assets in space and on the ground from space weather threats. See a general introduction and links to these documents in \citeA{Lanzerotti2015} and \citeA{Jonas2016}. \par 

  Recently, \citeA{Krauss2018} investigated the effects of magnetic storms on the subsequent drag forces on two LEO missions, CHAMP \cite<CHAllenge Mini-satellite Payload,>{Reigber2002a} and GRACE \cite<Gravity Recovery And Climate Experiment,>{Tapley2004a}. They found that satellite orbital decay rates and the subsequent time-integrated orbital decay show a strong dependence on the spacecraft's altitude. The authors reported that the orbital decay at CHAMP altitudes were approximately three times larger than the total orbital decay at GRACE altitudes during the severe storm of 24 August 2005. These authors showed that mild storms can cause orbital decays at CHAMP altitudes whose magnitudes are comparable to orbital decays at GRACE altitudes caused by much more intense storms. In addition, \citeA{Krauss2018} showed that the orbital decay of CHAMP and GRACE were strongly associated with minimum values of IMF $B_z$ and SYM-H recorded during storm main phases, where the lower $B_z$ and SYM-H, the more severe the orbital decay. In a previous work, \citeA{Krauss2015} showed that the orbital decay associated with the GRACE spacecraft during the most extreme magnetic storm in the CHAMP/GRACE era (20 November 2003) was $\sim$70 m approximately 60 hours after that storm main phase onset. \par

  Despite all these efforts, a statistical study that quantifies general statistical patterns of satellite orbital drag response to magnetic storms caused by CMEs, particularly with different storm intensity levels, still lacks in the literature. The main goal of this work is to fill in this gap. In addition, we will address the capability of an empirical thermosphere model of predicting satellite orbital decay effects during CME storms with different intensities. We will address the possible causes of uncertainties arising during storm times, which are of particular importance when computing probabilities and predictions for collisions of space objects during magnetically active times. \par

  In this paper, we present for the first time an SEA study of the spatiotemporal LEO satellite orbital drag effects with 217 magnetic storms caused by CMEs in the period May 2001 to December 2015. Our particular interest is in the spatial and temporal patterns of orbital drag effects induced by heating propagation from auroral latitudes toward equatorial latitudes, as well as effects introduced by CHAMP and GRACE altitudes. The thermosphere neutral mass density data were obtained from accelerometers on-board the CHAMP and GRACE missions. The background or quiet density was computed with a thermosphere empirical model. Data and model will be presented in section \ref{data_model}, as well as the techniques used for satellite orbital decay computations. The storm of 24 August 2005 will be used to illustrate our approach in section \ref{24Aug2005}. Section \ref{results} presents the main results. Finally, section \ref{conclusion} concludes the paper.

\section{Data, model and orbital drag computations}\label{data_model}

  \subsection{Catalogue of coronal mass ejections}

    We use the CME catalogue provided and maintained by Dr. Ian Richardson and Dr. Hillary Cane \cite{Richardson2010b}. Observation times in the solar wind at L1 upstream of the Earth of plasma and IMF data (next subsection), as well as shock-associated sudden impulse times and minimum Dst values are provided by the catalogue. Main information on techniques used to build and maintain this catalogue is outlined by \citeA{Richardson2010b}. \par

  \subsection{Interplanetary magnetic field, magnetic index, and sunspot number data}

    The IMF data are obtained from the OMNI database. IMF is represented in the geocentric solar magnetospheric system. The data are shifted to the magnetopause nose (at $X$ = 17$R_E$) and the techniques used to accomplish this were outlined by \citeA{King2005}. The IMF data has 1-min time resolution. \par

    In this work, magnetic activity is indicated by the SYM-H index delivered by the World Data Center in Kyoto, Japan. The SYM-H index is similar to the well-known 1-hour resolution Dst index, but it has resolution of 1 minute. We use the former as opposed to the latter because we are interested in thermosphere changes within the time domain of minutes. The SYM-H index was first suggested by \citeA{Iyemori1990}. \par

    Solar activity is indicated by the monthly-averaged SSN data. The SSN data are compiled by SILSO (Sunspot Index and Long-term Solar Observations) of the Royal Observatory of Belgium, in Brussels. An entire {\it Solar Physics} issue has been dedicated to the recent recalibration methods used to determine SSNs \cite{Clette2016b}.

  \subsection{Neutral mass density data}\label{neutral_density}

    We use data from two LEO satellites, namely CHAMP and GRACE. CHAMP was launched on 15 July 2000 and reentered on 19 September 2010. The mission started to orbit Earth at the initial altitude 456 km passing through Earth's poles in time intervals near 90 minutes with orbit inclination 87.25$^\circ$. CHAMP precessed around Earth completing a full longitudinal cycle around the planet in approximately 131 days. The mission was managed by the GeoForschungsZentrum (GFZ), the  German Research Centre for Geosciences, in Potsdam, Germany. CHAMP was equipped with the STAR (Spatial Triaxial Accelerometer for Research) accelerometer, whose acceleration precision was 3.0$\times10^{-9}$ m/s$^2$ within 0.1 Hz \cite{Bruinsma2004}.  \par

    The GRACE constellation consisted of a pair of satellites (GRACE-A flying $\sim$220 km ahead of GRACE-B) that were launched on 17 March 2002 at an altitude of near 500 km with orbit inclination 89.5$^\circ$. The spacecraft reentered on 10 March 2018 and 24 December 2017, respectively. GRACE's period was 95 minutes and it covered all local times in a time interval of 160 days. The accelerometer onboard GRACE was the Super-STAR instrument, with precision and cadence ten times larger than CHAMP's STAR accelerometer \cite{Flury2008}. The GRACE mission was operated by NASA and by the Deutsches Zentrum f\"ur Luft- und Raumfahrt e.V. (DLR), the German Aerospace Center. Given the similarities of both spacecraft observations, we use in this study only GRACE-A data, henceforth GRACE data. This choice is also justified by the higher data quality presented by GRACE-A particularly after 2007. \par

    This study is a continuation to the work of \citeA{Oliveira2017c}, who used the whole CHAMP data available and GRACE data from the mission's beginning up to September 2011. Now, with the addition of GRACE data from October 2011 to December 2015, it is possible to include over 50 storms in the current analysis. \par

    The whole CHAMP data set and the first GRACE data set portion were processed by \citeA{Sutton2008}. These data sets have been validated by many papers \cite<see, e.g.,>{Forbes2007,Prolss2011,Emmert2015,Zesta2016b,He2018}. The calibration method used in the latter part of the GRACE data set is described by \citeA{Klinger2016}. A validation of the October 2011 to December 2015 GRACE data has recently been performed by \citeA{Krauss2018}.

    \begin{figure}[t]
      \centering
      \includegraphics[width=1.00\textwidth]{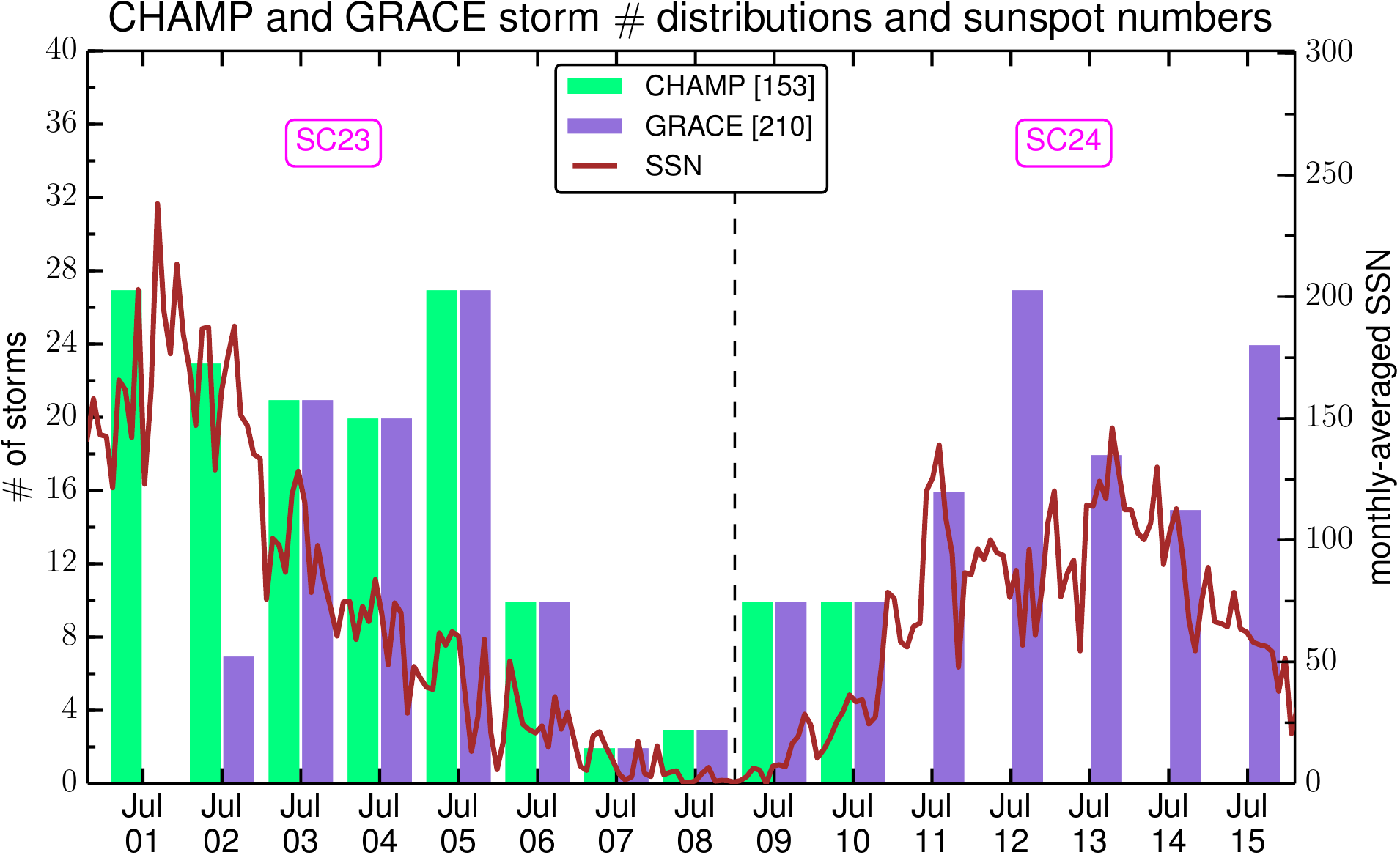}
      \caption{Yearly-averaged storm number distributions for CHAMP (light green bars) and GRACE (light purple bars) along with monthly-averaged SSNs (solid brown line) during the period 1 January 2001 to 31 December 2015. The vertical dashed line indicates the transition from SC23 to SC24 in January 2009. The numbers between brackets correspond to the number of events occurring during CHAMP and GRACE commission times.}
     \label{number_histogram}
    \end{figure}

  \subsection{Distribution of magnetic storm and sunspot numbers}

    Figure \ref{number_histogram} represents the annual number distributions of magnetic storms occurring during CHAMP's commission time (light green bars) and GRACE's commission time (light purple bars) during the interval May 2001 to December 2015, according to the CME catalogue. The solid brown line indicates the SILSO monthly-averaged SSNs for January 2001 to December 2015. \par

    \citeA{Oliveira2017c} presented results from the pre-maximum phase of solar cycle (SC) 23 until the beginning of SC24. SC24 began in January 2009 (dashed black line). The present work shows the additional period concerning the rising, maximum and most of the declining phase of SC24. Therefore, the current analysis uses 15 years of neutral mass density data, a period longer than a regular SC. SC24 is noticeably weaker in comparison to SC23 as for the smaller number of SSNs observed on the Sun's surface. The total number of storms observed by the spacecraft correlate well with SSNs, but the overall number of storms in SC24 is smaller than in SC23. \par

    \begin{figure}
      \centering
      \includegraphics[width=0.68\textwidth]{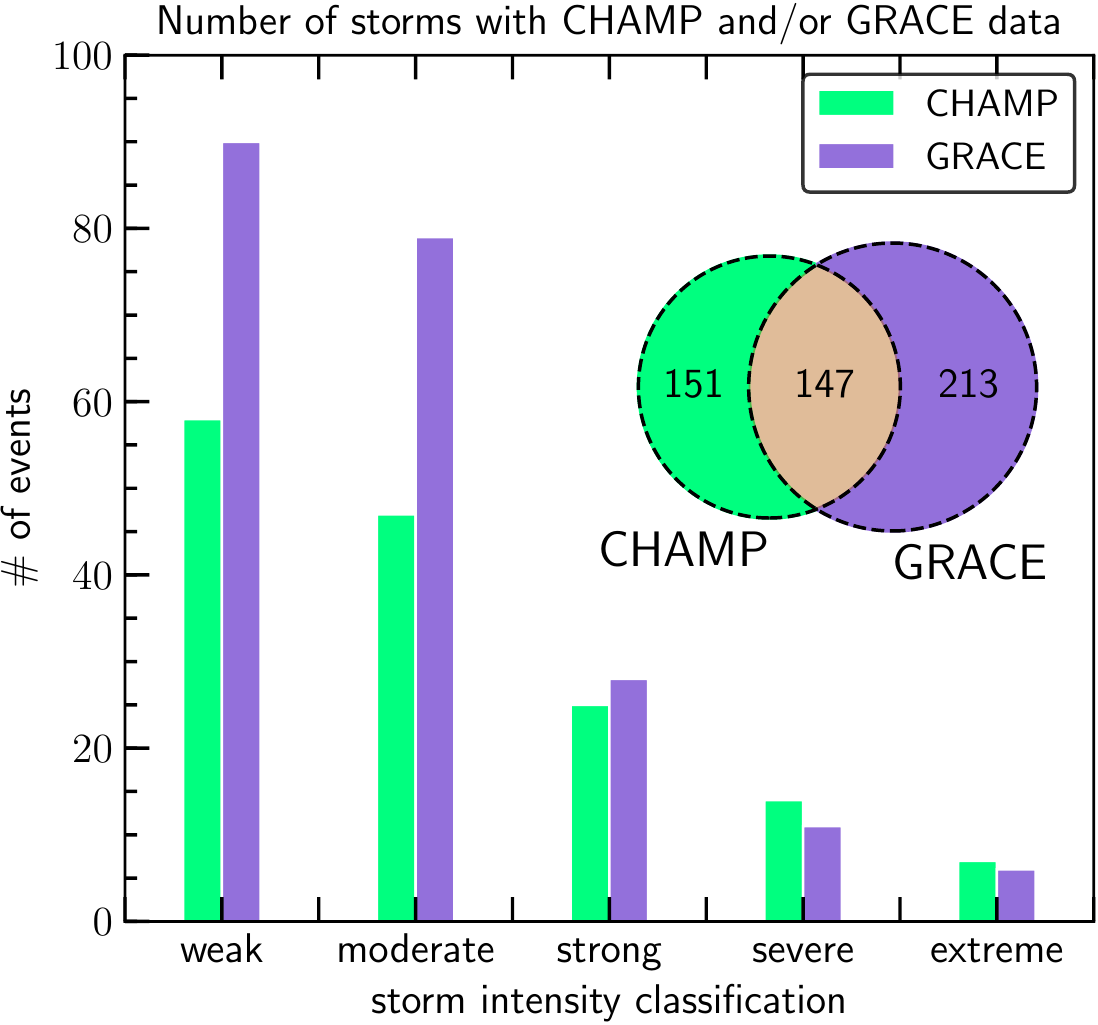}
      \caption{Histogram with the numbers of storms grouped by the intensity categories defined in Table \ref{table}. The Venn diagram documents the number of CHAMP and GRACE observations as well as the number of storms observed simultaneously by both missions.}
      \label{intensities}
    \end{figure}

    The National Oceanic and Atmospheric Administration (NOAA) provides space weather scales for magnetic storms (\url{https://www.swpc.noaa.gov/noaa-scales-explanation}). NOAA defines groups from 1 to 5, namely minor, moderate, strong, severe, and extreme, with respect to increasing values of the Kp index. We follow the same nomenclature as suggested by NOAA, but we define our groups in terms of the minimum value of the SYM-H index at the end of each individual storm main phase. The storm groups are summarized in Table \ref{table}. The link provided above also shows general information about the impact storms with different intensities can have on power transmission lines, satellite operations, and other technological systems. \par

    Figure \ref{intensities} shows results for the storm intensity distributions of CHAMP (light green bars) and GRACE (light purple bars), respectively. The Venn diagram in the inset plot shows that the number of storms with available data for CHAMP and GRACE are 151 and 213, respectively, whereas the total number of individual storms is 217, with 147 storms having data available from both missions. Table \ref{table} shows the number of events with data from either CHAMP or GRACE, or both, available for each category. \citeA{Zesta2018b} provided a detailed description of the dynamic response of all storms, with emphasis on the extreme events shown in Figure \ref{intensities}.

    \begin{table}
      \centering
      \begin{tabular}{l c c c}
          \hline
          Storm Intensity & Category$^\mathrm{a}$ & SYM-H interval$^\mathrm{b}$ & $\#$ of events$^\mathrm{c}$ \\
          \hline
          Minor    & G1 & SYM-H $\geq$ $-$50 nT             & 90\\
          Moderate & G2 & $-$100 $\leq$ SYM-H $<$ $-50$ nT  & 78\\
          Strong   & G3 & $-$150 $\leq$ SYM-H $<$ $-100$ nT & 28\\
          Severe   & G4 & $-$250 $\leq$ SYM-H $<$ $-150$ nT & 14\\
          Extreme  & G5 & SYM-H $<$ $-250$ nT               & 7 \\
          \hline
          & & & Total: 217 events \\
          \multicolumn{3}{l}{$^\mathrm{a}$ \footnotesize{Following NOAA's nomenclature.}} \\
          \multicolumn{3}{l}{$^\mathrm{b}$ \footnotesize{This definition of storm intensity  intervals is arbitrary.}} \\
          \multicolumn{3}{l}{$^\mathrm{c}$ \footnotesize{Number of events in each category with CHAMP and/or GRACE data.}} \\
      \end{tabular}
      \caption{Definitions of Storm Intensity Categories Used in this Work.}
      \label{table}
    \end{table}

  \subsection{CHAMP and GRACE semi-major axes}

    As described in section \ref{neutral_density}, GRACE was launched into a higher altitude ($h$) in comparison to CHAMP. This reflects on the altitude variations of both satellites when flying over equatorial and polar regions. In this work, we use the semi-major axes of both satellites, measured with respect to the Earth's center, taking into account the non-uniformity of the Earth's radius ($R_E$). Considering the Earth is flattened at the poles and bulged at the equator, and its equatorial radius $R_{Ee}$ = 6,378 km, and its polar radius $R_{Ep}$ = 6,357 km, we use a corrected expression for the Earth's non-uniform radius as a function of $R_{Ee}$, $R_{Ep}$, and geographic latitude $(\theta)$, provided by \citeA{Torge1980}. In addition, we include corrections ($a_{gp}$) to the semi-major axes resulting from gravitational perturbations as suggested by \citeA{Chen2012}. Therefore, the semi-major axes as a function of time ($t$), computed in this study are given by
    \begin{equation}\label{a}
      a(t) = R_E(R_{Ee},R_{Ep},\theta,t) + h(t) + a_{gp}(t)\,.
    \end{equation}

\begin{figure*}[t]
      \centering
      \includegraphics[width=0.95\textwidth]{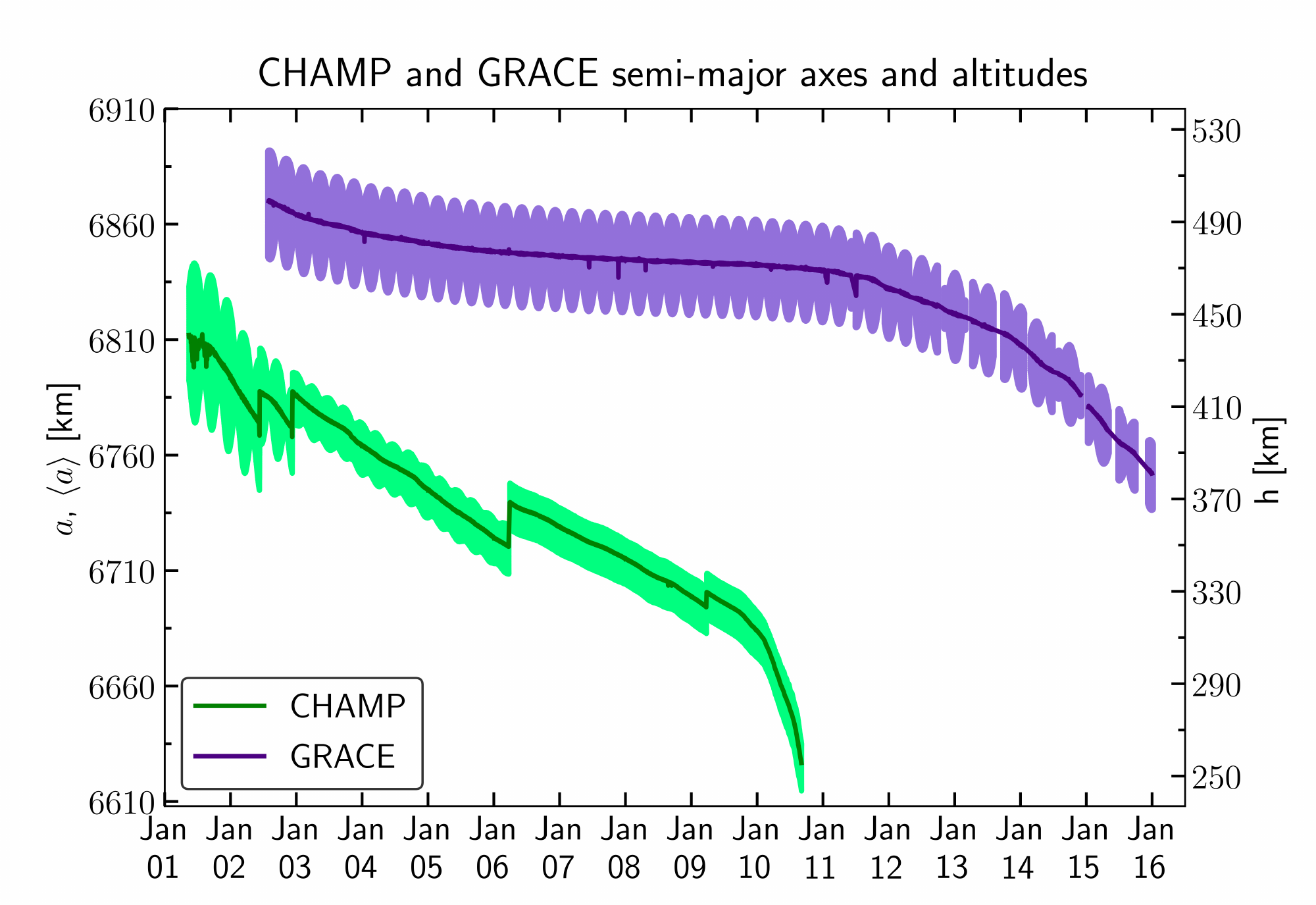}
      \caption{Semi-major axes $a$, in km, computed for CHAMP (light green line) and GRACE (light purple line) missions, with respect to the Earth's center, according to equation \ref{a}. The dark green and purple lines indicate semi-major axis daily averages $\langle a\rangle$ for CHAMP and GRACE, respectively. The secondary vertical axis indicates the spacecraft's altitudes with respect to the Earth's surface.}
      \label{semi_major}
    \end{figure*}

    The CHAMP (light green line) and GRACE (light purple line) semi-major axes are shown as a function of time in Figure \ref{semi_major}. The thick green and purple lines show the daily-averaged $\langle a\rangle$ semi-major axis for each mission. The sudden jumps in the CHAMP semi-major axis in 2002, 2003, 2006 and 2009 correspond to maneuver procedures used to correct the spacecraft's altitude \cite{Bruinsma2004}. On the other hand, GRACE presented a larger semi-major axis variation and a smaller decay rate because it operated at higher altitudes. See \citeA{Oliveira2017c} and \citeA{Krauss2018} for CHAMP's and GRACE's annual decay rates and altitude distributions, respectively.

  \subsection{Thermosphere empirical model}

    In this work, we use the Jacchia-Bowman 2008 empirical model \cite < >[hereafter JB2008]{Bowman2008} to compute modeled thermosphere neutral mass densities. Early versions of JB2008 used only Jacchia's diffusion equations \cite{Jacchia1970} and solar ultraviolet heating represented by the solar radio flux at wavelength 10.7 cm ($F_{10.7}$) index and its average $\langle F_{10.7}\rangle$ centered at 81-day intervals. It was noted that only taking ultraviolet (UV) heating would result in density errors with  periods of approximately 27 days, or a solar rotation. \citeA{Bowman2008} included semi-annual variation corrections that accounted for EUV and FUV radiations, namely extreme and far UV radiations, as well. In addition, the use of the Dst index as a replacement for the global ap index (3-hour time resolution) in the magnetic activity contribution also contributed to the improvement of the model accuracy. \par

    The JB2008 model computes thermosphere neutral mass density from a single parameter, the local exosphere temperature, here represented by $T_\infty$. This temperature is represented by
      \begin{equation}\label{temperature}
        T_\infty=T_\ell(\theta,\delta_\odot,\tau)+\Delta T_{LST}(\tau,\theta,h)+T_{UV}(\chi)+T_{MA}(Dst)\,,
      \end{equation}
    where $T_\ell$ corresponds to an empirical formula for the local exospheric temperature as a function of latitude $\theta$, solar declination $\delta_\odot$, and local time $\tau$; $\Delta T_{LST}$ is an altitude-dependent ($h$) local solar time correction; $T_{UV}$ is the solar contribution dependent on solar EUV and FUV irradiance as a function of solar indices $\chi$ including X-ray and Lyman-$\alpha$ wavelengths \cite<see review by>{He2018}; and $T_{MA}$ is a global correction resulting from magnetic activity and computed as a function of the Dst index according to an empirical formula provided by \citeA{Burke2009}.

  \subsection{Satellite orbital decay computation}\label{drag_tech}

    Nowadays, observations of LEO satellite drag effects are performed by either ground tracking radar or accelerometers on-board the spacecraft, as for the cases of CHAMP and GRACE (section \ref{neutral_density}). In the case of accelerometers, the focus of this paper, the drag acceleration is given by the expression known as the drag equation
      \begin{equation}\label{drag_equation}
        a_d=-\frac{1}{2}\rho C_D\frac{S}{m}V^2,\,\quad V = |\vec{V}_{s/c}-\vec{V}_{wind}|\,,
      \end{equation}
    where $S$ is the satellite area of contact with the external environment; $C_D$ is the drag coefficient; $m$ is the satellite's mass; $\rho$ is the neutral mass density; and $V$ is the magnitude of the relative speed between the spacecraft velocity $(\vec{V}_{s/c})$ and the local wind velocity ($\vec{V}_{wind}$). The wind velocity corresponds to the sum of the atmospheric rotational velocity and any residual winds at the satellite's location \cite{Marcos2010}. \par

    From all the parameters listed in equation \ref{drag_equation}, the spacecraft's area and mass history are assumed to be well known. The drag acceleration $a_d$ is known to be very accurate as measured by the STAR and Super-STAR instruments (see section 2.3). The relative velocity $V$ can sometimes be a large source of errors because the residual velocities in $\vec{V}_{wind}$ are not very well modeled and observed \cite{Marcos2010,Prieto2014,Zesta2016b}. As a result, the drag coefficient is most of the time the major source of density errors in neutral mass density determinations according to equation \ref{drag_equation}. \par

    The drag coefficient is a unitless macroscopic parameter that depends closely on the energy and momentum transfer between the atmospheric environment and the spacecraft. In addition, $C_D$ depends on the geometric properties of the spacecraft, the spacecraft orientation with respect to the gas flow (angle of atack), and the physical properties of the surfaces of contact \cite{Moe2005,Marcos2010,Zesta2016b}. In a review by \citeA{Vallado2014}, the authors pointed out that for LEO satellites such as CHAMP and GRACE $C_D$ can acquire values between 2 and 4. According to equation \ref{drag_equation}, this difference could lead to an error of 100\% in the measurement of neutral densities considering only errors introduced by the drag coefficient. \par

    Drag coefficients are usually obtained by numerical models that simulate the interaction between the atmosphere and the satellite through energy and momentum exchange \cite{Moe2005,Prieto2014}, and by fitting orbital drag parameters during operations \cite<e.g.,>{McLaughlin2011}. In this study, we use the drag coefficients calculated by \citeA{Sutton2009b}. He employed normalized coefficients specifically for the case of satellites with elongated shapes such as CHAMP and GRACE. By using the assumption of incident molecular flow under the regime of random thermal motion with diffuse reemission, \cite{Sentman1961,Moe2005}, \citeA{Sutton2009b} compared CHAMP data with High Accuracy Satellite Drag Model (HASDM) data \cite{Storz2002} and was able to reduce the density errors by nearly 36\% when these assumptions were not considered. \par

    The satellite storm-time orbital decay rate (ODR) is calculated according to the expression \cite{Chen2012}
      \begin{equation}\label{dadt}
        \frac{da}{dt}=-C_D\frac{S}{m}\sqrt{GM\langle a\rangle}\Delta\rho\,,
      \end{equation}
    where, respectively for each satellite, $C_D$, $S$ and $m$ are the same parameters shown in equation \ref{drag_equation}. In this equation, the plate surface areas are the projected values perpendicular to the in-track direction \cite{Doornbos2012,Krauss2012}. Both satellites' relevant areas and mass histories are obtained from \citeA{Bruinsma2003} and \citeA{Bettadpur2007}, respectively. $\Delta\rho$ = $\rho$ $-$ $\rho_0$ is the difference between the observed and modeled background densities (see section \ref{24Aug2005}). $G$ = 6.674$\times10^{-11}$ m$^3\cdot$kg$^{-1}\cdot$s$^{-2}$ is the gravitational constant, and $M$ = 5.972$\times10^{24}$ kg is the Earth's mass. $\langle a\rangle$ is the daily-averaged semi-major axis computed by equation \ref{a} and shown in Figure \ref{semi_major} for the particular time of observation. \par

    Since the background density effects have been removed, the storm-time orbital decay (SOD) is then computed by integrating ODR over time along the satellite's path as
    \begin{equation}\label{d}
      d(t)=\int\limits_{t_1}^{t_2}a(t)dt\,,
    \end{equation}
    where $t_1$ and $t_2$ are arbitrary times taken around some specific CME forcing event (CME impact time or onset of a magnetic storm main phase). \par

    In the next section we use the 24 August 2005 magnetic storm event to illustrate our methodology. Later on, we show results of superposed epoch analyses for CHAMP and GRACE to investigate satellite orbital drag effects and discuss the performance of the JB2008 model in predicting satellite orbital drag during our CME-driven storms.

  \section{Methodology}\label{24Aug2005}

    \subsection{The 24 August 2005 magnetic storm as an example}

\begin{figure*}[t]
        \centering
        \includegraphics[width=1.00\textwidth]{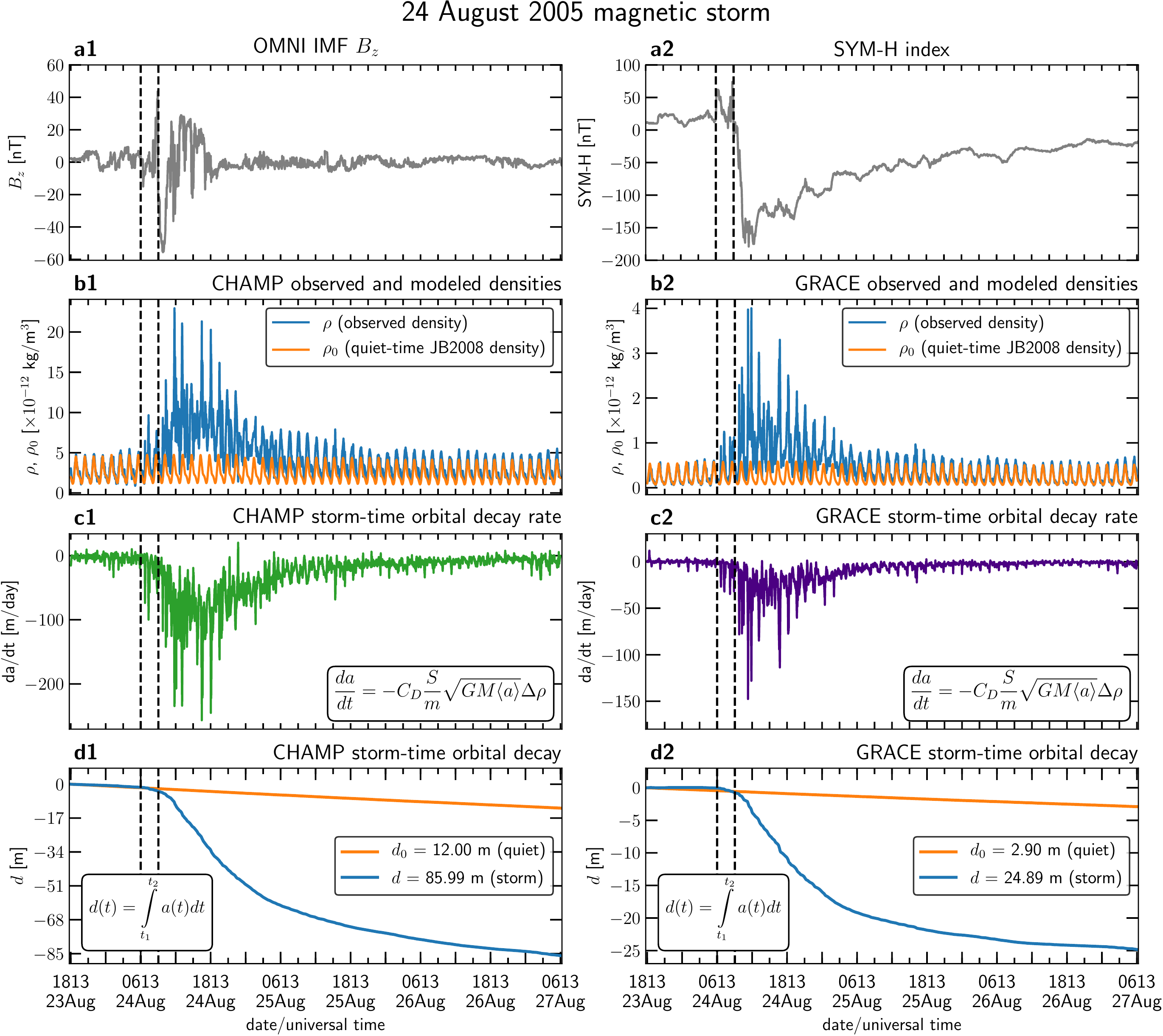}
        \vspace*{-0.5cm}
        \caption{IMF $B_z$, SYM-H, and CHAMP/GRACE neutral mass density data for the 24 August 2005 magnetic storm. In all panels, the first dashed vertical line corresponds to the CME-shock impact, whereas the second dashed vertical line indicates the storm main phase onset. (a1), IMF $B_z$ component; (a2), SYM-H index, both in nT. (b1-d1) and (b2-d2) indicate observed densities and quiet-time JB2008 densities, storm-time orbital decay rates ($da/dt$, equation \ref{dadt}), in m/day, and storm-time orbital decay ($d$, equation \ref{d}), in m, for both CHAMP and GRACE, respectively.}
        \label{decay_example}
      \end{figure*}

      Figure \ref{decay_example} documents results for the 24 August 2005 magnetic storm. Modeled and observed densities, as well as decay rates and storm-time orbital decays are shown for CHAMP (left column), and GRACE (right column). The top row shows results of IMF $B_z$ (Figure \ref{decay_example}a1), and SYM-H (Figure \ref{decay_example}a2). The horizontal axes indicate date and universal time. \par

      The first dashed vertical line indicates the time of CME/shock impact (leading edge), whereas the second one represents the time in which IMF $B_z$ abruptly turns southward (magnetic material). The first event is associated with the sudden impulse whose well-known signature is a sharp increase in ground magnetic measurements \cite{Oliveira2018b,Rudd2019}, while the second event usually coincides with the storm main phase onset, when Dst/SYM-H measurements become highly depressed due to ring current energization \cite{Gonzalez1987,Gonzalez1994}. Although compressions resulting from dynamic pressure enhancements can increase high-latitude neutral density measurements \cite{Shi2017,Ozturk2018}, storm time heating is responsible for larger and global effects on the thermospheric neutral mass density \cite<e.g.,>{Krauss2015,Oliveira2017c}. \par

      Before the CME/shock impact, IMF $B_z$ varies between $\pm$5 nT. During the shock impact, $B_z$ decreases to almost $-20$ nT. After the shock at 0613 UT, $B_z$ increases to values larger than 40 nT. That is the moment of IMF $B_z$ southward turning, at 0915 UT, when it abruptly turns from large positive values down to almost $-60$ nT. $B_z$ stays negative during the next 3 hours or so, reaching strong positive values, with a few excursions to negative values in the next 9 hours. Then, $B_z$ returns to the similar behavior it showed before the CME arrival. At the moment of shock impact, a sudden impulse signature is evident from the SYM-H profile. Then, at the end of the storm main phase, SYM-H arrives at the minimum value of $-178$ nT, which falls into the G4 (severe) storm category as shown in Table \ref{table}. The storm main phase lasted for only 3 hours during that storm, and the ring current took a few days to recover from that CME's driving effects. \par

      The techniques used to obtain the modeled quiet-time density used in this work have been described by \citeA{Oliveira2017c} and \citeA{Zesta2018b}, but here we provide a brief explanation. The ratio between observed and modeled background neutral density for the regime of low-magnetic activity (excluding storm and compression effects, i.e., $|$SYM-H $<$ 30nT) is fitted to a polynomial expansion to the 15$^{th}$ degree. The resulting fit function $f(t)$ is then interpolated and multiplied by the background JB2008 density with the term $T_{MA}$ in equation \ref{temperature} set to zero to exclude magnetic activity effects. The modeled quiet-time density is then the product between $f(t)$ and the background JB2008 density. \par

      The modeled quiet-time density, here represented by $\rho_0$, is indicated by the orange lines in Figure \ref{decay_example}b1 for CHAMP, and in Figure \ref{decay_example}b2 for GRACE. The solid blue lines indicated in both panels represent the CHAMP and GRACE observed density data, respectively. \par

      Before the CME arrival, the observed and the modeled densities agree remarkably well. This assures that the density ratio $\rho/\rho_0$ and the density difference $\Delta\rho=\rho-\rho_0$ are as much close to 1 and zero as possible, respectively. After the CME/shock arrival, due to CME forcing, particularly during the storm main phase, $\rho/\rho_0$ $>$ 1 and $\Delta\rho$ $>$ 0 indicate density enhancements due to magnetic activity only. This is evident for the periods between 0613 UT and 0915 UT, due to high-latitude compression effects, and after 0915 UT, due to storm effects, as seen in Figures \ref{decay_example}b1 and \ref{decay_example}b2. \par

      This is subsequently reflected on the ODRs, which are calculated by equation \ref{dadt} and shown in Figure \ref{decay_example}c1 (CHAMP) and \ref{decay_example}c2 (GRACE). The decay rates for both spacecraft oscillate around null values before shock impact in both cases, but they are slightly noisier in the CHAMP case. $da/dt$ starts to enhance for CHAMP and GRACE after the CME impact, but it reaches its minimum value after the storm main phase onset, leading to $da/dt$ = --250 m/day and $da/dt$ = --150 m/day for both spacecraft, respectively. The decay rate is more intense in the CHAMP case because GRACE was at higher altitudes during that storm (averaged 364.22 km and 481.95 km, respectively), leading to larger $\Delta\rho$ in the CHAMP case. Particularly for that storm, large $da/dt$ values for CHAMP were typically two times larger than GRACE $da/dt$ values. \par

      The SOD, as calculated by equation \ref{d}, is shown for CHAMP (Figure \ref{decay_example}d1) and GRACE (Figure \ref{decay_example}d2). In both panels, the solid orange lines indicate the orbital decays ($d_0$) of both satellites if there was no storm occurring, with $d_0$ = 12.00 m for CHAMP, and $d_0$ = 2.90 m for GRACE, at the end of the 72-hour storm interval. In contrast, the solid blue lines show the actual storm contribution to the orbital decay. Before CME arrival, $d(t)$ is close to zero for both spacecraft. It slightly decreases after the CME impact/compression, but it decreases abruptly after storm main phase onset. CHAMP's total decay is $\sim$ --86 m, while GRACE's decay is almost 3.5 times weaker, --25 m. This example clearly shows that storms can produce significant orbital drag effects on LEO satellites, which could be nearly 10 times larger than the drag effects during low or quiet magnetic conditions. Our drag effect results agree remarkably well with the results of \citeA{Krauss2018} for the very same magnetic storm within similar time intervals.\par

  \subsection{Modeled/observed data comparison for the 20 November 2003 magnetic storm} 

    In this subsection, we use the 20 November 2003 magnetic storm to investigate the JB2008 model performance in computing neutral mass density and predicting orbital drag effects. This storm had $-492$ nT as the minimum SYM-H value, and is hitherto the most extreme magnetic storm in the era of high-precision accelerometers on-board LEO satellites. \par
    
    \begin{figure}[t]
      \centering
      \includegraphics[width=1.00\textwidth]{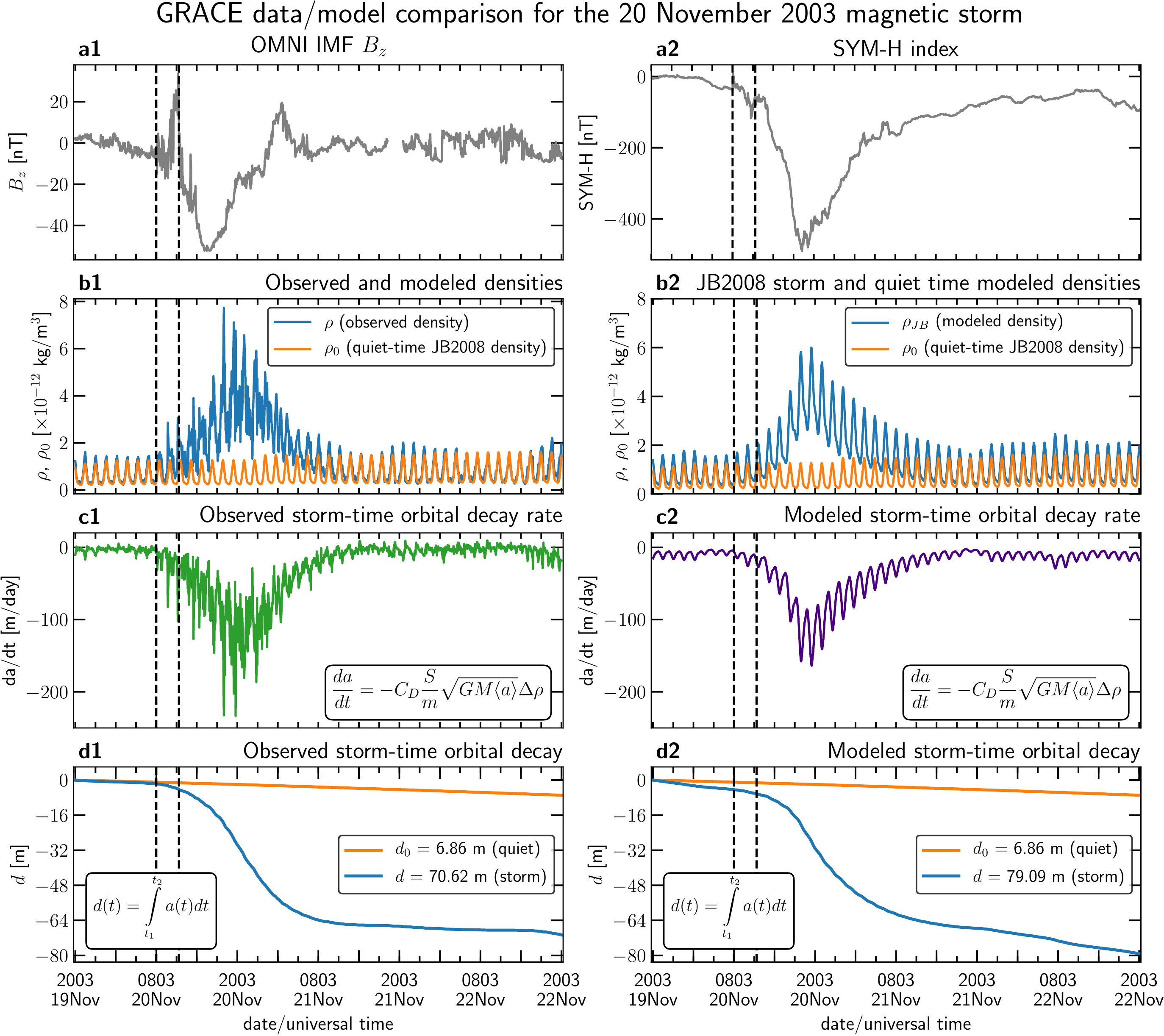}
      \caption{The same as in Figure \ref{decay_example}, but for GRACE observed data (left column) and GRACE modeled data (right column) for the 20 November 2003 magnetic storm. This is not only the most extreme storm of SC23, but also the most extreme storm of the era of high-accuracy satellite data provided by the CHAMP and GRACE missions.}
      \label{grace_data_comparison}
    \end{figure}

    The results are shown in Figure \ref{grace_data_comparison}, displayed in the same fashion as in Figure \ref{decay_example}, but with two differences. First, data (left column) and model (right column) results are shown for GRACE only; second, the modeled storm-time density data is computed by using the full magnetic activity contribution to the exospheric temperature provided by the JB2008 model, meaning the term $T_{MA}$ is not zero in equation \ref{temperature}. \par

    The general IMF $B_z$ and SYM-H index dynamics are very similar to the 24 August 2005 magnetic storm. Minimum $B_z$ (Figure \ref{grace_data_comparison}a1) reached almost $-60$ nT in both storms, but it stayed southward-oriented for nearly 12 hours in the 2003 storm. This reflects on the duration of the main phase of the 2003 storm since it lasted longer and the minimum SYM-H value was much lower than in the case of the 2005 storm (Figure \ref{grace_data_comparison}a2); however, the ring current during the 20 November 2003 storm recovered faster in comparison to the 24 August 2005 magnetic storm. This result is consistent with the works of \citeA{Aguado2010} and \citeA{Cid2013}, who performed modeling and experimental studies to conclude that, in general, the more intense storms are associated with faster magnetospheric recovery times. \par

    Figure \ref{grace_data_comparison}b1 shows the same trends as seen for the observed cases for CHAMP and GRACE in Figure \ref{decay_example}b1 and \ref{decay_example}b2. However, the modeled density shown in Figure \ref{grace_data_comparison}b2 shows a slightly different behavior. In spite of the fact that JB2008 captured the density dynamics remarkably well, it overestimated densities during quiet times, underestimated densities during storm times, particularly the largest values, and overestimated densities again during storm recovery phase. The density overestimation of the pre-storm data may be due to carbon dioxide (CO$_2$) cooling effects during quiet times \cite{Mlynczak2014}. On the other hand, the density underestimation during storm times may be due to the inability of the model to capture the spikes, but corrections to the exospheric temperature estimation may alleviate this deficiency \cite{Sutton2018}. Finally, the density overestimation during storm recovery phase may be a result of the super-production of nitric oxide (NO), that has a high cooling power effect on the thermosphere through infrared emissions \cite{Kockarts1980,Mlynczak2003,Knipp2017a}. These effects are also seen in the observed and modeled decay rates (Figures \ref{grace_data_comparison}c1 and \ref{grace_data_comparison}c2), where the most enhanced values of $da/dt$ computed by the JB2008 model can be 100\% smaller than the observed $da/dt$ values. \par

    The SOD for GRACE during the 20 November 2003 storm was --70.62 m during the 60-hour storm interval shown in Figure \ref{grace_data_comparison}d1. This SOD also agrees well with the results of \citeA{Krauss2015} for the same event. This is more than 10 times larger than the orbital decay considering the same conditions during quiet times ($d_0$ = --6.85 m). In contrast, as shown in Figure \ref{grace_data_comparison}d2, the modeled SOD is --79.09 m, which shows a moderate overall agreement with the observations, with relative error of $\sim$ 12\%. This is the result of the error superposition produced by the overestimation and underestimation intervals discussed above. This result shows that our method applied to long-term modeled orbital decay computations produces good results for predicting LEO satellite orbital decay during the most extreme magnetic storm in our data set, but it is not appropriate to reproduce the storm hour-by-hour dynamics correctly. This suggests the model needs to be updated in order to predict and forecast satellite orbital decay in short-time intervals with more precision.

  \section{Statistical results}\label{results}

    In this paper, we use the technique of SEA to investigate statistical patterns and dynamics of ODRs and SODs during magnetic storms with different intensities. The SEA technique is used when a quantity is highly dynamic in a variable system. Apparently this technique was introduced in our community by \citeA{Chree1913}, who superposed SSN frequency and the daily range of magnetic declination observed at the Kew Observatory. Here, we superpose satellite orbital drag effect variations by lining up the density data with respect to the storm main phase onset for each storm. This choice of zero epoch time (ZET) coincides with the IMF $B_z$ southward turning similar to those seen in Figures \ref{decay_example} and \ref{grace_data_comparison}. \par

    However, as pointed out by a reviewer, mixing together events where a satellite's orbit varies significantly from event to event can cause unintended weighting; storms occurring when the satellites are at lower altitudes (later in the satellite mission) are overemphasized while those occurring when the satellite is at higher altitudes (earlier in the satellite mission) are underemphasized in the average. Although this may be a caveat, we will proceed with the analysis given the limited number of events with high-quality density data, particularly the ones occurring during the most extreme magnetic storms.

    \subsection{Superposed epoch analysis - storm-time orbital decay rate (ODR)}\label{section_dadt}

      \subsubsection{CHAMP storm-time ODR results}

\begin{figure*}[t]
          \centering
          \includegraphics[width=1.00\textwidth]{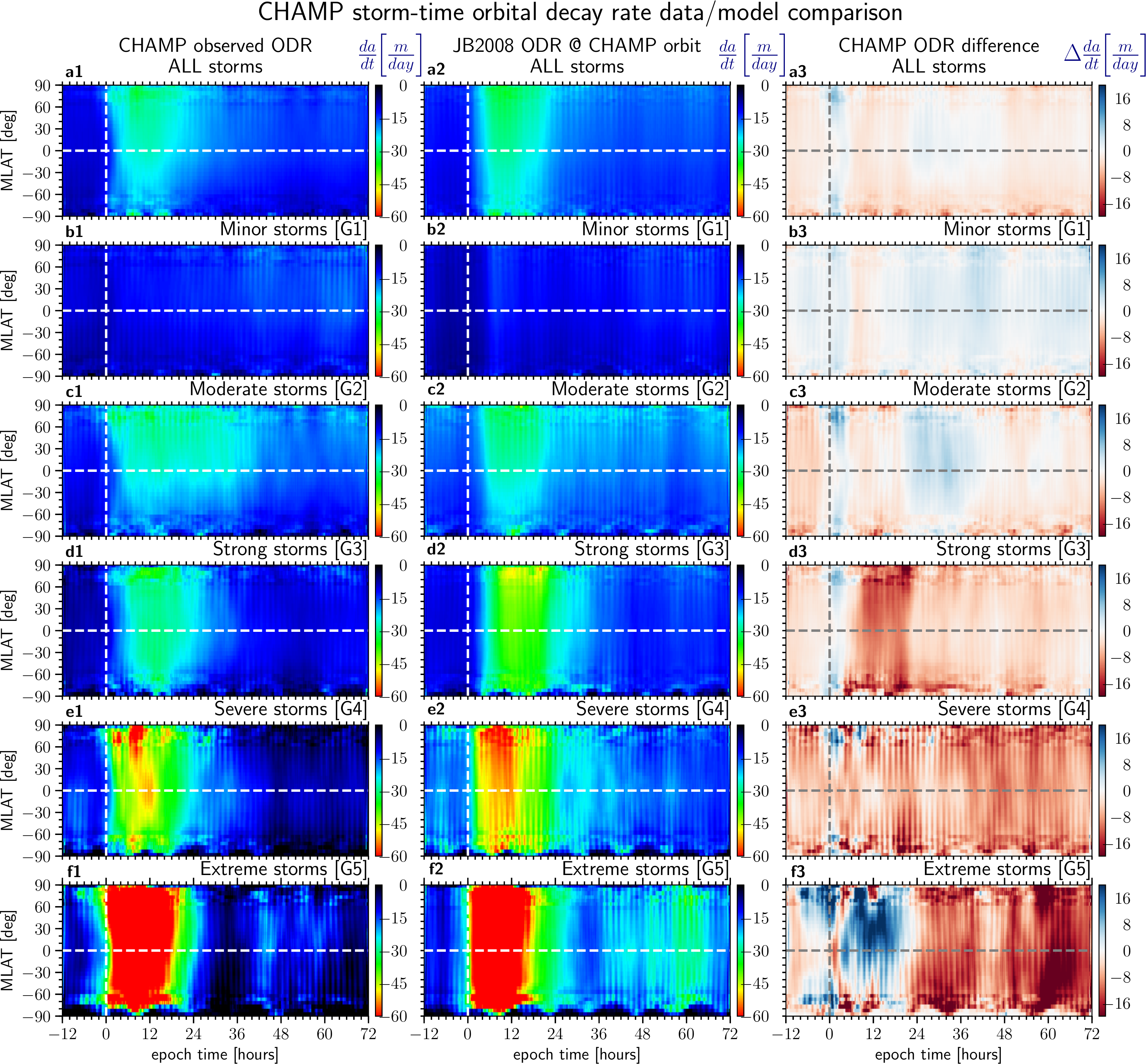}
          \caption{Results of superposed epoch analysis of CHAMP storm-time orbital decay rate $da/dt$, in m/day, calculated according to equation \ref{dadt}. Left column, observed $da/dt$; middle column, JB2008 $da/dt$; and right column, difference $\Delta X$ = $X_{mod}$ $-$ $X_{obs}$, with $X$ = $da/dt$. First row indicates results for all events, while the second to the sixth rows indicate results for storms with increasing intensities (Table \ref{table}). The latitudinal data are binned in 5-degree bins, whereas the epoch time data are binned in 0.25-hour bins.}
          \label{champ_dadt}
        \end{figure*}

        Figure \ref{champ_dadt} shows results for the storm-time ODRs associated with the CHAMP satellite. The left column indicates results for the observed $da/dt$ (as discussed for the 24 August 2005 magnetic storm), and the middle column represents results for the modeled $da/dt$ (as discussed for the 20 November 2003 magnetic storm). The right column shows the difference $\Delta X$ = $X_{mod}$ $-$ $X_{obs}$, with $X$ = $da/dt$, between the modeled and observed ODRs. This means that if $\Delta X$ $>/<$ 0, the model underestimates/overestimates ODR, respectively. The first row shows SEA results for all storms, whereas the subsequent rows show results for the storm categories with increasing intensities (Table \ref{table}). All panels in Figure \ref{champ_dadt} are centered at 12 hours and 72 hours around ZET, marked by the dashed vertical lines. The vertical axes indicate magnetic latitude (MLAT), in degrees. MLAT data are allocated in 5-degree bins, whereas epoch time data are allocated in 15-minute bins. \par

        For all storms, observed $da/dt$ in Figure \ref{champ_dadt}a1 shows similar patterns as reported by \citeA{Oliveira2017c} for latitudinal thermosphere heating: observed $da/dt$ is first enhanced at high latitudes and later enhanced at equatorial latitudes within approximately 3 hours. This is a result of heating propagation from auroral regions towards very low latitude regions due to global wind surges and TADs \cite{Fuller-Rowell1994,Bruinsma2007,Sutton2009a,Oliveira2017c}. $da/dt$ enhancements are slightly stronger in the Northern Hemisphere (NH) because of altitudinal effects introduced by the inclined orbit of CHAMP. ODRs remain moderate ($\sim$ --30 m/day) until 18 hours after ZET, when the thermosphere cools first in the Southern Hemisphere (SH). As shown in Figure \ref{champ_dadt}a2, the model fails to reproduce the latitudinal patterns of the $da/dt$ enhancements seen in the previous description and reported by \citeA{Oliveira2017c}. As a result, JB2008 underestimates $\Delta X$ by approximately --12 m/day during 0 $<$ t $<$ 3 hours, particularly at high latitudes, and later overestimates $\Delta X$ by the same amount, with some weak underestimation around t = 24 hour and t = 42 hours (Figure \ref{champ_dadt}a3). \par

        The satellite drag effects for the minor storms (G1) are very weak, as seen for the observed (Figure \ref{champ_dadt}b1) and modeled (Figure \ref{champ_dadt}b2) ODRs. Figure \ref{champ_dadt}b3 shows that JB2008 overestimates $da/dt$ ($\Delta X$ around 6 m/day) for the G1 storms, with exception for an underestimation interval ($\Delta X$ around --6 m/day) between t = 6 hours and t = 12 hours. Figure \ref{champ_dadt}c1 shows that the moderate observed $da/dt$ resembles the $da/dt$ behavior for all storms until approximately t = 18 hours, but it remains enhanced for more 18 hours mainly in the NH. The model results look similar, but the latitudinal heating distribution does not appear in the model results (Figure \ref{champ_dadt}c2). There is some overestimation within 3 hours after ZET for the G2 storms (Figure \ref{champ_dadt}c3), but the overall model performance is of overestimation, with some worse points at both hemispheres' high latitudes, particularly in the SH. However, orbital decay rates tend to persist longer in this category. For the case of strong storms (G3, Figure \ref{champ_dadt}d1), the observed ODR is strong between 0 $<$ t $<$ 22 hours, particularly in the NH. The modeled $da/dt$ levels stay strong until t = 26 hours, but are stronger than the observed results particularly in the NH (Figure \ref{champ_dadt}d2). The model does not reproduce latitudinal heating propagation due to TADs, and the model overestimates ODR during 6 $<$ t $<$ 22 hours, and later particularly for the SH high latitudes (Figure \ref{champ_dadt}d3). Observed and modeled $da/dt$ results for the severe storms (G4) are similar to the G3 storm results, but are clearly more intense, particularly in the NH (Figures \ref{champ_dadt}e1 and \ref{champ_dadt}e2). Clearly G4 storm orbital drag effects are overestimated by JB2008, and $\Delta X$ $<$ --10 m/day in all latitudes with $\Delta X$ $<$ --16 m/day appearing every other 6 hours  (Figure \ref{champ_dadt}e3). \par

        The results shown for the G5 (extreme) storms in the last row of Figure \ref{champ_dadt} indicate a significantly distinct behavior in comparison to all storms in the other groups, or for all storms combined. Some observed $da/dt$ enhancements are seen at high and middle latitudes before ZET (Figure \ref{champ_dadt}f1) due to the CME/shock impacts preceding the onset of the storm main phases \cite{Oliveira2017c,Shi2017,Ozturk2018}. In addition, right at t = 0, ODRs are strongly enhanced at high- and mid- latitude to values much more intense than those shown in the other storm intensity categories, with $da/dt$ $<$ --60 m/day. $da/dt$ becomes intensely enhanced at equatorial latitudes due to TAD propagation, which occurs in less than 1.5 hour, half the average time of 3.0 hours found by \citeA{Oliveira2017c} for all storms. As for the modeled $da/dt$, Figure \ref{champ_dadt}f2, the model satisfactorily reproduces the observed $da/dt$ behavior until t = 18 hours (except the TAD propagation after ZET), but the model overestimates orbital drag effects throughout the storm recovery phase. Perhaps the model, due to its statistical and empirical nature, did not capture the overall heating by TAD propagation as seen in the observation case because the model uses the Dst index, with time resolution of 1 hour, for the magnetic activity contribution (equation \ref{temperature}). \par

        \begin{figure*}[t]
          \centering
          \includegraphics[width=1.00\textwidth]{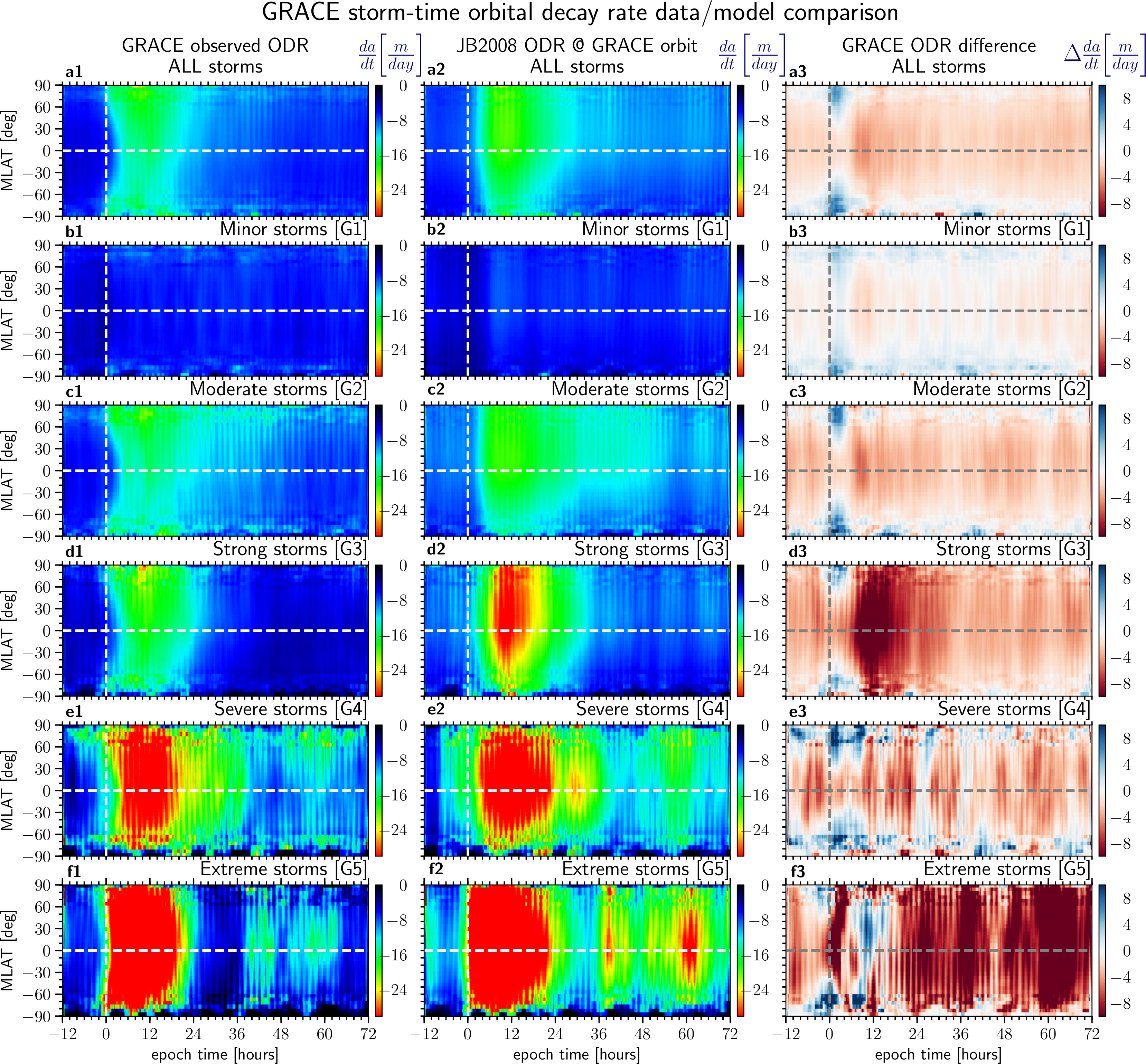}
          \caption{Superposition of GRACE storm-time orbital decay rate $da/dt$ performed in the same fashion as in Figure \ref{champ_dadt}.}
          \label{grace_dadt}
       \end{figure*}

        Yet for the extreme events, observations show a first cooling of the thermosphere at around t = 18 hours in the SH and 3 hours later in the NH, where $da/dt$ increases from less than --60 m/day to --40 m/day. At t = 24 hours, a secondary cooling occurs in both hemispheres, and slightly later at middle and low latitudes in the NH, leading the thermosphere to $da/dt$ levels similar to pre-storm levels. This is probably due to enhancements of NO at CHAMP's altitude, leading to a very fast and effective thermostat effect resulting from the competition between the storm time heating and the NO cooling \cite{Mlynczak2003,Knipp2017a,Zesta2018b}. However, this sudden cooling is not evident in the JB2008 results. The model in fact overestimates $da/dt$ particularly after t = 23 hours, but the cooling occurs more symmetrically in all latitudes. The comparison between observation and model results (Figure \ref{champ_dadt}f3) shows that the model underestimates ODR  during 0 $<$ t $<$ 12 hours, slightly underestimating $da/dt$ at middle and low latitudes and some overestimation at high latitudes. However, the model significantly overestimates ODRs ($\Delta X$ $<$ --18 m/day) at high latitudes around t = 20 hours and almost at all latitudes at t $>$ 24 hours. This clearly shows that the model cannot capture NO cooling effects during the recovery phases of extreme storms at this time \cite{Bowman2008,Knipp2013,Knipp2017a,Zesta2018b}, although early efforts have been undertaken in order to capture JB2008 averaged cooling effects for CHAMP and GRACE for larger time spans \cite{Weimer2010}. \par

        \subsubsection{GRACE storm-time ODR results}

          Figure \ref{grace_dadt} shows ODR results for GRACE plotted in the same way as in Figure \ref{champ_dadt}. One of the main differences corresponds to the smaller $da/dt$ levels (around half) in comparison to CHAMP, because GRACE was always at higher altitudes than CHAMP. This confirms results by \citeA{Chen2012} and \citeA{Krauss2018}. In this case the general JB2008 model performance for all storms is worse than the model performance in the CHAMP case (Figures \ref{grace_dadt}a1-a3). JB2008 overestimates orbital drag effects at almost all latitudes, being worse at mid and low latitudes. Even for the minor storms the model performance is worse (Figure \ref{grace_dadt}b3). Figure \ref{grace_dadt}c1 shows that the NH $da/dt$ of GRACE G2 storms persist longer in the NH in comparison to storms in the other categories. Similarly to CHAMP's strong storms (group G3), $da/dt$ during storms in the same group in the case of GRACE are highly (although strongly) overestimated at low and equatorial latitudes after ZET during 6 $<$ t $<$ 18 hours. For the severe storms (G4), the model overestimates ODRs in low/middle high latitudes and underestimates ODRs in high latitudes (Figures \ref{grace_dadt}e1-e3). \par

          In all cases for GRACE so far, $da/dt$ latitudinal heating caused by TAD propagation is poorly reproduced by the model. That is the same for the extreme (G5) storms. The data show rapid and intense heating propagation from auroral to equatorial zones in a similar way as for the CHAMP case, but the thermosphere cooling effects (t = 18 hours) are slightly different. As shown in Figure \ref{grace_dadt}f1, the observed $da/dt$ decreases first at high latitudes being followed by mid and low latitudes within 1.5 hour. The observed results look very similar to the JB2008 results at that time. However, ODRs are overestimated between 36 hours and 46 hours and again between 48 hours and 66 hours, with highly overestimated $da/dt$ levels at t = 62 hours at low latitudes. As a result, as seen in Figure \ref{grace_dadt}f3, the $da/dt$ difference for the G5 storms are very high after t = 20 hours for all latitude regions, in addition to high overestimation during 0 $<$ t $<$ 3 hours at high latitudes particularly in the NH. Again, the $da/dt$ overestimation during the recovery phases of the extreme storms observed by GRACE result from the lack of NO cooling effects computed by the model \cite{Knipp2013,Knipp2017a}. \par  

          \subsection{Superposed epoch analysis - storm-time orbital decay (SOD)}\label{section_total_decay}

          \subsubsection{CHAMP SOD results}

\begin{figure}[t]
              \centering
              \includegraphics[width=1.00\textwidth]{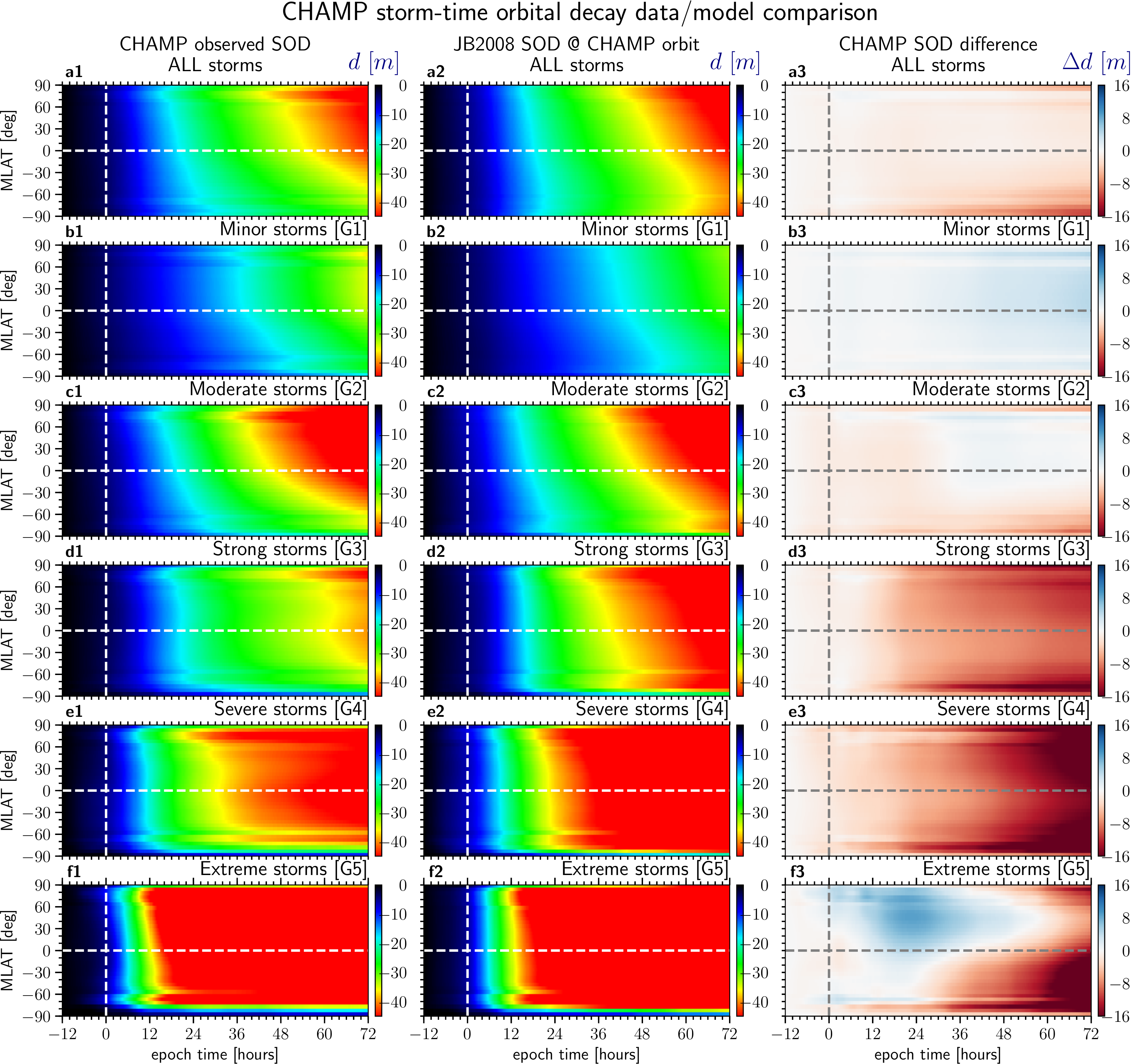}
              \caption{Results of CHAMP superposed epoch analysis for the storm-time orbital decay $d(t)$ (equation \ref{d}), in m. All panels correspond to similar analyses as in Figure \ref{champ_dadt}. Since the time resolution of the data is the size of each bin, or 15 minutes, $d(t)$ in each latitudinal bin corresponds to the temporal cumulative sum along each latitude bin.}
              \label{champ_d}
            \end{figure}

            SODs as calculated from equation \ref{d} are shown for the CHAMP SEA in Figure \ref{champ_d}. The general format, time range, and data binning are the same as in Figure \ref{champ_dadt}. Since the time intervals are the same for all latitude bins, $d$ is the cumulative sum of $da/dt$ along all latitude bins over time. That is the reason why SODs decrease in all latitude regions as time progresses. \par

            For all storms (first row), observations show that significant orbital decay ($d$ $<$ --30 m) occurs first at high latitudes in the NH around t = 14 hours and then gradually intensifies towards SH high latitudes in the next 6 hours or so (Figure \ref{champ_d}a1). This results from the general high NH levels shown by $da/dt$ in Figure \ref{champ_dadt}a1 which lead to high $d$ values at high latitudes. As shown in Figure \ref{champ_d}a2, the model performs relatively well for all storms, with low difference $\Delta d$ levels ($\sim$ --6 m) in SH high latitudes during recovery phase (Figure \ref{champ_d}a3). Figures \ref{champ_d}b1-b3 show that, in general, the same occurs for the minor (G1) storms, but total decay levels are smaller and occur later. However, the model mostly underestimates SODs during storm and recovery times. The overall behavior of storms in the moderate (G2, Figure \ref{champ_d}c1) category resembles the general behavior of all storms, with $d$ being more intense later during storm recovery phase, particularly at the NH high latitudes. The model $\Delta d$ performance is very similar to the model $\Delta d$ for all storms (Figure \ref{champ_d}a3). In the case of strong (G3) storms, the model overestimates SODs particularly at the NH high latitudes. As seen in Figure \ref{champ_d}e1, the severe (G4) storms are highly overestimated by JB2008. Data observations show that $d$ reaches --30 m around t = 12 hours, and $d$ near --40 m is reached around t = 18 hours at a small high-latitude range in the NH and at other latitudes around 26 hours. In contrast, the model overestimates $d$ for these storms, where they reach values smaller than --60 m around t = 12 hours first at NH high latitudes and later increasingly at lower latitudes. The overall $\Delta d$ is relatively high for severe storms observed by CHAMP. The observed moderate storm effects are more intense than the observed strong events because $da/dt$ stayed enhanced longer in the moderate storm group as shown in Figure \ref{champ_dadt}c3. \par

            \begin{figure}[t]
              \centering
              \includegraphics[width=1.00\textwidth]{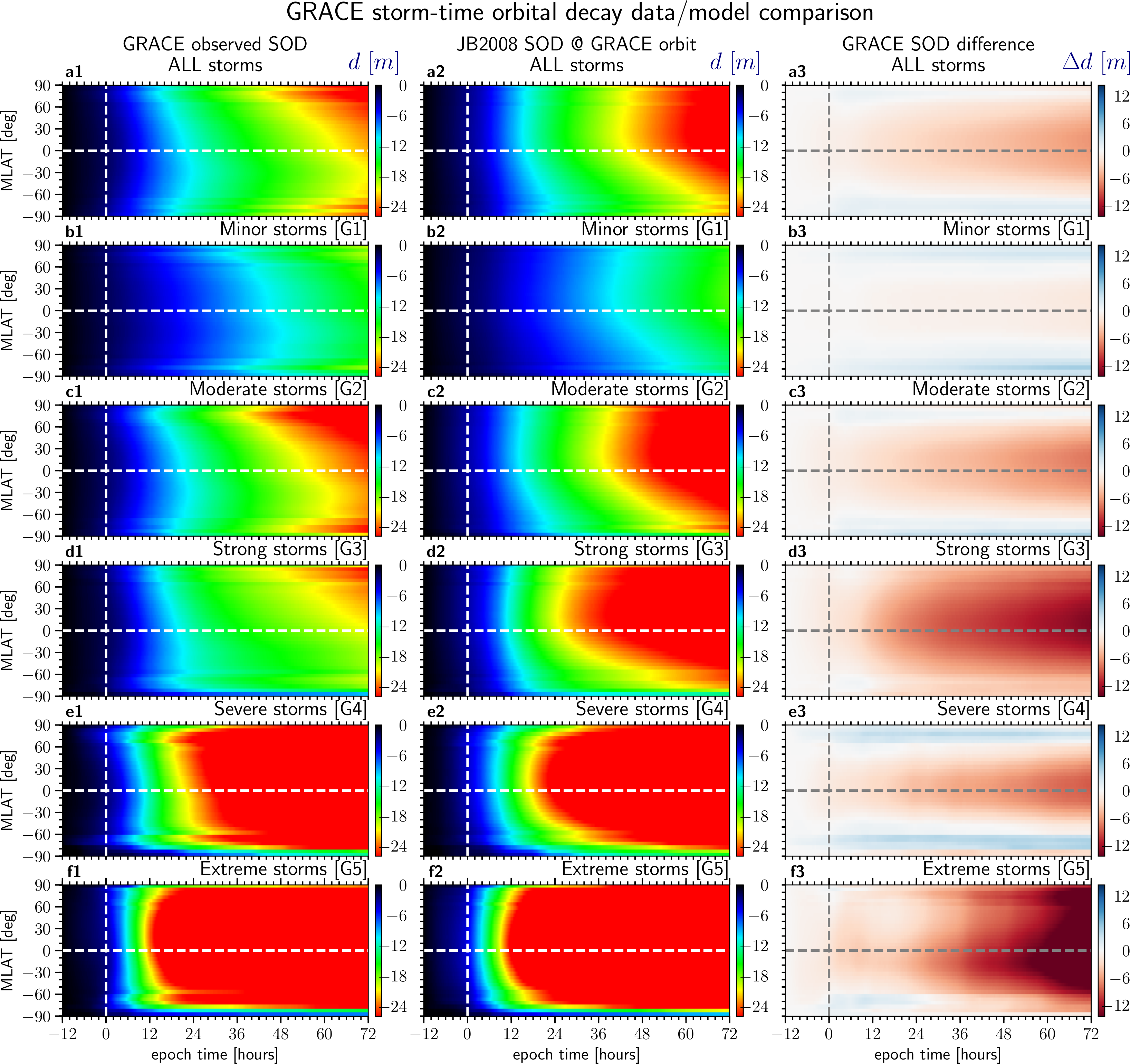}
              \caption{Superposition of GRACE storm-time orbital decay $d(t)$ performed in the same fashion as in Figure \ref{champ_d}, and calculated from data plotted in Figure \ref{grace_dadt}.}
              \label{grace_d}
            \end{figure}

            As for the extreme storms, data show that satellites decay much faster in comparison to the other storm categories or all storms combined, with the SH being led by the NH by $\sim6$ hours (Figure \ref{champ_d}f1). Observed SOD reaches --30 m and values smaller than --40 m around 3 hours and 12 hours after ZET, and becomes even smaller at all latitudes during storm main phase and storm recovery phase. The model reproduces the general observed $d$ patterns quite well, except for the high-to-low TAD propagation (Figure \ref{champ_d}f2). As a result and shown in Figure \ref{champ_d}f3, JB2008 moderately underestimates SODs between ZET and t = 36 hours, particularly for the NH, while it overestimates SOD in the SH after t = 30 hours, more strongly for t $>$ 60 hours. \par 

          \subsubsection{GRACE SOD results}

            Figure \ref{grace_d} shows the latitudinal distribution for the SODs occurring during storms observed by GRACE. Here, $d$ is calculated and plotted in the same way as in Figure \ref{champ_d}. The GRACE $d$ levels are smaller than the CHAMP $d$ levels as a consequence of the higher GRACE altitude. For all storms, the total observed $d$ response occurs first at high latitudes, being slightly earlier in the NH, and then their effects spread toward low and equatorial latitudes. At least in comparison to the SODs at CHAMP altitudes, the GRACE SODs are more susceptible to energy propagation effects due to TADs at GRACE altitudes. This pattern is similar for all storm categories from G1 to G4 categories, with the satellite decaying faster with increasing storm intensity. The exception is for the moderate storms (Figure \ref{grace_d}c1), which show more intense $d$ values in comparison to strong storms (Figure \ref{grace_d}d1). However, the model fails to capture TAD effects in these categories because total orbital decay becomes more intense at low and middle latitudes before the high latitudes. As a result, the SOD differences are high in the low latitude regions, particularly for the moderate (G2) and strong (G3) storm categories (Figures \ref{grace_d}c3 and \ref{grace_d}d3). \par

            \begin{figure}[t]
              \centering
              \includegraphics[width=0.80\textwidth]{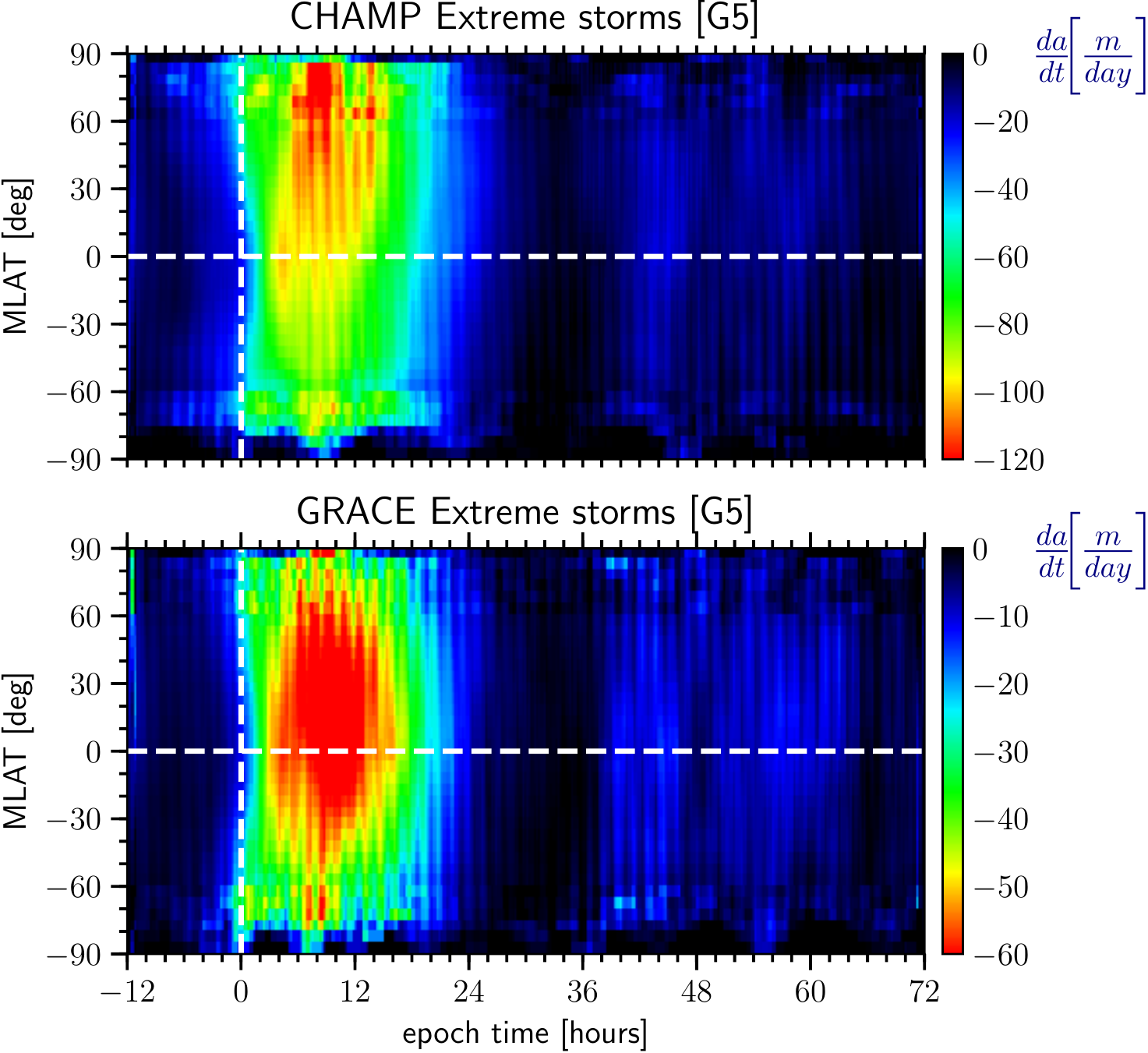}
              \caption{Observed CHAMP (top) and GRACE (bottom) storm-time ODRs plotted similarly to Figures \ref{champ_dadt}f1 and \ref{grace_dadt}f1, respectively, but with increased colorbar ranges. Strong $da/dt$ enhancements are seen at equatorial regions for CHAMP and GRACE, but they are more intense in the case of the latter spacecraft.}
              \label{ppef}
            \end{figure}

            Surprisingly, Figure \ref{grace_d}f1 indicates that the SOD observed by GRACE during extreme storms show a different behavior in comparison to the other storm categories. This behavior is different from the behavior of the extreme storms in the case of CHAMP. $d$ reaches --15 m 6 hours after ZET at the equator and then gradually decreases toward high latitudes of both hemispheres in the following 6 hours (see discussion below). Although the model performance when reproducing SODs is remarkably good between ZET and 24 hours for the extreme storms (Figure \ref{grace_d}f2), later $\Delta d$'s increase at mid- and low latitudes in the SH and at high and middle latitudes in the NH hemisphere (Figure \ref{grace_d}f3). This is a result of density overestimation by JB2008 during recovery phases of extreme storms, possibly associated with the lack of NO cooling effects accounted for by the model, as shown in Figures \ref{decay_example} and \ref{grace_data_comparison}. \par

            \remove{The effect mentioned above is probably due to high levels of prompt penetration electric fields (PPEFs) at the magnetic equator during extreme magnetic storms. PPEFs intensify $\vec{E}\times\vec{B}$ forces which cause the uplift of the dayside equatorial ionosphere, a phenomenon well-known as dayside ``super-fountain effect" (Mannucci et al., 2004; Tsurutani et al., 2007; Verkhoglyadova et al., 2007). Tsurutani et al., (2007) showed with simulations for the 30 October 2003 ``Halloween" storm that dayside ionospheric PPEFs cause rapid oxygen ion motions which in turn cause the uplift of neutral oxygen to altitudes $>$ 400 km from the reference altitude $\sim$ 340 km. Later, Lakhina and Tsurutani (2017) showed that the subsequent and rapid upward of oxygen neutrals and ions to higher altitudes can severely cause satellite drag effects around altitudes as high as 850 km.}

            \begin{figure}[t]
              \centering
              \includegraphics[width=1.00\textwidth]{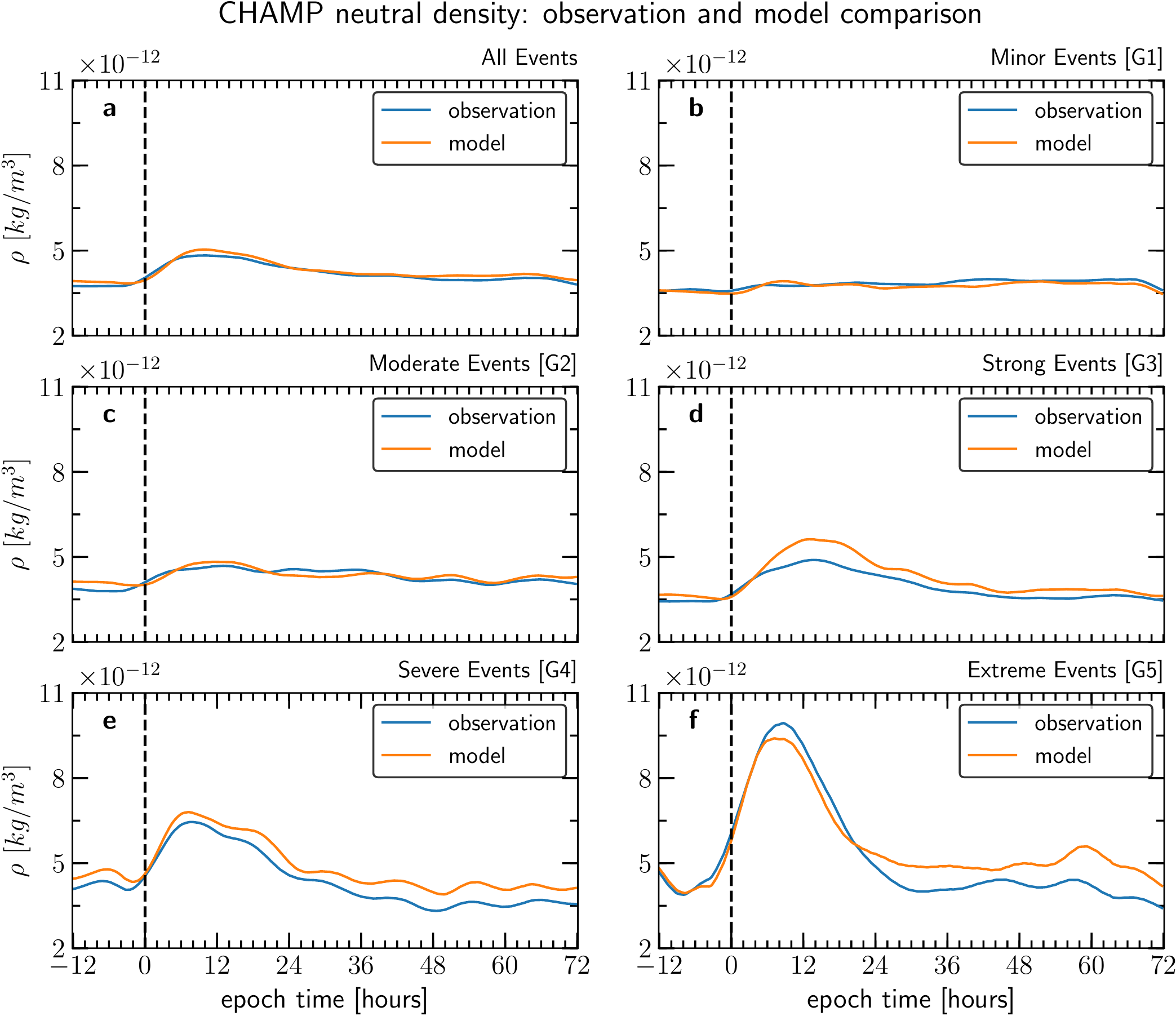}
              \caption{Superposition of observed CHAMP densities (solid blue lines) and for the modeled CHAMP densities (solid orange lines). (a) indicates results for all storms, while (b-f) indicate results for storms in the groups G1-G5 (Table \ref{table}). Results are shown in the time interval ($t_1$, $t_2$) = ($-12$, 72) hours around zero epoch time. The Monte Carlo/Savitzky-Golay smoothing scheme of \citeA{Oliveira2017c} was used in the computation.}
              \label{champ_tot_d}
            \end{figure}

            \remove{In order to properly understand the PPEF effects on CHAMP and GRACE extreme orbital drag, we show their}{The corresponding CHAMP and GRACE $da/dt$ results for the G5 storms (Figures \ref{champ_dadt}f1 and \ref{grace_dadt}f1), plotted for an enlarged colorbar range\change{. This is}{\protect{, are}} shown with more detail in Figure \ref{ppef} for CHAMP (top) and GRACE (bottom). CHAMP ODRs are enhanced first at high latitudes and propagate equatorward within 1.5 hour ($da/dt$ $\sim$ --75 m/day), but stronger ODR enhancements ($da/dt$ $\sim$ --100 m/day) are seen occurring during early storm main phase at equatorial latitudes around the second orbit. In the case of GRACE, this effect is more evident because strong ODR effects ($da/dt$ $<$ --60 m/day) are observed at equatorial latitudes nearly 3 hours after ZET. As a result, this explains the strong $d$ response for GRACE occurring first at equatorial latitudes which are observed first at the NH high latitudes for CHAMP (Figures \ref{champ_d}f1 and \ref{grace_d}f1, respectively). \change{Since the approximate average altitudes for CHAMP and GRACE were, respectively, 390 km and 490 km during the G5 events, most likely these $da/dt$ enhancements are associated with the upward motion of oxygen neutrals caused by rapid oxygen ion motions enhanced by strong PPEFs (Tsurutani et al., 2007; Verkhoglyadova et al. 2007; Lakhina and Tsurutani, 2017).}{\protect{Such signatures are consistent with either the global redistribution of high-latitude Joule heating \cite<see, e.g.,>{Fuller-Rowell1994,Richmond2000,Oliveira2017c} or possibly from the direct uplift of the thermosphere through ion/neutral interactions as postulated by \citeA{Tsurutani2007} and \citeA{Lakhina2017}. However, as pointed by the latter authors, the last hypothesis has yet to be tested by nonlinear simulations that take into account the coupling between gravity, pressure gradients, viscosity, and advection of neutal atom flow effects, along with the heating and expansion during the uplift process.}}

  \subsection{Comparisons of modeled and observed densities}\label{section_error}

    The statistical results for observed (solid blue line) and modeled (solid orange line) neutral mas densities are obtained for CHAMP (Figures \ref{champ_tot_d}) and GRACE (Figure \ref{grace_tot_d}), respectively. They are computed by averaging $\rho$ over all latitude bins for each particular 15-min temporal bin. Then since density assumes always positive values, a Monte Carlo/Savitzky-Golay filter scheme is applied to the obtained time series in order to smooth $\rho$ locally. This method eliminates local spikes and yields more accurate averages for each temporal bin. See details in \citeA{Oliveira2017c}. \par

    Figure \ref{champ_tot_d}a shows results for all CHAMP-observed storms, while (b-f) respectively show results for the same storms in the groups G1 to G5 as shown in Table \ref{table}. The dashed vertical lines indicate ZET, and the time range of the plots spans around 12 hours before and 72 hours after ZET. In general, both spacecraft and JB2008 densities show that peak values occur the earlier the more intense the storm category. In addition, results show that the model follows the same dynamic response as shown by the observations, with the model underestimating and overestimating densities during intervals of 3-6 hours for all storms (Figure \ref{champ_tot_d}a), minor storms (b) and moderate storms (c). \par

    \begin{figure}[t]
      \centering
      \includegraphics[width=1.00\textwidth]{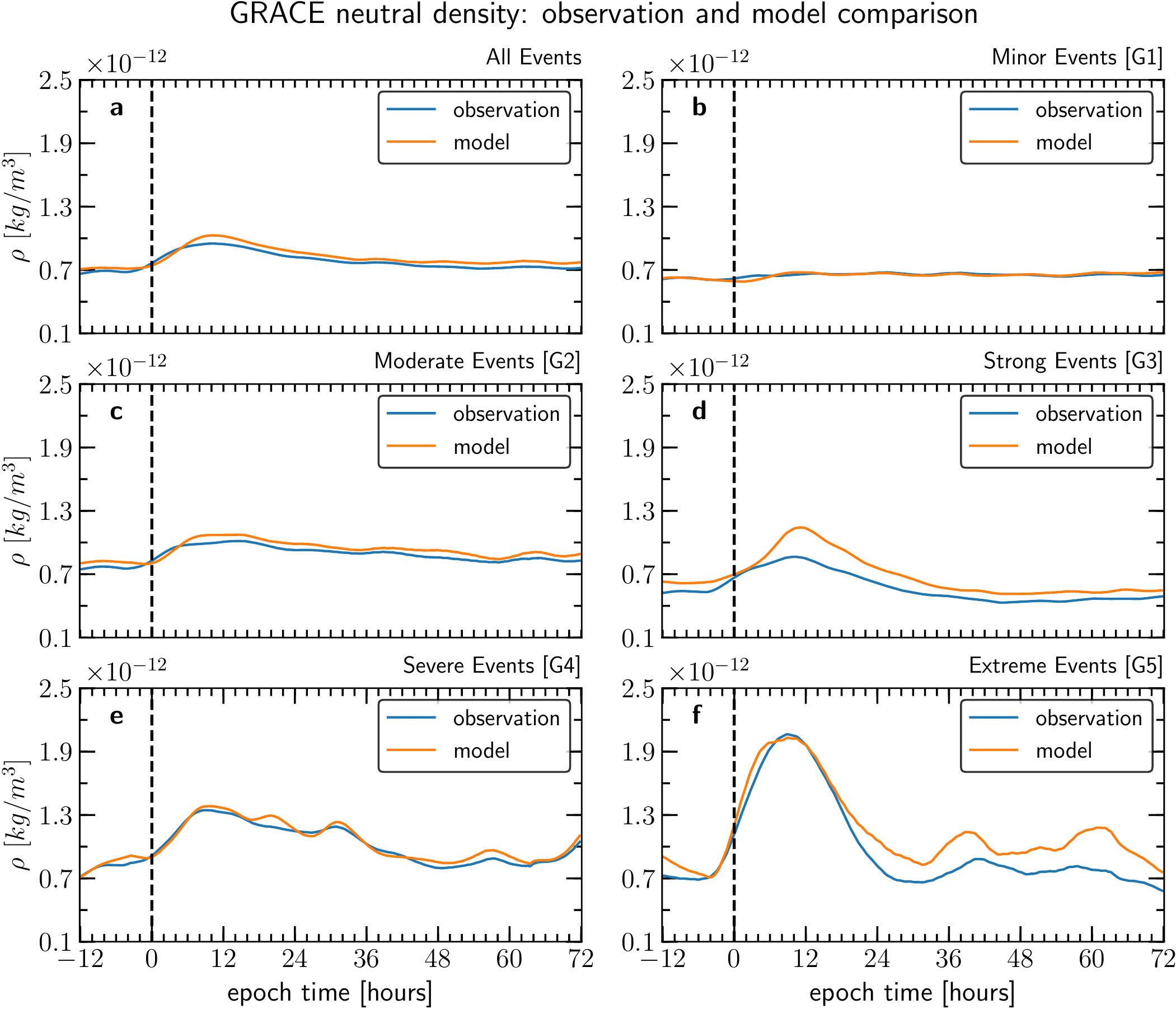}
      \caption{The same as for Figure \ref{champ_tot_d}, but for GRACE.}
      \label{grace_tot_d}
    \end{figure}

    CHAMP results for the strong (G3) storm category, Figure \ref{champ_tot_d}d, show that JB2008 highly overestimated densities in the interval 3 $<$ t $<$ 24 hours, which explains the high ODR levels shown in Figure \ref{champ_dadt}d3 and the large SOD difference for t $>$ 24 hours in Figure \ref{champ_d}d3. For the severe (G4) storms, JB2008 overestimates densities for t $>$ 6 hours and the difference becomes stronger as time progresses. This is the reason why $\Delta d$ becomes stronger during late storm recovery phase (Figure \ref{champ_d}e3). \par

    Extreme (G5) storm results show that the densities computed by JB2008 are smaller than the densities observed by CHAMP during the interval 6 $<$ t $<$ 21 hours, and the situation is inverted after t = 24 hours. This explains the underestimated $d$ values and the overestimated $d$ values for the extreme storms shown in Figure \ref{champ_d}f3. \par

    As represented in Figure \ref{grace_tot_d}, similar trends are seen for GRACE. However, JB2008 density computations seem to agree with observations more in the case of GRACE with respect to the case of CHAMP. This is more evident for the case of severe storms (Figure \ref{grace_tot_d}e), except for the strong storms. Another difference occurs for the extreme (G5) storms, as indicated by Figure \ref{grace_tot_d}f where density is slightly overestimated in the interval 0 $<$ t $<$ 6 hours and then density agrees well with observations in the interval 6 $<$ t $<$ 18 hours. Then, density becomes highly overestimated for t $>$ 18 hours. This modeled density results explain the highly overestimated SOD shown in Figure \ref{grace_d}f3. \par

    \begin{figure}[t]
      \centering
      \includegraphics[width=1.00\textwidth]{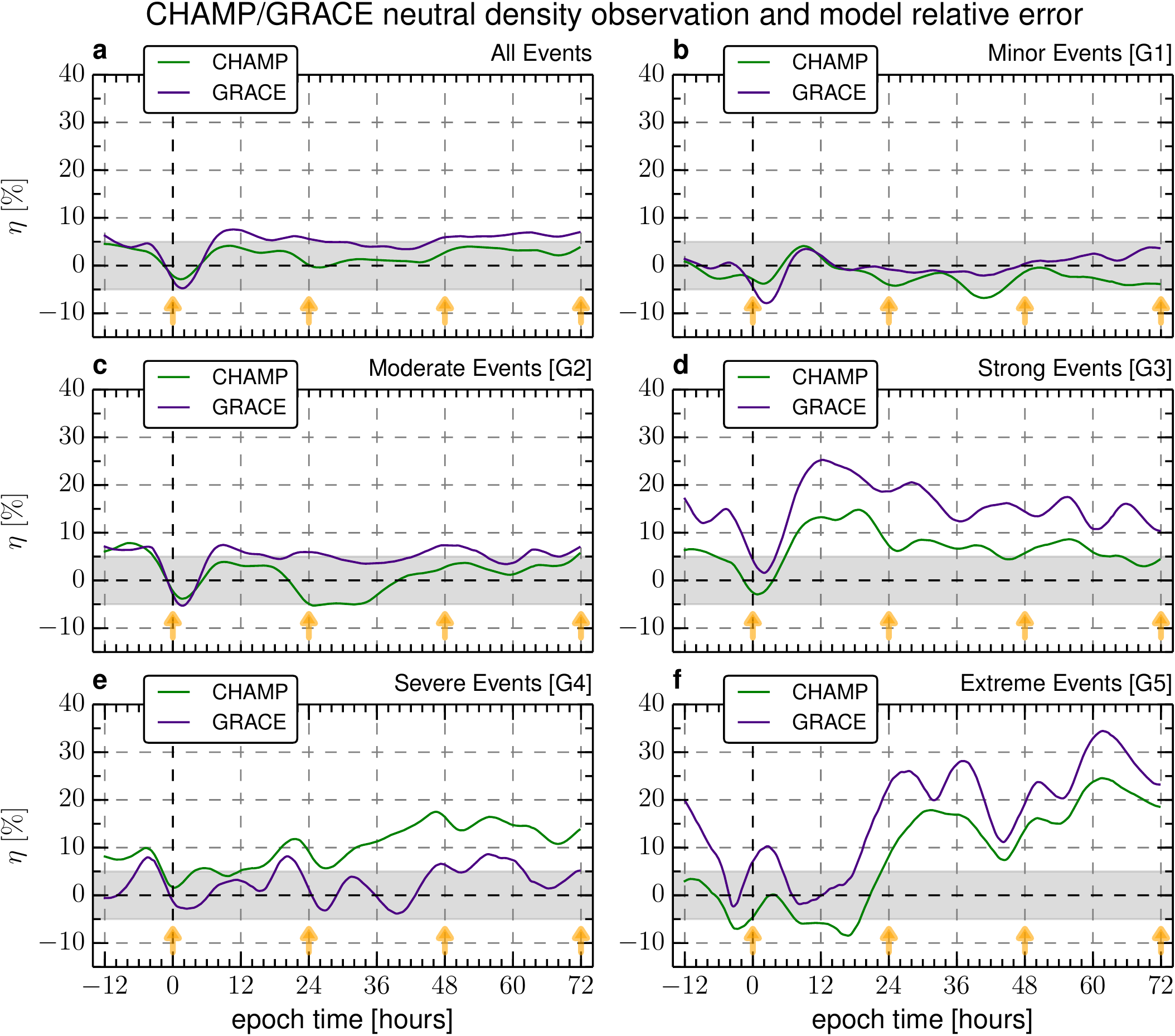}
      \caption{Uncertainties of densities computed by JB2008 for CHAMP (solid green lines) and GRACE (solid purple lines). Panel (a) indicates results for all storms, while panels (b-f) indicate results for the storms G1-G5 shown in Table \ref{table}.}
      \label{data_model_error}
    \end{figure}

    \subsection{Storm-time CHAMP and GRACE uncertainties}

    Now we investigate uncertainties associated with JB2008 when computing densities at CHAMP and GRACE orbits. Figure \ref{data_model_error} shows results for the relative error $\eta$ = ($\rho_{mod}$ - $\rho_{obs}$)/$\rho_{mod}$ of the model/observation comparison calculated from the last two plots for CHAMP (Figure \ref{champ_tot_d}) and GRACE (Figure \ref{grace_tot_d}). The solid green lines indicate CHAMP results, whereas the solid purple lines indicate GRACE results. Panel (a) indicates results for all events, and panels (b-f) indicate results for storms in the categories with increasing storm intensities. The shaded light gray area corresponds to the $\pm$5\% confidence interval (CI) recommended by the U.S. Air Force for acceptable orbital decay predictions, and the light orange arrows correspond to key times for the accurate knowledge of storm-time satellite orbital drag effects \cite{Lewis2019}. \par

    For all storms (Figure \ref{data_model_error}a), relative errors of the CHAMP satellite behave well because they stay inside the $\pm$5\% CI during storm times. The model uncertainties are higher in the GRACE case. Relative errors stay about half of the time inside the CI, being outside of it during storm main phase and late storm recovery phase. JB2008 underestimates data at CHAMP's orbit during early storm main phase and overestimates density afterwards. The same occurs for the GRACE case. Figure \ref{data_model_error}b shows that in the case of minor (G1) storms CHAMP and GRACE uncertainties were most of the time inside the CI, with JB2008 underestimating data for CHAMP's orbits. The case of moderate storms (Figure \ref{data_model_error}c) is very similar to the case of all storms superposed, except that JB2008 underestimates density for CHAMP's orbits during the interval 21 $<$ t $<$ 35 hours. \par

    The model behavior for the case of strong (G3) storms is quite different from the previous storm categories, as shown in Figure \ref{data_model_error}d. CHAMP and GRACE density uncertainties stay outside the CI with the GRACE uncertainties being higher than the CHAMP uncertainties. Maximum relative errors are $\eta$ $\sim$ 15\% for CHAMP (t = 18 hours) and $\eta$ $\sim$ 25\% for GRACE (t = 12 hours). Figure \ref{data_model_error}e shows that the situation is different in the case of the severe (G4) events, where GRACE's uncertainties are smaller than CHAMP's uncertainties. CHAMP relative errors are inside the CI during the first 12 storm-time hours, when they increase later during storm recovery phase reaching values up to 18\%. GRACE uncertainties stay inside the CI most of the time, being outside of it only during the intervals 18 $<$ t $<$ 22 hours and 44 $<$ t $<$ 62 hours. \par

    Finally, for the extreme (G5) events (Figure \ref{data_model_error}f), relative errors show that the model performance for GRACE density predictions is worse than the model performance for CHAMP density predictions. In the case of CHAMP, densities are underestimated until 22 hours after ZET when it becomes highly overestimated afterwards with relative errors as high as 25\% during late storm recovery phase. In the case of GRACE, $\eta$ is within the CI only during the interval 6 $<$ t $<$ 18 hours, when later it takes very high values around 35\% at t = 62 hours during late recovery phase. \par

    \definecolor{light-gray}{gray}{0.90}
    \begin{table}
      \centering
      \begin{tabular}{r | c c c | c c c}
                  \hline
                   & & CHAMP & & & GRACE \\
                  \cline{2-7} 
                  \hline
                  Storm category   &  25Q  &  $\langle\eta\rangle$ & 75Q &  25Q  &  $\langle\eta\rangle$ & 75Q \\
                  \hline
                        All Events &   0.98 &   2.09 &   3.38 &   4.02 &   4.72 &  6.16 \\
                 Minor Events [G1] & --3.69 & --2.18 & --1.01 & --1.24 & --0.15 &  1.23 \\
              Moderate Events [G2] & --1.49 &   1.22 &   3.25 &   4.11 &   4.81 &  6.40 \\
                Strong Events [G3] &   4.86 &   6.77 &   8.36 &  12.73 &  15.37 & 18.69 \\
                Severe Events [G4] &   6.22 &  10.15 &  13.88 &   0.02 &   2.69 &  5.47 \\
               Extreme Events [G5] & --2.11 &   8.30 &  16.97 &   7.84 &  16.60 & 24.97 \\
                \hline
      \end{tabular}
      \caption{Statistical values for the CHAMP and GRACE errors, for each storm intensity category, shown in Figure \ref{data_model_error}. These values are calculated for the interval ($t_1$, $t_2$) = (--12, 72) hours. Q25 and Q75 correspond to the 25\% and 75\% quartiles, respectively.}
      \label{error_table}
    \end{table}

    Statistical values for the relative errors, plotted in Figure \ref{data_model_error}, for both satellites, are shown in Table \ref{error_table}. These values are computed over the same time spans shown in that figure. The values Q25 and Q75 correspond to the 25\% and 75\% quartiles of the uncertainties, respectively. \par

    The relative errors shown in Figure \ref{data_model_error} for both CHAMP and GRACE explain why satellite orbital drag effects are overestimated during late recovery phases of the most intense storm categories, particularly in the case of the extreme events. This effect is caused by high density values produced by JB2008, which impact the computation of orbital drag effects by increasing their uncertainties as shown by equations (4) and (5). Such overestimation effects arise because the JB2008 model does not take NO cooling effects into account when computing densities mainly during late storm main phase and subsequent storm recovery phase \cite{Mlynczak2003,Bowman2008,Knipp2017a,Zesta2018b}. The model should be adjusted accordingly in order to incorporate NO cooling effects into the density computation physics. \par

\section{Conclusion}\label{conclusion}

  In this paper, we investigated the effects of atmospheric air drag forces on the orbital decays of spacecraft in low-Earth orbit during a time span longer than a regular solar cycle. We used data from two LEO satellites, CHAMP and GRACE, which flew in the upper atmosphere in two different regions separated by $\sim$100 km altitude, being GRACE at higher altitudes. We used a data base with 217 distinct CME-driven magnetic storms, with 151 CHAMP observations and 213 GRACE observations, totalizing 364 individual satellite observations. We computed orbital decay rates (ODRs) and storm-time orbital decays (SODs) for each satellite for all magnetic storms by using a novel method suggested by \citeA{Oliveira2017c} that captures density dynamics during storms. We then assessed the performance of a well-known thermosphere mass density model, namely the Jacchia-Bowman 2008 model (JB2008), by comparing observed and modeled background neutral densities, used to compute satellite orbital drag effects, by means of superposed epoch analyses with storms of different intensities. The main conclusions of this work are summarized as follows:

  \begin{enumerate}

    \item As a continuation to a previous work \cite{Oliveira2017c}, we increased, with the extension of the time span from October 2011 to December 2015, our previous work by nearly 50 magnetic storms. This includes the rising, maximum, and almost all of the declining phase of SC24. We found that the overall SSNs during SC24 are smaller than SSNs in SC23. This reflects on the relative smaller number of CMEs arriving at Earth and the subsequent number of CME-driven magnetic storms following the CME arrivals. When grouping the storms with respect to storm intensities, we found that the more intense the storm category, the fewer the number of storms in that category.

    \item We found that storm-time orbital decay rates depend heavily on altitude, as suggested by previous studies \cite{Chen2012,Krauss2018}. SEA results show that CHAMP ODRs are twice larger than the GRACE ODRs. $da/dt$ increases with storm intensity, being strongest for the extreme storm category (G5 group). Particularly for G5 storms, CHAMP and GRACE show a sudden decrease in $da/dt$ effects probably due to high production of NO, a very efficient cooling agent in the thermosphere \cite{Kockarts1980,Mlynczak2003,Knipp2017a,Zesta2018b}. The model overestimates $da/dt$ particularly during storm recovery phase, but for the extreme storms a slight $da/dt$ underestimation and a strong $da/dt$ overestimation is seen during storm main phase and storm recovery phase for both missions.

    \item The total orbital decay for CHAMP occurs first at NH high latitudes and then spread out steadily to SH high latitudes. Generally, storm-time satellite orbital drag effects increase with storm intensities, being more intense and occurring first during the strongest storms. GRACE's SOD response is very similar to CHAMP's response, but $d$ is enhanced first at high latitudes, with the NH leading the SH by a few hours, and later storm effects become enhanced at both hemispheres' middle and low latitudes. This behavior is consistent with energy and heating distribution by global wind surge patterns and TADs \cite{Fuller-Rowell1994,Bruinsma2007,Oliveira2017c}. However, the extreme storms observed by GRACE show a different pattern: SOD effects at equatorial and low latitudes are more strongly enhanced with respect to the high latitudes. \change{We attributed this effect to the upward motion of oxygen neutrals caused by the super-fountain effect which in turn is caused by PPEFs in equatorial ionospheric regions (Mannucci et al., 2004; Tsurutani et al., 2007; Verkhoglyadova et al., 2007), which increase satellite drag effects at high altitudes as in the case of GRACE and higher (Lakhina $\&$ Tsurutani, 2017)}{\protect{These effects are consistent with Joule heating propagation from high latitudes to low/equatorial latitudes \cite{Fuller-Rowell1994,Richmond2000,Oliveira2017c}}}.

    \item When assessing performance evaluation of density predictions for each satellite orbits, we found that uncertainty levels are low for the combination of all storms and the minor (G1) and moderate (G2) storm categories. In contrast, for both satellites, we found that the storms in the strong (G3) category show high model uncertainties during late storm main phase but they decrease during storm recovery phase. For the case of severe (G4) storms, CHAMP's uncertainties are higher than GRACE's uncertainties during. CHAMP's uncertainties increase after late storm main phase and during storm recovery phase. In the case of extreme (G5) storms, uncertainties are the largest of all storm categories particularly after t = 20 hours. GRACE's uncertainties are larger than CHAMP's uncertainties. We suggest that such prediction errors arise from the lack of NO cooling effects in the JB2008 model \cite{Mlynczak2003,Bowman2008,Knipp2017a,Zesta2018b}. Such model deficiencies should be addressed in the future for the improvement of density and satellite orbital drag predictions. 

  \end{enumerate}

  The results of this work are consistent with the results found by \citeA{Bussy-Virat2018}, who computed probabilities of collision between objects in space affected by thermospheric density uncertainties, as the ones introduced by magnetic storms. They found that during a severe storm (Table \ref{table}), the collision probability increased by 50\%. Despite the fact that probability led to a false alarm according to their collision prediction scheme, \citeA{Bussy-Virat2018} demonstrated the importance of predicting neutral densities during magnetic storms for satellite orbital drag predictions and forecasting. The results of this work point out the possible source of errors for satellite orbital drag predictions during magnetic storms, namely CO$_2$ and NO cooling effects relevant during pre-storm and recovery phase times, respectively. In addition, this is relevant to our community efforts in communicating uncertainties of space weather data and models for further improving space weather prediction and forecasting capabilities \cite{Knipp2018a}.

  \acknowledgments
    D.M.O. acknowledges the NASA grant HISFM18-HIF (Heliophysics Innovation Fund). E.Z. was supported by the NASA Heliophysics Internal Scientist Funding Model through the grants HISFM18-0009, HISFM18-0006 and HISFM18-HIF. The authors acknowledge the opportunities to use the CME list, SSN data, OMNI data, SYM-H data, CHAMP and GRACE data, and the JB2008 empirical model. The CME list is available at \url{http://www.srl.caltech.edu/ACE/ASC/DATA/level3/icmetable2.html}. The SSN data were downloaded from the website \url{http://www.sidc.be/silso/datafiles}. The OMNI database is located at \url{https://omniweb.gsfc.nasa.gov/}. The SYM-H data are available at \url{http://wdc.kugi.kyoto-u.ac.jp/}. CHAMP and GRACE data were obtained through access provided by the Information System and Data Center (ISDC) in Potsdam, Germany (\url{isdc.gfz-potsdam.de}). The JB2008 thermosphere empirical model is available for download at \url{http://sol.spacenvironment.net/jb2008/}. D.M.O. also acknowledges the NSF-AGU Travel Grant for the support to attend the 2019 Chapman on Scientific Challenges Pertaining to Space Weather Forecasting Including Extremes in Pasadena, CA, where part of this work was presented.


\begin{thebibliography}{}

\bibitem [\protect \citeauthoryear {%
Aguado%
, Cid%
, Saiz%
\BCBL {}\ \BBA {} Cerrato%
}{%
Aguado%
\ \protect \BOthers {.}}{%
{\protect \APACyear {2010}}%
}]{%
Aguado2010}
\APACinsertmetastar {%
Aguado2010}%
\begin{APACrefauthors}%
Aguado, J.%
, Cid, C.%
, Saiz, E.%
\BCBL {}\ \BBA {} Cerrato, Y.%
\end{APACrefauthors}%
\unskip\
\newblock
\APACrefYearMonthDay{2010}{}{}.
\newblock
{\BBOQ}\APACrefatitle {Hyperbolic decay of the {D}st Index during the recovery
  phase of intense geomagnetic storms} {Hyperbolic decay of the {D}st index
  during the recovery phase of intense geomagnetic storms}.{\BBCQ}
\newblock
\APACjournalVolNumPages{Journal\ of\ Geophysical\ Research}{115}{A7}{}.
\newblock
\begin{APACrefDOI} \doi{10.1029/2009JA014658} \end{APACrefDOI}
\PrintBackRefs{\CurrentBib}

\bibitem [\protect \citeauthoryear {%
Bettadpur%
}{%
Bettadpur%
}{%
{\protect \APACyear {2007}}%
}]{%
Bettadpur2007}
\APACinsertmetastar {%
Bettadpur2007}%
\begin{APACrefauthors}%
Bettadpur, S.%
\end{APACrefauthors}%
\unskip\
\newblock
\APACrefYearMonthDay{2007}{}{}.
\newblock
\APACrefbtitle {{GRACE 327-720 (CSR-GR-03-02) Gravity Recovery and Climate
  Experiment}} {{GRACE 327-720 (CSR-GR-03-02) Gravity Recovery and Climate
  Experiment}}\ \APACbVolEdTR{}{\BTR{}\ \BNUMS\ {Product Specification Document
  (Rev 4.6 -- May 29, 2012)}}.
\newblock
\APACaddressInstitution{Austin, Texas}{The University of Texas at Austin,
  Center for Space Research}.
\PrintBackRefs{\CurrentBib}

\bibitem [\protect \citeauthoryear {%
Bowman%
\ \protect \BOthers {.}}{%
Bowman%
\ \protect \BOthers {.}}{%
{\protect \APACyear {2008}}%
}]{%
Bowman2008}
\APACinsertmetastar {%
Bowman2008}%
\begin{APACrefauthors}%
Bowman, B\BPBI R.%
, Tobiska, W\BPBI K.%
, Marcos, F\BPBI A.%
, Huang, C\BPBI Y.%
, Lin, C\BPBI S.%
\BCBL {}\ \BBA {} Burke, W\BPBI J.%
\end{APACrefauthors}%
\unskip\
\newblock
\APACrefYearMonthDay{2008}{}{}.
\newblock
{\BBOQ}\APACrefatitle {A new empirical thermospheric density model {JB2008}
  using new solar and geomagnetic indices} {A new empirical thermospheric
  density model {JB2008} using new solar and geomagnetic indices}.{\BBCQ}
\newblock
\BIn{} \APACrefbtitle {{AIAA/AAS Astrodynamics Specialist Conference, AIAA
  2008--6438}.} {{AIAA/AAS Astrodynamics Specialist Conference, AIAA
  2008--6438}.}
\newblock
\APACaddressPublisher{Honolulu, HI}{}.
\PrintBackRefs{\CurrentBib}

\bibitem [\protect \citeauthoryear {%
Bruinsma%
\ \BBA {} Biancale%
}{%
Bruinsma%
\ \BBA {} Biancale%
}{%
{\protect \APACyear {2003}}%
}]{%
Bruinsma2003}
\APACinsertmetastar {%
Bruinsma2003}%
\begin{APACrefauthors}%
Bruinsma, S.%
\BCBT {}\ \BBA {} Biancale, R.%
\end{APACrefauthors}%
\unskip\
\newblock
\APACrefYearMonthDay{2003}{}{}.
\newblock
{\BBOQ}\APACrefatitle {{Total Densities Derived from Accelerometer Data}}
  {{Total Densities Derived from Accelerometer Data}}.{\BBCQ}
\newblock
\APACjournalVolNumPages{Journal\ of\ Spacecraft\ and\ Rockets}{40}{2}{230-236}.
\newblock
\begin{APACrefDOI} \doi{10.2514/2.3937} \end{APACrefDOI}
\PrintBackRefs{\CurrentBib}

\bibitem [\protect \citeauthoryear {%
S.~Bruinsma%
, Sutton%
, Solomon%
, Fuller-Rowell%
\BCBL {}\ \BBA {} Fedrizzi%
}{%
S.~Bruinsma%
\ \protect \BOthers {.}}{%
{\protect \APACyear {2018}}%
}]{%
Bruinsma2018}
\APACinsertmetastar {%
Bruinsma2018}%
\begin{APACrefauthors}%
Bruinsma, S.%
, Sutton, E.%
, Solomon, S\BPBI C.%
, Fuller-Rowell, T.%
\BCBL {}\ \BBA {} Fedrizzi, M.%
\end{APACrefauthors}%
\unskip\
\newblock
\APACrefYearMonthDay{2018}{}{}.
\newblock
{\BBOQ}\APACrefatitle {Space Weather Modeling Capabilities Assessment: Neutral
  Density for Orbit Determination at low Earth orbit} {Space weather modeling
  capabilities assessment: Neutral density for orbit determination at low earth
  orbit}.{\BBCQ}
\newblock
\APACjournalVolNumPages{Space\ Weather}{16}{11}{1806-1816}.
\newblock
\begin{APACrefDOI} \doi{10.1029/2018SW002027} \end{APACrefDOI}
\PrintBackRefs{\CurrentBib}

\bibitem [\protect \citeauthoryear {%
S\BPBI L.~Bruinsma%
}{%
S\BPBI L.~Bruinsma%
}{%
{\protect \APACyear {2015}}%
}]{%
Bruinsma2015}
\APACinsertmetastar {%
Bruinsma2015}%
\begin{APACrefauthors}%
Bruinsma, S\BPBI L.%
\end{APACrefauthors}%
\unskip\
\newblock
\APACrefYearMonthDay{2015}{}{}.
\newblock
{\BBOQ}\APACrefatitle {{The DTM-2013 thermosphere model}} {{The DTM-2013
  thermosphere model}}.{\BBCQ}
\newblock
\APACjournalVolNumPages{Journal\ of\ Space\ Weather\ and\ Space\
  Climate}{5}{A1}{}.
\newblock
\begin{APACrefDOI} \doi{10.1051/swsc/2015001} \end{APACrefDOI}
\PrintBackRefs{\CurrentBib}

\bibitem [\protect \citeauthoryear {%
Bruinsma%
\ \BBA {} Forbes%
}{%
Bruinsma%
\ \BBA {} Forbes%
}{%
{\protect \APACyear {2007}}%
}]{%
Bruinsma2007}
\APACinsertmetastar {%
Bruinsma2007}%
\begin{APACrefauthors}%
Bruinsma, S\BPBI L.%
\BCBT {}\ \BBA {} Forbes, J\BPBI M.%
\end{APACrefauthors}%
\unskip\
\newblock
\APACrefYearMonthDay{2007}{}{}.
\newblock
{\BBOQ}\APACrefatitle {Global observations of traveling atmospheric
  disturbances {(TADs)} in the thermosphere} {Global observations of traveling
  atmospheric disturbances {(TADs)} in the thermosphere}.{\BBCQ}
\newblock
\APACjournalVolNumPages{Geophysical\ Research\ Letters}{34}{L14103}{}.
\newblock
\begin{APACrefDOI} \doi{10.1029/2007GL030243} \end{APACrefDOI}
\PrintBackRefs{\CurrentBib}

\bibitem [\protect \citeauthoryear {%
Bruinsma%
, Tamagnan%
\BCBL {}\ \BBA {} Biancale%
}{%
Bruinsma%
\ \protect \BOthers {.}}{%
{\protect \APACyear {2004}}%
}]{%
Bruinsma2004}
\APACinsertmetastar {%
Bruinsma2004}%
\begin{APACrefauthors}%
Bruinsma, S\BPBI L.%
, Tamagnan, D.%
\BCBL {}\ \BBA {} Biancale, R.%
\end{APACrefauthors}%
\unskip\
\newblock
\APACrefYearMonthDay{2004}{}{}.
\newblock
{\BBOQ}\APACrefatitle {Atmospheric densities derived from {CHAMP/STAR}
  accelerometer observations} {Atmospheric densities derived from {CHAMP/STAR}
  accelerometer observations}.{\BBCQ}
\newblock
\APACjournalVolNumPages{Planetary\ and\ Space\ Science}{62}{4}{297--312}.
\newblock
\begin{APACrefDOI} \doi{10.1016/j.pss.2003.11.004} \end{APACrefDOI}
\PrintBackRefs{\CurrentBib}

\bibitem [\protect \citeauthoryear {%
Burke%
\ \protect \BOthers {.}}{%
Burke%
\ \protect \BOthers {.}}{%
{\protect \APACyear {2009}}%
}]{%
Burke2009}
\APACinsertmetastar {%
Burke2009}%
\begin{APACrefauthors}%
Burke, W\BPBI J.%
, Lin, C\BPBI S.%
, Hagan, M\BPBI P.%
, Huang, C\BPBI Y.%
, Weimer, D\BPBI R.%
, Wise, J\BPBI O.%
\BDBL {}Marcos, F\BPBI A.%
\end{APACrefauthors}%
\unskip\
\newblock
\APACrefYearMonthDay{2009}{}{}.
\newblock
{\BBOQ}\APACrefatitle {Storm time global thermosphere: A driven-dissipative
  thermodynamic system} {Storm time global thermosphere: A driven-dissipative
  thermodynamic system}.{\BBCQ}
\newblock
\APACjournalVolNumPages{Journal\ of\ Geophysical\ Research}{114}{A6}{}.
\newblock
\begin{APACrefDOI} \doi{10.1029/2008JA013848} \end{APACrefDOI}
\PrintBackRefs{\CurrentBib}

\bibitem [\protect \citeauthoryear {%
{Bussy-Virat}%
, Ridley%
\BCBL {}\ \BBA {} Getchius%
}{%
{Bussy-Virat}%
\ \protect \BOthers {.}}{%
{\protect \APACyear {2018}}%
}]{%
Bussy-Virat2018}
\APACinsertmetastar {%
Bussy-Virat2018}%
\begin{APACrefauthors}%
{Bussy-Virat}, C\BPBI D.%
, Ridley, A\BPBI J.%
\BCBL {}\ \BBA {} Getchius, J\BPBI W.%
\end{APACrefauthors}%
\unskip\
\newblock
\APACrefYearMonthDay{2018}{}{}.
\newblock
{\BBOQ}\APACrefatitle {Effects of Uncertainties in the Atmospheric Density on
  the Probability of Collision Between Space Objects} {Effects of uncertainties
  in the atmospheric density on the probability of collision between space
  objects}.{\BBCQ}
\newblock
\APACjournalVolNumPages{Space\ Weather}{16}{5}{519-537}.
\newblock
\begin{APACrefDOI} \doi{10.1029/2017SW001705} \end{APACrefDOI}
\PrintBackRefs{\CurrentBib}

\bibitem [\protect \citeauthoryear {%
Chen%
, Xu%
, Wang%
, Lei%
\BCBL {}\ \BBA {} Burns%
}{%
Chen%
\ \protect \BOthers {.}}{%
{\protect \APACyear {2012}}%
}]{%
Chen2012}
\APACinsertmetastar {%
Chen2012}%
\begin{APACrefauthors}%
Chen, G\BHBI m.%
, Xu, J.%
, Wang, W.%
, Lei, J.%
\BCBL {}\ \BBA {} Burns, A\BPBI G.%
\end{APACrefauthors}%
\unskip\
\newblock
\APACrefYearMonthDay{2012}{}{}.
\newblock
{\BBOQ}\APACrefatitle {A comparison of the effects of {CIR--} and
  {CME--induced} geomagnetic activity on thermospheric densities and spacecraft
  orbits: {C}ase studies} {A comparison of the effects of {CIR--} and
  {CME--induced} geomagnetic activity on thermospheric densities and spacecraft
  orbits: {C}ase studies}.{\BBCQ}
\newblock
\APACjournalVolNumPages{Journal\ of\ Geophysical\ Research}{117}{A8}{}.
\newblock
\begin{APACrefDOI} \doi{10.1029/2012JA017782} \end{APACrefDOI}
\PrintBackRefs{\CurrentBib}

\bibitem [\protect \citeauthoryear {%
Chopra%
}{%
Chopra%
}{%
{\protect \APACyear {1961}}%
}]{%
Chopra1961}
\APACinsertmetastar {%
Chopra1961}%
\begin{APACrefauthors}%
Chopra, K\BPBI P.%
\end{APACrefauthors}%
\unskip\
\newblock
\APACrefYearMonthDay{1961}{}{}.
\newblock
{\BBOQ}\APACrefatitle {Interactions of Rapidly Moving Bodies in Terrestrial
  Atmosphere} {Interactions of rapidly moving bodies in terrestrial
  atmosphere}.{\BBCQ}
\newblock
\APACjournalVolNumPages{Reviews\ of\ Modern\ Physics}{33}{2}{153-189}.
\newblock
\begin{APACrefDOI} \doi{10.1103/RevModPhys.33.153} \end{APACrefDOI}
\PrintBackRefs{\CurrentBib}

\bibitem [\protect \citeauthoryear {%
Chree%
}{%
Chree%
}{%
{\protect \APACyear {1913}}%
}]{%
Chree1913}
\APACinsertmetastar {%
Chree1913}%
\begin{APACrefauthors}%
Chree, C.%
\end{APACrefauthors}%
\unskip\
\newblock
\APACrefYearMonthDay{1913}{}{}.
\newblock
{\BBOQ}\APACrefatitle {Some Phenomena of Sunspots and of Terrestrial Magnetism
  at {K}ew Observatory} {Some phenomena of sunspots and of terrestrial
  magnetism at {K}ew observatory}.{\BBCQ}
\newblock
\APACjournalVolNumPages{Philosophical\ Transactions\ of\ the\ Royal\ Society\
  of\ London}{212}{484-496}{75-116}.
\newblock
\begin{APACrefDOI} \doi{10.1098/rsta.1913.0003} \end{APACrefDOI}
\PrintBackRefs{\CurrentBib}

\bibitem [\protect \citeauthoryear {%
Cid%
\ \protect \BOthers {.}}{%
Cid%
\ \protect \BOthers {.}}{%
{\protect \APACyear {2013}}%
}]{%
Cid2013}
\APACinsertmetastar {%
Cid2013}%
\begin{APACrefauthors}%
Cid, C.%
, Palacios, J.%
, Saiz, E.%
, Cerrato, Y.%
, Aguado, J.%
\BCBL {}\ \BBA {} Guerrero, A.%
\end{APACrefauthors}%
\unskip\
\newblock
\APACrefYearMonthDay{2013}{}{}.
\newblock
{\BBOQ}\APACrefatitle {Modeling the recovery phase of extreme geomagnetic
  storms} {Modeling the recovery phase of extreme geomagnetic storms}.{\BBCQ}
\newblock
\APACjournalVolNumPages{Journal\ of\ Geophysical\ Research}{118}{7}{4352-4359}.
\newblock
\begin{APACrefDOI} \doi{10.1002/jgra.50409} \end{APACrefDOI}
\PrintBackRefs{\CurrentBib}

\bibitem [\protect \citeauthoryear {%
Clette%
\ \BBA {} Lef\`evre%
}{%
Clette%
\ \BBA {} Lef\`evre%
}{%
{\protect \APACyear {2016}}%
}]{%
Clette2016b}
\APACinsertmetastar {%
Clette2016b}%
\begin{APACrefauthors}%
Clette, F.%
\BCBT {}\ \BBA {} Lef\`evre, L.%
\end{APACrefauthors}%
\unskip\
\newblock
\APACrefYearMonthDay{2016}{}{}.
\newblock
{\BBOQ}\APACrefatitle {{The New Sunspot Number: Assembling All Corrections}}
  {{The New Sunspot Number: Assembling All Corrections}}.{\BBCQ}
\newblock
\APACjournalVolNumPages{Solar\ Physics}{291}{9-10}{2629-2651}.
\newblock
\begin{APACrefDOI} \doi{10.1007/s11207-016-1014-y} \end{APACrefDOI}
\PrintBackRefs{\CurrentBib}

\bibitem [\protect \citeauthoryear {%
Doornbos%
}{%
Doornbos%
}{%
{\protect \APACyear {2012}}%
}]{%
Doornbos2012}
\APACinsertmetastar {%
Doornbos2012}%
\begin{APACrefauthors}%
Doornbos, E.%
\end{APACrefauthors}%
\unskip\
\newblock
\APACrefYear{2012}.
\newblock
\APACrefbtitle {Thermospheric Density and Wind Determination from Satellite
  Dynamics} {Thermospheric density and wind determination from satellite
  dynamics}.
\newblock
\APACaddressPublisher{New York,\ NY}{Springer}.
\newblock
\begin{APACrefDOI} \doi{10.1007/978-3-642-25129-0} \end{APACrefDOI}
\PrintBackRefs{\CurrentBib}

\bibitem [\protect \citeauthoryear {%
Eddy%
}{%
Eddy%
}{%
{\protect \APACyear {1976}}%
}]{%
Eddy1976}
\APACinsertmetastar {%
Eddy1976}%
\begin{APACrefauthors}%
Eddy, J\BPBI A.%
\end{APACrefauthors}%
\unskip\
\newblock
\APACrefYearMonthDay{1976}{}{}.
\newblock
{\BBOQ}\APACrefatitle {{The Maunder Minimum}} {{The Maunder Minimum}}.{\BBCQ}
\newblock
\APACjournalVolNumPages{Science}{192}{4245}{1189--1202}.
\newblock
\begin{APACrefDOI} \doi{10.1126/science.192.4245.1189} \end{APACrefDOI}
\PrintBackRefs{\CurrentBib}

\bibitem [\protect \citeauthoryear {%
Emmert%
}{%
Emmert%
}{%
{\protect \APACyear {2015}}%
}]{%
Emmert2015}
\APACinsertmetastar {%
Emmert2015}%
\begin{APACrefauthors}%
Emmert, J\BPBI T.%
\end{APACrefauthors}%
\unskip\
\newblock
\APACrefYearMonthDay{2015}{}{}.
\newblock
{\BBOQ}\APACrefatitle {Thermospheric mass density: {A} review} {Thermospheric
  mass density: {A} review}.{\BBCQ}
\newblock
\APACjournalVolNumPages{Advances in Space\ {Research}}{56}{5}{773--824}.
\newblock
\begin{APACrefDOI} \doi{10.1016/j.asr.2015.05.038} \end{APACrefDOI}
\PrintBackRefs{\CurrentBib}

\bibitem [\protect \citeauthoryear {%
Flury%
, Bettadpur%
\BCBL {}\ \BBA {} Tapley%
}{%
Flury%
\ \protect \BOthers {.}}{%
{\protect \APACyear {2008}}%
}]{%
Flury2008}
\APACinsertmetastar {%
Flury2008}%
\begin{APACrefauthors}%
Flury, J.%
, Bettadpur, S.%
\BCBL {}\ \BBA {} Tapley, B\BPBI D.%
\end{APACrefauthors}%
\unskip\
\newblock
\APACrefYearMonthDay{2008}{}{}.
\newblock
{\BBOQ}\APACrefatitle {{Precise accelerometry onboard the GRACE gravity field
  satellite mission}} {{Precise accelerometry onboard the GRACE gravity field
  satellite mission}}.{\BBCQ}
\newblock
\APACjournalVolNumPages{Advances in Space\ {Research}}{42}{8}{1414-1423}.
\newblock
\begin{APACrefDOI} \doi{10.1016/j.asr.2008.05.004} \end{APACrefDOI}
\PrintBackRefs{\CurrentBib}

\bibitem [\protect \citeauthoryear {%
Forbes%
}{%
Forbes%
}{%
{\protect \APACyear {2007}}%
}]{%
Forbes2007}
\APACinsertmetastar {%
Forbes2007}%
\begin{APACrefauthors}%
Forbes, J\BPBI M.%
\end{APACrefauthors}%
\unskip\
\newblock
\APACrefYearMonthDay{2007}{}{}.
\newblock
{\BBOQ}\APACrefatitle {Dynamics of the Thermosphere} {Dynamics of the
  thermosphere}.{\BBCQ}
\newblock
\APACjournalVolNumPages{Journal\ of\ the\ Meteorological\ Society\ of\
  Japan}{85B}{}{193--213}.
\newblock
\begin{APACrefDOI} \doi{10.2151/jmsj.85B.193} \end{APACrefDOI}
\PrintBackRefs{\CurrentBib}

\bibitem [\protect \citeauthoryear {%
Fuller-Rowell%
, Codrescu%
, Moffett%
\BCBL {}\ \BBA {} Quegan%
}{%
Fuller-Rowell%
\ \protect \BOthers {.}}{%
{\protect \APACyear {1994}}%
}]{%
Fuller-Rowell1994}
\APACinsertmetastar {%
Fuller-Rowell1994}%
\begin{APACrefauthors}%
Fuller-Rowell, T\BPBI J.%
, Codrescu, M\BPBI V.%
, Moffett, R\BPBI J.%
\BCBL {}\ \BBA {} Quegan, S.%
\end{APACrefauthors}%
\unskip\
\newblock
\APACrefYearMonthDay{1994}{}{}.
\newblock
{\BBOQ}\APACrefatitle {Response of the thermosphere and ionosphere to
  geomagnetic storms} {Response of the thermosphere and ionosphere to
  geomagnetic storms}.{\BBCQ}
\newblock
\APACjournalVolNumPages{Journal\ of\ Geophysical\
  Research}{99}{A3}{3893--3914}.
\newblock
\begin{APACrefDOI} \doi{10.1029/93JA02015} \end{APACrefDOI}
\PrintBackRefs{\CurrentBib}

\bibitem [\protect \citeauthoryear {%
Gonzalez%
\ \protect \BOthers {.}}{%
Gonzalez%
\ \protect \BOthers {.}}{%
{\protect \APACyear {1994}}%
}]{%
Gonzalez1994}
\APACinsertmetastar {%
Gonzalez1994}%
\begin{APACrefauthors}%
Gonzalez, W\BPBI D.%
, Joselyn, J\BPBI A.%
, Kamide, Y.%
, Kroehl, H\BPBI W.%
, Rostoker, G.%
, Tsurutani, B\BPBI T.%
\BCBL {}\ \BBA {} Vasyli\={u}nas, V\BPBI M.%
\end{APACrefauthors}%
\unskip\
\newblock
\APACrefYearMonthDay{1994}{}{}.
\newblock
{\BBOQ}\APACrefatitle {What is a geomagnetic storm?} {What is a geomagnetic
  storm?}{\BBCQ}
\newblock
\APACjournalVolNumPages{Journal\ of\ Geophysical\
  Research}{99}{A4}{5771--5792}.
\newblock
\begin{APACrefDOI} \doi{10.1029/93JA02867} \end{APACrefDOI}
\PrintBackRefs{\CurrentBib}

\bibitem [\protect \citeauthoryear {%
Gonzalez%
\ \BBA {} Tsurutani%
}{%
Gonzalez%
\ \BBA {} Tsurutani%
}{%
{\protect \APACyear {1987}}%
}]{%
Gonzalez1987}
\APACinsertmetastar {%
Gonzalez1987}%
\begin{APACrefauthors}%
Gonzalez, W\BPBI D.%
\BCBT {}\ \BBA {} Tsurutani, B\BPBI T.%
\end{APACrefauthors}%
\unskip\
\newblock
\APACrefYearMonthDay{1987}{}{}.
\newblock
{\BBOQ}\APACrefatitle {Criteria of interplanetary parameters causing intense
  magnetic storms ({D}st {$<$} $-$100 n{T})} {Criteria of interplanetary
  parameters causing intense magnetic storms ({D}st {$<$} $-$100 n{T})}.{\BBCQ}
\newblock
\APACjournalVolNumPages{Planetary\ and\ Space\ Science}{35}{9}{1101-1109}.
\newblock
\begin{APACrefDOI} \doi{10.1016/0032-0633(87)90015-8} \end{APACrefDOI}
\PrintBackRefs{\CurrentBib}

\bibitem [\protect \citeauthoryear {%
Gosling%
, McComas%
, Phillips%
\BCBL {}\ \BBA {} Bame%
}{%
Gosling%
\ \protect \BOthers {.}}{%
{\protect \APACyear {1991}}%
}]{%
GOsling1991}
\APACinsertmetastar {%
GOsling1991}%
\begin{APACrefauthors}%
Gosling, J\BPBI T.%
, McComas, D\BPBI J.%
, Phillips, J\BPBI L.%
\BCBL {}\ \BBA {} Bame, S\BPBI J.%
\end{APACrefauthors}%
\unskip\
\newblock
\APACrefYearMonthDay{1991}{}{}.
\newblock
{\BBOQ}\APACrefatitle {Geomagnetic activity associated with earth passage of
  interplanetary shock disturbances and coronal mass ejections} {Geomagnetic
  activity associated with earth passage of interplanetary shock disturbances
  and coronal mass ejections}.{\BBCQ}
\newblock
\APACjournalVolNumPages{Journal\ of\ Geophysical\ Research}{96}{A5}{7831-7839}.
\newblock
\begin{APACrefDOI} \doi{10.1029/91JA00316} \end{APACrefDOI}
\PrintBackRefs{\CurrentBib}

\bibitem [\protect \citeauthoryear {%
He%
\ \protect \BOthers {.}}{%
He%
\ \protect \BOthers {.}}{%
{\protect \APACyear {2018}}%
}]{%
He2018}
\APACinsertmetastar {%
He2018}%
\begin{APACrefauthors}%
He, C.%
, Yang, Y.%
, Carter, B.%
, Kerr, E.%
, Wu, S.%
, Deleflie, F.%
\BDBL {}Norman, R.%
\end{APACrefauthors}%
\unskip\
\newblock
\APACrefYearMonthDay{2018}{}{}.
\newblock
{\BBOQ}\APACrefatitle {Review and comparison of empirical thermospheric mass
  density models} {Review and comparison of empirical thermospheric mass
  density models}.{\BBCQ}
\newblock
\APACjournalVolNumPages{Progress\ in\ Aerospace\ Sciences}{103}{}{31-51}.
\newblock
\begin{APACrefDOI} \doi{10.1016/j.paerosci.2018.10.003} \end{APACrefDOI}
\PrintBackRefs{\CurrentBib}

\bibitem [\protect \citeauthoryear {%
Hocke%
\ \BBA {} Schlegel%
}{%
Hocke%
\ \BBA {} Schlegel%
}{%
{\protect \APACyear {1996}}%
}]{%
Hocke1996}
\APACinsertmetastar {%
Hocke1996}%
\begin{APACrefauthors}%
Hocke, K.%
\BCBT {}\ \BBA {} Schlegel, K.%
\end{APACrefauthors}%
\unskip\
\newblock
\APACrefYearMonthDay{1996}{}{}.
\newblock
{\BBOQ}\APACrefatitle {A review of atmospheric gravity waves and travelling
  ionospheric disturbances: 1982--1995} {A review of atmospheric gravity waves
  and travelling ionospheric disturbances: 1982--1995}.{\BBCQ}
\newblock
\APACjournalVolNumPages{Annales\ Geophysicae}{14}{9}{917--940}.
\newblock
\begin{APACrefDOI} \doi{10.1007/s00585-996-0917-6} \end{APACrefDOI}
\PrintBackRefs{\CurrentBib}

\bibitem [\protect \citeauthoryear {%
Hunsucker%
}{%
Hunsucker%
}{%
{\protect \APACyear {1982}}%
}]{%
Hunsucker1982}
\APACinsertmetastar {%
Hunsucker1982}%
\begin{APACrefauthors}%
Hunsucker, R\BPBI D.%
\end{APACrefauthors}%
\unskip\
\newblock
\APACrefYearMonthDay{1982}{}{}.
\newblock
{\BBOQ}\APACrefatitle {Atmospheric gravity waves generated in the
  high--latitude ionosphere: A review} {Atmospheric gravity waves generated in
  the high--latitude ionosphere: A review}.{\BBCQ}
\newblock
\APACjournalVolNumPages{Reviews\ of\ Geophysics}{20}{2}{}.
\newblock
\begin{APACrefDOI} \doi{10.1029/RG020i002p00293} \end{APACrefDOI}
\PrintBackRefs{\CurrentBib}

\bibitem [\protect \citeauthoryear {%
Illing%
\ \BBA {} Hundhausen%
}{%
Illing%
\ \BBA {} Hundhausen%
}{%
{\protect \APACyear {1985}}%
}]{%
Illing1985}
\APACinsertmetastar {%
Illing1985}%
\begin{APACrefauthors}%
Illing, R\BPBI M\BPBI E.%
\BCBT {}\ \BBA {} Hundhausen, A\BPBI J.%
\end{APACrefauthors}%
\unskip\
\newblock
\APACrefYearMonthDay{1985}{}{}.
\newblock
{\BBOQ}\APACrefatitle {Observation of a coronal transient from 1.2 to 6 solar
  radii} {Observation of a coronal transient from 1.2 to 6 solar radii}.{\BBCQ}
\newblock
\APACjournalVolNumPages{Journal\ of\ Geophysical\ Research}{90}{A1}{275--282}.
\newblock
\begin{APACrefDOI} \doi{10.1029/JA090iA01p00275} \end{APACrefDOI}
\PrintBackRefs{\CurrentBib}

\bibitem [\protect \citeauthoryear {%
Iyemori%
}{%
Iyemori%
}{%
{\protect \APACyear {1990}}%
}]{%
Iyemori1990}
\APACinsertmetastar {%
Iyemori1990}%
\begin{APACrefauthors}%
Iyemori, T.%
\end{APACrefauthors}%
\unskip\
\newblock
\APACrefYearMonthDay{1990}{}{}.
\newblock
{\BBOQ}\APACrefatitle {Storm-time magnetospheric currents inferred from
  mid--latitude geomagnetic field variations} {Storm-time magnetospheric
  currents inferred from mid--latitude geomagnetic field variations}.{\BBCQ}
\newblock
\APACjournalVolNumPages{Journal\ of\ Geomagnetism\ and\
  Geoelectricity}{42}{11}{1249--1265}.
\newblock
\begin{APACrefDOI} \doi{10.5636/jgg.42.1249} \end{APACrefDOI}
\PrintBackRefs{\CurrentBib}

\bibitem [\protect \citeauthoryear {%
Jacchia%
}{%
Jacchia%
}{%
{\protect \APACyear {1959}}%
}]{%
Jacchia1959}
\APACinsertmetastar {%
Jacchia1959}%
\begin{APACrefauthors}%
Jacchia, L\BPBI G.%
\end{APACrefauthors}%
\unskip\
\newblock
\APACrefYearMonthDay{1959}{}{}.
\newblock
{\BBOQ}\APACrefatitle {Corpuscular Radiation and the Acceleration of Artificial
  Satellites} {Corpuscular radiation and the acceleration of artificial
  satellites}.{\BBCQ}
\newblock
\APACjournalVolNumPages{Nature}{183}{526}{1662--1663}.
\newblock
\begin{APACrefDOI} \doi{10.1038/1831662a0} \end{APACrefDOI}
\PrintBackRefs{\CurrentBib}

\bibitem [\protect \citeauthoryear {%
Jacchia%
}{%
Jacchia%
}{%
{\protect \APACyear {1970}}%
}]{%
Jacchia1970}
\APACinsertmetastar {%
Jacchia1970}%
\begin{APACrefauthors}%
Jacchia, L\BPBI G.%
\end{APACrefauthors}%
\unskip\
\newblock
\APACrefYearMonthDay{1970}{}{}.
\newblock
{\BBOQ}\APACrefatitle {New static models of the thermosphere and exosphere with
  empirical temperature profiles} {New static models of the thermosphere and
  exosphere with empirical temperature profiles}.{\BBCQ}
\newblock
\BIn{} \APACrefbtitle {Spec. Rep. 313.} {Spec. rep. 313.}
\newblock
\APACaddressPublisher{Cambridge, Massachusetts}{Smithson, Astrophys. Obs.}
\PrintBackRefs{\CurrentBib}

\bibitem [\protect \citeauthoryear {%
Jonas%
\ \BBA {} McCarron%
}{%
Jonas%
\ \BBA {} McCarron%
}{%
{\protect \APACyear {2016}}%
}]{%
Jonas2016}
\APACinsertmetastar {%
Jonas2016}%
\begin{APACrefauthors}%
Jonas, S.%
\BCBT {}\ \BBA {} McCarron, E\BPBI D.%
\end{APACrefauthors}%
\unskip\
\newblock
\APACrefYearMonthDay{2016}{}{}.
\newblock
{\BBOQ}\APACrefatitle {{White House Releases National Space Weather Strategy
  and Action Plan}} {{White House Releases National Space Weather Strategy and
  Action Plan}}.{\BBCQ}
\newblock
\APACjournalVolNumPages{Space Weather}{14}{2}{54-55}.
\newblock
\begin{APACrefDOI} \doi{10.1002/2015SW001357} \end{APACrefDOI}
\PrintBackRefs{\CurrentBib}

\bibitem [\protect \citeauthoryear {%
{Kalafatoglu Eyiguler}%
, Kaymaz%
, Frissell%
, Ruohoniemi%
\BCBL {}\ \BBA {} Rast\"atter%
}{%
{Kalafatoglu Eyiguler}%
\ \protect \BOthers {.}}{%
{\protect \APACyear {2018}}%
}]{%
KalafatogluEyiguler2018}
\APACinsertmetastar {%
KalafatogluEyiguler2018}%
\begin{APACrefauthors}%
{Kalafatoglu Eyiguler}, E\BPBI C.%
, Kaymaz, Z.%
, Frissell, N\BPBI A.%
, Ruohoniemi, J\BPBI M.%
\BCBL {}\ \BBA {} Rast\"atter, L.%
\end{APACrefauthors}%
\unskip\
\newblock
\APACrefYearMonthDay{2018}{}{}.
\newblock
{\BBOQ}\APACrefatitle {Investigating Upper Atmospheric {J}oule Heating Using
  Cross-Combination of Data for Two Moderate Substorm Cases} {Investigating
  upper atmospheric {J}oule heating using cross-combination of data for two
  moderate substorm cases}.{\BBCQ}
\newblock
\APACjournalVolNumPages{Space\ Weather}{16}{8}{987-1012}.
\newblock
\begin{APACrefDOI} \doi{10.1029/2018SW001956} \end{APACrefDOI}
\PrintBackRefs{\CurrentBib}

\bibitem [\protect \citeauthoryear {%
Kessler%
\ \BBA {} {Cour-Palais}%
}{%
Kessler%
\ \BBA {} {Cour-Palais}%
}{%
{\protect \APACyear {1978}}%
}]{%
Kessler1978}
\APACinsertmetastar {%
Kessler1978}%
\begin{APACrefauthors}%
Kessler, D\BPBI J.%
\BCBT {}\ \BBA {} {Cour-Palais}, B\BPBI G.%
\end{APACrefauthors}%
\unskip\
\newblock
\APACrefYearMonthDay{1978}{}{}.
\newblock
{\BBOQ}\APACrefatitle {Collision frequency of artificial satellites: {T}he
  creation of a debris belt} {Collision frequency of artificial satellites:
  {T}he creation of a debris belt}.{\BBCQ}
\newblock
\APACjournalVolNumPages{Journal\ of\ Geophysical\ Research}{83}{A6}{2637-2646}.
\newblock
\begin{APACrefDOI} \doi{10.1029/JA083iA06p02637} \end{APACrefDOI}
\PrintBackRefs{\CurrentBib}

\bibitem [\protect \citeauthoryear {%
Kilpua%
, Lugaz%
, Mays%
\BCBL {}\ \BBA {} Temmer%
}{%
Kilpua%
\ \protect \BOthers {.}}{%
{\protect \APACyear {2019}}%
}]{%
Kilpua2019}
\APACinsertmetastar {%
Kilpua2019}%
\begin{APACrefauthors}%
Kilpua, E\BPBI K\BPBI J.%
, Lugaz, N.%
, Mays, M\BPBI L.%
\BCBL {}\ \BBA {} Temmer, M.%
\end{APACrefauthors}%
\unskip\
\newblock
\APACrefYearMonthDay{2019}{}{}.
\newblock
{\BBOQ}\APACrefatitle {Forecasting the Structure and Orientation of Earthbound
  Coronal Mass Ejections} {Forecasting the structure and orientation of
  earthbound coronal mass ejections}.{\BBCQ}
\newblock
\APACjournalVolNumPages{Space\ Weather}{17}{4}{498-526}.
\newblock
\begin{APACrefDOI} \doi{10.1029/2018SW001944} \end{APACrefDOI}
\PrintBackRefs{\CurrentBib}

\bibitem [\protect \citeauthoryear {%
{King--Hele}%
}{%
{King--Hele}%
}{%
{\protect \APACyear {1987}}%
}]{%
King-Hele1987}
\APACinsertmetastar {%
King-Hele1987}%
\begin{APACrefauthors}%
{King--Hele}, D.%
\end{APACrefauthors}%
\unskip\
\newblock
\APACrefYear{1987}.
\newblock
\APACrefbtitle {Satellite Orbits in an Atmosphere: Theory and Applications}
  {Satellite orbits in an atmosphere: Theory and applications}.
\newblock
\APACaddressPublisher{Glasgow, Scotland}{Blackie and Son Ltd.}
\PrintBackRefs{\CurrentBib}

\bibitem [\protect \citeauthoryear {%
King%
\ \BBA {} Papitashvili%
}{%
King%
\ \BBA {} Papitashvili%
}{%
{\protect \APACyear {2005}}%
}]{%
King2005}
\APACinsertmetastar {%
King2005}%
\begin{APACrefauthors}%
King, J\BPBI H.%
\BCBT {}\ \BBA {} Papitashvili, N\BPBI E.%
\end{APACrefauthors}%
\unskip\
\newblock
\APACrefYearMonthDay{2005}{}{}.
\newblock
{\BBOQ}\APACrefatitle {Solar wind spatial scales in and comparisons of hourly
  {Wind} and {ACE} plasma and magnetic field data} {Solar wind spatial scales
  in and comparisons of hourly {Wind} and {ACE} plasma and magnetic field
  data}.{\BBCQ}
\newblock
\APACjournalVolNumPages{Journal\ of\ Geophysical\ Research}{110}{A2}{1--9}.
\newblock
\begin{APACrefDOI} \doi{10.1029/2004JA010649} \end{APACrefDOI}
\PrintBackRefs{\CurrentBib}

\bibitem [\protect \citeauthoryear {%
Klinger%
\ \BBA {} {Mayer-G\"urr}%
}{%
Klinger%
\ \BBA {} {Mayer-G\"urr}%
}{%
{\protect \APACyear {2016}}%
}]{%
Klinger2016}
\APACinsertmetastar {%
Klinger2016}%
\begin{APACrefauthors}%
Klinger, B.%
\BCBT {}\ \BBA {} {Mayer-G\"urr}, T.%
\end{APACrefauthors}%
\unskip\
\newblock
\APACrefYearMonthDay{2016}{}{}.
\newblock
{\BBOQ}\APACrefatitle {{The role of accelerometer data calibration within GRACE
  gravity field recovery: Results from ITSG-Grace2016}} {{The role of
  accelerometer data calibration within GRACE gravity field recovery: Results
  from ITSG-Grace2016}}.{\BBCQ}
\newblock
\APACjournalVolNumPages{Advances in Space\ {Research}}{58}{9}{1597-1609}.
\newblock
\begin{APACrefDOI} \doi{10.1016/j.asr.2016.08.007} \end{APACrefDOI}
\PrintBackRefs{\CurrentBib}

\bibitem [\protect \citeauthoryear {%
Knipp%
, Hapgood%
\BCBL {}\ \BBA {} Welling%
}{%
Knipp%
\ \protect \BOthers {.}}{%
{\protect \APACyear {2018}}%
}]{%
Knipp2018a}
\APACinsertmetastar {%
Knipp2018a}%
\begin{APACrefauthors}%
Knipp, D\BPBI J.%
, Hapgood, M\BPBI A.%
\BCBL {}\ \BBA {} Welling, D.%
\end{APACrefauthors}%
\unskip\
\newblock
\APACrefYearMonthDay{2018}{}{}.
\newblock
{\BBOQ}\APACrefatitle {Communicating Uncertainty and Reliability in Space
  Weather Data, Models, and Applications} {Communicating uncertainty and
  reliability in space weather data, models, and applications}.{\BBCQ}
\newblock
\APACjournalVolNumPages{Space\ Weather}{16}{10}{1453-1454}.
\newblock
\begin{APACrefDOI} \doi{10.1029/2018SW002083} \end{APACrefDOI}
\PrintBackRefs{\CurrentBib}

\bibitem [\protect \citeauthoryear {%
Knipp%
\ \protect \BOthers {.}}{%
Knipp%
\ \protect \BOthers {.}}{%
{\protect \APACyear {2013}}%
}]{%
Knipp2013}
\APACinsertmetastar {%
Knipp2013}%
\begin{APACrefauthors}%
Knipp, D\BPBI J.%
, Kilcommons, L.%
, Hunt, L.%
, Mlynczak, M.%
, Pilipenko, V.%
, Bowman, B.%
\BDBL {}Drake, K.%
\end{APACrefauthors}%
\unskip\
\newblock
\APACrefYearMonthDay{2013}{}{}.
\newblock
{\BBOQ}\APACrefatitle {Thermospheric damping response to sheath-enhanced
  geospace storms} {Thermospheric damping response to sheath-enhanced geospace
  storms}.{\BBCQ}
\newblock
\APACjournalVolNumPages{Geophysical\ Research\ Letters}{40}{7}{1263--1267}.
\newblock
\begin{APACrefDOI} \doi{10.1002/grl.50197} \end{APACrefDOI}
\PrintBackRefs{\CurrentBib}

\bibitem [\protect \citeauthoryear {%
Knipp%
\ \protect \BOthers {.}}{%
Knipp%
\ \protect \BOthers {.}}{%
{\protect \APACyear {2017}}%
}]{%
Knipp2017a}
\APACinsertmetastar {%
Knipp2017a}%
\begin{APACrefauthors}%
Knipp, D\BPBI J.%
, Pette, D\BPBI V.%
, Kilcommons, L\BPBI M.%
, Isaacs, T\BPBI L.%
, Cruz, A\BPBI A.%
, Mlynczak, M\BPBI G.%
\BDBL {}Lin, C\BPBI Y.%
\end{APACrefauthors}%
\unskip\
\newblock
\APACrefYearMonthDay{2017}{}{}.
\newblock
{\BBOQ}\APACrefatitle {Thermospheric Nitric Oxide Response to Shock-led Storms}
  {Thermospheric nitric oxide response to shock-led storms}.{\BBCQ}
\newblock
\APACjournalVolNumPages{Space\ Weather}{15}{2}{325-342}.
\newblock
\begin{APACrefDOI} \doi{10.1002/2016SW001567} \end{APACrefDOI}
\PrintBackRefs{\CurrentBib}

\bibitem [\protect \citeauthoryear {%
Kockarts%
}{%
Kockarts%
}{%
{\protect \APACyear {1980}}%
}]{%
Kockarts1980}
\APACinsertmetastar {%
Kockarts1980}%
\begin{APACrefauthors}%
Kockarts, G.%
\end{APACrefauthors}%
\unskip\
\newblock
\APACrefYearMonthDay{1980}{}{}.
\newblock
{\BBOQ}\APACrefatitle {Nitric oxide cooling in the terrestrial thermosphere}
  {Nitric oxide cooling in the terrestrial thermosphere}.{\BBCQ}
\newblock
\APACjournalVolNumPages{Geophysical\ Research\ Letters}{7}{2}{137-140}.
\newblock
\begin{APACrefDOI} \doi{10.1029/GL007i002p00137} \end{APACrefDOI}
\PrintBackRefs{\CurrentBib}

\bibitem [\protect \citeauthoryear {%
Krauss%
\ \protect \BOthers {.}}{%
Krauss%
\ \protect \BOthers {.}}{%
{\protect \APACyear {2012}}%
}]{%
Krauss2012}
\APACinsertmetastar {%
Krauss2012}%
\begin{APACrefauthors}%
Krauss, S.%
, Fichtinger, B.%
, Lammer, H.%
, amd Yu. N.~Kulikov, W\BPBI H.%
, Ribas, I.%
, Shematovich, V\BPBI I.%
\BDBL {}Hanslmeier, A.%
\end{APACrefauthors}%
\unskip\
\newblock
\APACrefYearMonthDay{2012}{}{}.
\newblock
{\BBOQ}\APACrefatitle {{Solar flares as proxy for the young Sun: satellite
  observed thermosphere response to an X17.2 flare of Earth's upper
  atmosphere}} {{Solar flares as proxy for the young Sun: satellite observed
  thermosphere response to an X17.2 flare of Earth's upper atmosphere}}.{\BBCQ}
\newblock
\APACjournalVolNumPages{Annales\ Geophysicae}{30}{}{1129-1141}.
\newblock
\begin{APACrefDOI} \doi{10.5194/angeo-30-1129-2012} \end{APACrefDOI}
\PrintBackRefs{\CurrentBib}

\bibitem [\protect \citeauthoryear {%
Krauss%
, Temmer%
\BCBL {}\ \BBA {} Vennerstrom%
}{%
Krauss%
\ \protect \BOthers {.}}{%
{\protect \APACyear {2018}}%
}]{%
Krauss2018}
\APACinsertmetastar {%
Krauss2018}%
\begin{APACrefauthors}%
Krauss, S.%
, Temmer, M.%
\BCBL {}\ \BBA {} Vennerstrom, S.%
\end{APACrefauthors}%
\unskip\
\newblock
\APACrefYearMonthDay{2018}{}{}.
\newblock
{\BBOQ}\APACrefatitle {{Multiple satellite analysis of the Earth's thermosphere
  and interplanetary magnetic field variations due to ICME/CIR events during
  2003-2015}} {{Multiple satellite analysis of the Earth's thermosphere and
  interplanetary magnetic field variations due to ICME/CIR events during
  2003-2015}}.{\BBCQ}
\newblock
\APACjournalVolNumPages{Journal\ of\ Geophysical\ Research:\ Space\
  Physics}{123}{10}{8884-8894}.
\newblock
\begin{APACrefDOI} \doi{10.1029/2018JA025778} \end{APACrefDOI}
\PrintBackRefs{\CurrentBib}

\bibitem [\protect \citeauthoryear {%
Krauss%
, Temmer%
, Veronig%
, Baur%
\BCBL {}\ \BBA {} Lammer%
}{%
Krauss%
\ \protect \BOthers {.}}{%
{\protect \APACyear {2015}}%
}]{%
Krauss2015}
\APACinsertmetastar {%
Krauss2015}%
\begin{APACrefauthors}%
Krauss, S.%
, Temmer, M.%
, Veronig, A.%
, Baur, O.%
\BCBL {}\ \BBA {} Lammer, H.%
\end{APACrefauthors}%
\unskip\
\newblock
\APACrefYearMonthDay{2015}{}{}.
\newblock
{\BBOQ}\APACrefatitle {Thermosphere and geomagnetic response to interplanetary
  coronal mass ejections observed by {ACE} and {GRACE}: {S}tatistical results.}
  {Thermosphere and geomagnetic response to interplanetary coronal mass
  ejections observed by {ACE} and {GRACE}: {S}tatistical results.}{\BBCQ}
\newblock
\APACjournalVolNumPages{Journal\ of\ Geophysical\ Research:\ Space\
  Physics}{120}{10}{8848--8860}.
\newblock
\begin{APACrefDOI} \doi{10.1002/2015JA021702} \end{APACrefDOI}
\PrintBackRefs{\CurrentBib}

\bibitem [\protect \citeauthoryear {%
Lakhina%
\ \BBA {} Tsurutani%
}{%
Lakhina%
\ \BBA {} Tsurutani%
}{%
{\protect \APACyear {2017}}%
}]{%
Lakhina2017}
\APACinsertmetastar {%
Lakhina2017}%
\begin{APACrefauthors}%
Lakhina, G\BPBI S.%
\BCBT {}\ \BBA {} Tsurutani, B\BPBI T.%
\end{APACrefauthors}%
\unskip\
\newblock
\APACrefYearMonthDay{2017}{}{}.
\newblock
{\BBOQ}\APACrefatitle {Satellite drag effects due to uplifted oxygen neutrals
  during super magnetic storms} {Satellite drag effects due to uplifted oxygen
  neutrals during super magnetic storms}.{\BBCQ}
\newblock
\APACjournalVolNumPages{Nonlinear\ Processes\ in\ Geophysics}{24}{}{745-750}.
\newblock
\begin{APACrefDOI} \doi{10.5194/npg-24-745-2017} \end{APACrefDOI}
\PrintBackRefs{\CurrentBib}

\bibitem [\protect \citeauthoryear {%
Lanzerotti%
}{%
Lanzerotti%
}{%
{\protect \APACyear {2015}}%
}]{%
Lanzerotti2015}
\APACinsertmetastar {%
Lanzerotti2015}%
\begin{APACrefauthors}%
Lanzerotti, L\BPBI J.%
\end{APACrefauthors}%
\unskip\
\newblock
\APACrefYearMonthDay{2015}{}{}.
\newblock
{\BBOQ}\APACrefatitle {{Space Weather Strategy and Action Plan: The National
  Program Is Rolled Out}} {{Space Weather Strategy and Action Plan: The
  National Program Is Rolled Out}}.{\BBCQ}
\newblock
\APACjournalVolNumPages{Space\ Weather}{13}{12}{824-825}.
\newblock
\begin{APACrefDOI} \doi{10.1002/2015SW001334} \end{APACrefDOI}
\PrintBackRefs{\CurrentBib}

\bibitem [\protect \citeauthoryear {%
Lewis%
}{%
Lewis%
}{%
{\protect \APACyear {2019}}%
}]{%
Lewis2019}
\APACinsertmetastar {%
Lewis2019}%
\begin{APACrefauthors}%
Lewis, C\BPBI D.%
\end{APACrefauthors}%
\unskip\
\newblock
\APACrefYearMonthDay{2019}{}{}.
\newblock
{\BBOQ}\APACrefatitle {{The DARPA Space Environment Exploitation (SEE)
  Program}} {{The DARPA Space Environment Exploitation (SEE) Program}}.{\BBCQ}
\newblock
\BIn{} \APACrefbtitle {{2019 AGU Chapman Conference on Scientific Challenges
  Pertaining to Space Weather Forecasting Including Extremes}.} {{2019 AGU
  Chapman Conference on Scientific Challenges Pertaining to Space Weather
  Forecasting Including Extremes}.}
\newblock
\APACaddressPublisher{Pasadena, CA, 11-15 February}{}.
\PrintBackRefs{\CurrentBib}

\bibitem [\protect \citeauthoryear {%
Liu%
\ \BBA {} L\"uhr%
}{%
Liu%
\ \BBA {} L\"uhr%
}{%
{\protect \APACyear {2005}}%
}]{%
Liu2005a}
\APACinsertmetastar {%
Liu2005a}%
\begin{APACrefauthors}%
Liu, H.%
\BCBT {}\ \BBA {} L\"uhr, H.%
\end{APACrefauthors}%
\unskip\
\newblock
\APACrefYearMonthDay{2005}{}{}.
\newblock
{\BBOQ}\APACrefatitle {Strong disturbance of the upper thermospheric density
  due to magnetic storms: {CHAMP} observations} {Strong disturbance of the
  upper thermospheric density due to magnetic storms: {CHAMP}
  observations}.{\BBCQ}
\newblock
\APACjournalVolNumPages{Journal\ of\ Geophysical\ Research}{110}{A9}{1--9}.
\newblock
\begin{APACrefDOI} \doi{10.1029/2004JA010908} \end{APACrefDOI}
\PrintBackRefs{\CurrentBib}

\bibitem [\protect \citeauthoryear {%
Lu%
, Richmond%
, L\"{u}hr%
\BCBL {}\ \BBA {} Paxton%
}{%
Lu%
\ \protect \BOthers {.}}{%
{\protect \APACyear {2016}}%
}]{%
Lu2016a}
\APACinsertmetastar {%
Lu2016a}%
\begin{APACrefauthors}%
Lu, G.%
, Richmond, A\BPBI D.%
, L\"{u}hr, H.%
\BCBL {}\ \BBA {} Paxton, L.%
\end{APACrefauthors}%
\unskip\
\newblock
\APACrefYearMonthDay{2016}{}{}.
\newblock
{\BBOQ}\APACrefatitle {High-latitude energy input and its impact on the
  thermosphere} {High-latitude energy input and its impact on the
  thermosphere}.{\BBCQ}
\newblock
\APACjournalVolNumPages{Journal\ of\ Geophysical\ Research:\ Space\
  Physics}{121}{7}{7108--7124}.
\newblock
\begin{APACrefDOI} \doi{10.1002/2015JA022294} \end{APACrefDOI}
\PrintBackRefs{\CurrentBib}

\bibitem [\protect \citeauthoryear {%
Lugaz%
\ \protect \BOthers {.}}{%
Lugaz%
\ \protect \BOthers {.}}{%
{\protect \APACyear {2016}}%
}]{%
Lugaz2016}
\APACinsertmetastar {%
Lugaz2016}%
\begin{APACrefauthors}%
Lugaz, N.%
, Farrugia, C\BPBI J.%
, Winslow, R\BPBI M.%
, Al-Haddad, N.%
, Kilpua, E\BPBI K\BPBI J.%
\BCBL {}\ \BBA {} Riley, P.%
\end{APACrefauthors}%
\unskip\
\newblock
\APACrefYearMonthDay{2016}{}{}.
\newblock
{\BBOQ}\APACrefatitle {Factors Affecting the Geo-effectiveness of Shocks and
  Sheaths at 1 {AU}} {Factors affecting the geo-effectiveness of shocks and
  sheaths at 1 {AU}}.{\BBCQ}
\newblock
\APACjournalVolNumPages{Journal\ of\ Geophysical\ Research:\ Space\
  Physics}{120}{11}{10,861--10,879}.
\newblock
\begin{APACrefDOI} \doi{10.1002/2016JA023100} \end{APACrefDOI}
\PrintBackRefs{\CurrentBib}

\bibitem [\protect \citeauthoryear {%
Marcos%
, Delay%
\BCBL {}\ \BBA {} Sutton%
}{%
Marcos%
\ \protect \BOthers {.}}{%
{\protect \APACyear {2010}}%
}]{%
Marcos2010}
\APACinsertmetastar {%
Marcos2010}%
\begin{APACrefauthors}%
Marcos, F\BPBI A.%
, Delay, S\BPBI H.%
\BCBL {}\ \BBA {} Sutton, E\BPBI K.%
\end{APACrefauthors}%
\unskip\
\newblock
\APACrefYearMonthDay{2010}{}{}.
\newblock
{\BBOQ}\APACrefatitle {Toward next level satellite drag modelling} {Toward next
  level satellite drag modelling}.{\BBCQ}
\newblock
\BIn{} \APACrefbtitle {{AIAA Guidance, Navigation, and Control Conference}.}
  {{AIAA Guidance, Navigation, and Control Conference}.}
\newblock
\APACaddressPublisher{Toronto, Ontario Canada}{}.
\PrintBackRefs{\CurrentBib}

\bibitem [\protect \citeauthoryear {%
Mayr%
\ \protect \BOthers {.}}{%
Mayr%
\ \protect \BOthers {.}}{%
{\protect \APACyear {1990}}%
}]{%
Mayr1990}
\APACinsertmetastar {%
Mayr1990}%
\begin{APACrefauthors}%
Mayr, H\BPBI G.%
, Harris, I.%
, Herrero, F\BPBI A.%
, Spencer, N\BPBI W.%
, Varosi, F.%
\BCBL {}\ \BBA {} Pesnell, W\BPBI D.%
\end{APACrefauthors}%
\unskip\
\newblock
\APACrefYearMonthDay{1990}{}{}.
\newblock
{\BBOQ}\APACrefatitle {Thermospheric gravity waves: Observations and
  interpretation using the transfer function model ({TFM})} {Thermospheric
  gravity waves: Observations and interpretation using the transfer function
  model ({TFM})}.{\BBCQ}
\newblock
\APACjournalVolNumPages{Space\ Science\ Reviews}{54}{3}{297--375}.
\newblock
\begin{APACrefDOI} \doi{10.1007/BF00177800} \end{APACrefDOI}
\PrintBackRefs{\CurrentBib}

\bibitem [\protect \citeauthoryear {%
{McLaughlin}%
, Mance%
\BCBL {}\ \BBA {} Lichtenberg%
}{%
{McLaughlin}%
\ \protect \BOthers {.}}{%
{\protect \APACyear {2011}}%
}]{%
McLaughlin2011}
\APACinsertmetastar {%
McLaughlin2011}%
\begin{APACrefauthors}%
{McLaughlin}, C\BPBI A.%
, Mance, S.%
\BCBL {}\ \BBA {} Lichtenberg, T.%
\end{APACrefauthors}%
\unskip\
\newblock
\APACrefYearMonthDay{2011}{}{}.
\newblock
{\BBOQ}\APACrefatitle {{Drag Coefficient Estimation in Orbit Determination}}
  {{Drag Coefficient Estimation in Orbit Determination}}.{\BBCQ}
\newblock
\APACjournalVolNumPages{The\ Journal\ of\ the\ Astronautical\
  Sciences}{58}{3}{513-530}.
\newblock
\begin{APACrefDOI} \doi{10.1007/BF03321183} \end{APACrefDOI}
\PrintBackRefs{\CurrentBib}

\bibitem [\protect \citeauthoryear {%
Mehta%
, Linares%
\BCBL {}\ \BBA {} Sutton%
}{%
Mehta%
\ \protect \BOthers {.}}{%
{\protect \APACyear {2018}}%
}]{%
Mehta2018}
\APACinsertmetastar {%
Mehta2018}%
\begin{APACrefauthors}%
Mehta, P\BPBI M.%
, Linares, R.%
\BCBL {}\ \BBA {} Sutton, E\BPBI K.%
\end{APACrefauthors}%
\unskip\
\newblock
\APACrefYearMonthDay{2018}{}{}.
\newblock
{\BBOQ}\APACrefatitle {A Quasi-Physical Dynamic Reduced Order Model for
  Thermospheric Mass Density via Hermitian Space-Dynamic Mode Decomposition} {A
  quasi-physical dynamic reduced order model for thermospheric mass density via
  hermitian space-dynamic mode decomposition}.{\BBCQ}
\newblock
\APACjournalVolNumPages{Space\ Weather}{16}{5}{569-588}.
\newblock
\begin{APACrefDOI} \doi{10.1029/2018SW001840} \end{APACrefDOI}
\PrintBackRefs{\CurrentBib}

\bibitem [\protect \citeauthoryear {%
Mlynczak%
\ \protect \BOthers {.}}{%
Mlynczak%
\ \protect \BOthers {.}}{%
{\protect \APACyear {2014}}%
}]{%
Mlynczak2014}
\APACinsertmetastar {%
Mlynczak2014}%
\begin{APACrefauthors}%
Mlynczak, M\BPBI G.%
, Hunt, L\BPBI A.%
, Mertens, C\BPBI J.%
, Thomas~Marshall, B.%
, Russell~III, J\BPBI M.%
, Woods, T.%
\BDBL {}Gordley, L\BPBI L.%
\end{APACrefauthors}%
\unskip\
\newblock
\APACrefYearMonthDay{2014}{}{}.
\newblock
{\BBOQ}\APACrefatitle {Influence of solar variability on the infrared radiative
  cooling of the thermosphere from 2002 to 2014} {Influence of solar
  variability on the infrared radiative cooling of the thermosphere from 2002
  to 2014}.{\BBCQ}
\newblock
\APACjournalVolNumPages{Geophysical\ Research\ Letters}{41}{7}{2508-2513}.
\newblock
\begin{APACrefDOI} \doi{10.1002/2014GL059556} \end{APACrefDOI}
\PrintBackRefs{\CurrentBib}

\bibitem [\protect \citeauthoryear {%
Mlynczak%
\ \protect \BOthers {.}}{%
Mlynczak%
\ \protect \BOthers {.}}{%
{\protect \APACyear {2003}}%
}]{%
Mlynczak2003}
\APACinsertmetastar {%
Mlynczak2003}%
\begin{APACrefauthors}%
Mlynczak, M\BPBI G.%
, {Martin-Torres}, F\BPBI J.%
, Russell, J.%
, Beaumont, K.%
, Jacobson, S.%
, Kozyra, J.%
\BDBL {}Paxton, L.%
\end{APACrefauthors}%
\unskip\
\newblock
\APACrefYearMonthDay{2003}{}{}.
\newblock
{\BBOQ}\APACrefatitle {{The natural thermostat of nitric oxide emission at 5.3
  $\mu$m in the thermosphere observed during the solar storms of April 2002}}
  {{The natural thermostat of nitric oxide emission at 5.3 $\mu$m in the
  thermosphere observed during the solar storms of April 2002}}.{\BBCQ}
\newblock
\APACjournalVolNumPages{Geophysical\ Research\ Letters}{30}{21}{}.
\newblock
\begin{APACrefDOI} \doi{10.1029/2003GL017693} \end{APACrefDOI}
\PrintBackRefs{\CurrentBib}

\bibitem [\protect \citeauthoryear {%
Moe%
\ \BBA {} Moe%
}{%
Moe%
\ \BBA {} Moe%
}{%
{\protect \APACyear {2005}}%
}]{%
Moe2005}
\APACinsertmetastar {%
Moe2005}%
\begin{APACrefauthors}%
Moe, K.%
\BCBT {}\ \BBA {} Moe, M\BPBI M.%
\end{APACrefauthors}%
\unskip\
\newblock
\APACrefYearMonthDay{2005}{}{}.
\newblock
{\BBOQ}\APACrefatitle {Gas-surface interactions and satellite drag
  coefficients} {Gas-surface interactions and satellite drag
  coefficients}.{\BBCQ}
\newblock
\APACjournalVolNumPages{Planetary\ and\ Space\ Science}{53}{8}{793-801}.
\newblock
\begin{APACrefDOI} \doi{10.1016/j.pss.2005.03.005} \end{APACrefDOI}
\PrintBackRefs{\CurrentBib}

\bibitem [\protect \citeauthoryear {%
Oliveira%
\ \protect \BOthers {.}}{%
Oliveira%
\ \protect \BOthers {.}}{%
{\protect \APACyear {2018}}%
}]{%
Oliveira2018b}
\APACinsertmetastar {%
Oliveira2018b}%
\begin{APACrefauthors}%
Oliveira, D\BPBI M.%
, Arel, D.%
, Raeder, J.%
, Zesta, E.%
, Ngwira, C\BPBI M.%
, Carter, B\BPBI A.%
\BDBL {}Gjerloev, J\BPBI W.%
\end{APACrefauthors}%
\unskip\
\newblock
\APACrefYearMonthDay{2018}{}{}.
\newblock
{\BBOQ}\APACrefatitle {Geomagnetically induced currents caused by
  interplanetary shocks with different impact angles and speeds}
  {Geomagnetically induced currents caused by interplanetary shocks with
  different impact angles and speeds}.{\BBCQ}
\newblock
\APACjournalVolNumPages{Space\ Weather}{16}{6}{636-647}.
\newblock
\begin{APACrefDOI} \doi{10.1029/2018SW001880} \end{APACrefDOI}
\PrintBackRefs{\CurrentBib}

\bibitem [\protect \citeauthoryear {%
Oliveira%
, Zesta%
, Schuck%
\BCBL {}\ \BBA {} Sutton%
}{%
Oliveira%
\ \protect \BOthers {.}}{%
{\protect \APACyear {2017}}%
}]{%
Oliveira2017c}
\APACinsertmetastar {%
Oliveira2017c}%
\begin{APACrefauthors}%
Oliveira, D\BPBI M.%
, Zesta, E.%
, Schuck, P\BPBI W.%
\BCBL {}\ \BBA {} Sutton, E\BPBI K.%
\end{APACrefauthors}%
\unskip\
\newblock
\APACrefYearMonthDay{2017}{}{}.
\newblock
{\BBOQ}\APACrefatitle {Thermosphere global time response to geomagnetic storms
  caused by coronal mass ejections} {Thermosphere global time response to
  geomagnetic storms caused by coronal mass ejections}.{\BBCQ}
\newblock
\APACjournalVolNumPages{Journal\ of\ Geophysical\ Research:\ Space\
  Physics}{122}{10}{10,762-10,782}.
\newblock
\begin{APACrefDOI} \doi{10.1002/2017JA024006} \end{APACrefDOI}
\PrintBackRefs{\CurrentBib}

\bibitem [\protect \citeauthoryear {%
Ozturk%
, Zou%
, Ridley%
\BCBL {}\ \BBA {} Slavin%
}{%
Ozturk%
\ \protect \BOthers {.}}{%
{\protect \APACyear {2018}}%
}]{%
Ozturk2018}
\APACinsertmetastar {%
Ozturk2018}%
\begin{APACrefauthors}%
Ozturk, D\BPBI S.%
, Zou, S.%
, Ridley, A\BPBI J.%
\BCBL {}\ \BBA {} Slavin, J\BPBI A.%
\end{APACrefauthors}%
\unskip\
\newblock
\APACrefYearMonthDay{2018}{}{}.
\newblock
{\BBOQ}\APACrefatitle {{Modeling study of geospace system response to the solar
  wind dynamic pressure enhancement on March 17, 2015.}} {{Modeling study of
  geospace system response to the solar wind dynamic pressure enhancement on
  March 17, 2015.}}{\BBCQ}
\newblock
\APACjournalVolNumPages{Journal\ of\ Geophysical\ Research:\ Space\
  Physics}{123}{4}{2974-2989}.
\newblock
\begin{APACrefDOI} \doi{10.1002/2017JA025099} \end{APACrefDOI}
\PrintBackRefs{\CurrentBib}

\bibitem [\protect \citeauthoryear {%
Pardini%
\ \BBA {} Anselmo%
}{%
Pardini%
\ \BBA {} Anselmo%
}{%
{\protect \APACyear {2009}}%
}]{%
Pardini2009}
\APACinsertmetastar {%
Pardini2009}%
\begin{APACrefauthors}%
Pardini, C.%
\BCBT {}\ \BBA {} Anselmo, L.%
\end{APACrefauthors}%
\unskip\
\newblock
\APACrefYearMonthDay{2009}{}{}.
\newblock
{\BBOQ}\APACrefatitle {{Assessment of the consequences of the Fengyun-1C
  breakup in low Earth orbit}} {{Assessment of the consequences of the
  Fengyun-1C breakup in low Earth orbit}}.{\BBCQ}
\newblock
\APACjournalVolNumPages{Advances in Space\ {Research}}{44}{5}{545-557}.
\newblock
\begin{APACrefDOI} \doi{10.1016/j.asr.2009.04.014} \end{APACrefDOI}
\PrintBackRefs{\CurrentBib}

\bibitem [\protect \citeauthoryear {%
Picone%
, Hedin%
, Drob%
\BCBL {}\ \BBA {} Aikin%
}{%
Picone%
\ \protect \BOthers {.}}{%
{\protect \APACyear {2002}}%
}]{%
Picone2002}
\APACinsertmetastar {%
Picone2002}%
\begin{APACrefauthors}%
Picone, J\BPBI M.%
, Hedin, A\BPBI E.%
, Drob, D\BPBI P.%
\BCBL {}\ \BBA {} Aikin, A\BPBI C.%
\end{APACrefauthors}%
\unskip\
\newblock
\APACrefYearMonthDay{2002}{}{}.
\newblock
{\BBOQ}\APACrefatitle {{NRLMSISE-00} empirical model of the atmosphere:
  {S}tatistical comparisons and scientific issues} {{NRLMSISE-00} empirical
  model of the atmosphere: {S}tatistical comparisons and scientific
  issues}.{\BBCQ}
\newblock
\APACjournalVolNumPages{Journal\ of\ Geophysical\ Research}{107}{A12}{SIA
  15-1--SIA 15-16}.
\newblock
\begin{APACrefDOI} \doi{10.1029/2002JA009430} \end{APACrefDOI}
\PrintBackRefs{\CurrentBib}

\bibitem [\protect \citeauthoryear {%
Prieto%
, Graziano%
\BCBL {}\ \BBA {} Roberts%
}{%
Prieto%
\ \protect \BOthers {.}}{%
{\protect \APACyear {2014}}%
}]{%
Prieto2014}
\APACinsertmetastar {%
Prieto2014}%
\begin{APACrefauthors}%
Prieto, D\BPBI M.%
, Graziano, B\BPBI P.%
\BCBL {}\ \BBA {} Roberts, P\BPBI C\BPBI E.%
\end{APACrefauthors}%
\unskip\
\newblock
\APACrefYearMonthDay{2014}{}{}.
\newblock
{\BBOQ}\APACrefatitle {Spacecraft drag modelling} {Spacecraft drag
  modelling}.{\BBCQ}
\newblock
\APACjournalVolNumPages{Progress\ in\ Aerospace\ Sciences}{64}{}{56-65}.
\newblock
\begin{APACrefDOI} \doi{10.1016/j.paerosci.2013.09.001} \end{APACrefDOI}
\PrintBackRefs{\CurrentBib}

\bibitem [\protect \citeauthoryear {%
Pr\"olss%
}{%
Pr\"olss%
}{%
{\protect \APACyear {2011}}%
}]{%
Prolss2011}
\APACinsertmetastar {%
Prolss2011}%
\begin{APACrefauthors}%
Pr\"olss, G.%
\end{APACrefauthors}%
\unskip\
\newblock
\APACrefYearMonthDay{2011}{}{}.
\newblock
{\BBOQ}\APACrefatitle {Density Perturbations in the Upper Atmosphere Caused by
  the Dissipation of Solar Wind Energy} {Density perturbations in the upper
  atmosphere caused by the dissipation of solar wind energy}.{\BBCQ}
\newblock
\APACjournalVolNumPages{Surveys\ in\ Geophysics}{32}{2}{101--195}.
\newblock
\begin{APACrefDOI} \doi{10.1007/s10712-010-9104-0} \end{APACrefDOI}
\PrintBackRefs{\CurrentBib}

\bibitem [\protect \citeauthoryear {%
Ramesh%
}{%
Ramesh%
}{%
{\protect \APACyear {2010}}%
}]{%
Ramesh2010}
\APACinsertmetastar {%
Ramesh2010}%
\begin{APACrefauthors}%
Ramesh, K\BPBI B.%
\end{APACrefauthors}%
\unskip\
\newblock
\APACrefYearMonthDay{2010}{}{}.
\newblock
{\BBOQ}\APACrefatitle {Coronal mass ejections and sunspots -- {A} perspective}
  {Coronal mass ejections and sunspots -- {A} perspective}.{\BBCQ}
\newblock
\APACjournalVolNumPages{The\ Astrophysical\ Journal\
  Letters}{712}{1}{L77--L80}.
\newblock
\begin{APACrefDOI} \doi{10.1088/2041-8205/712/1/L77} \end{APACrefDOI}
\PrintBackRefs{\CurrentBib}

\bibitem [\protect \citeauthoryear {%
Reigber%
, L\"uhr%
\BCBL {}\ \BBA {} Schwintzer%
}{%
Reigber%
\ \protect \BOthers {.}}{%
{\protect \APACyear {2002}}%
}]{%
Reigber2002a}
\APACinsertmetastar {%
Reigber2002a}%
\begin{APACrefauthors}%
Reigber, C.%
, L\"uhr, H.%
\BCBL {}\ \BBA {} Schwintzer, P.%
\end{APACrefauthors}%
\unskip\
\newblock
\APACrefYearMonthDay{2002}{}{}.
\newblock
{\BBOQ}\APACrefatitle {{CHAMP} mission status} {{CHAMP} mission status}.{\BBCQ}
\newblock
\APACjournalVolNumPages{Advances in Space\ {Research}}{30}{2}{129-134}.
\newblock
\begin{APACrefDOI} \doi{10.1016/S0273-1177(02)00276-4} \end{APACrefDOI}
\PrintBackRefs{\CurrentBib}

\bibitem [\protect \citeauthoryear {%
Richardson%
\ \BBA {} Cane%
}{%
Richardson%
\ \BBA {} Cane%
}{%
{\protect \APACyear {2010}}%
}]{%
Richardson2010b}
\APACinsertmetastar {%
Richardson2010b}%
\begin{APACrefauthors}%
Richardson, I\BPBI G.%
\BCBT {}\ \BBA {} Cane, H\BPBI V.%
\end{APACrefauthors}%
\unskip\
\newblock
\APACrefYearMonthDay{2010}{}{}.
\newblock
{\BBOQ}\APACrefatitle {Interplanetary circumstances of quasi-perpendicular
  interplanetary shocks in {1996-2005}} {Interplanetary circumstances of
  quasi-perpendicular interplanetary shocks in {1996-2005}}.{\BBCQ}
\newblock
\APACjournalVolNumPages{Journal\ of\ Geophysical\ Research}{115}{A7}{}.
\newblock
\begin{APACrefDOI} \doi{10.1029/2009JA015039} \end{APACrefDOI}
\PrintBackRefs{\CurrentBib}

\bibitem [\protect \citeauthoryear {%
A.~Richmond%
\ \BBA {} Lu%
}{%
A.~Richmond%
\ \BBA {} Lu%
}{%
{\protect \APACyear {2000}}%
}]{%
Richmond2000}
\APACinsertmetastar {%
Richmond2000}%
\begin{APACrefauthors}%
Richmond, A.%
\BCBT {}\ \BBA {} Lu, G.%
\end{APACrefauthors}%
\unskip\
\newblock
\APACrefYearMonthDay{2000}{}{}.
\newblock
{\BBOQ}\APACrefatitle {{Upper-atmospheric effects of magnetic storms: A brief
  tutorial}} {{Upper-atmospheric effects of magnetic storms: A brief
  tutorial}}.{\BBCQ}
\newblock
\APACjournalVolNumPages{Journal\ of\ Atmospheric\ and\ Solar-Terrestrial\
  Physics}{62}{12}{1115-1127}.
\newblock
\begin{APACrefDOI} \doi{10.1016/S1364-6826(00)00094-8} \end{APACrefDOI}
\PrintBackRefs{\CurrentBib}

\bibitem [\protect \citeauthoryear {%
Richmond%
\ \BBA {} Matsushita%
}{%
Richmond%
\ \BBA {} Matsushita%
}{%
{\protect \APACyear {1975}}%
}]{%
Richmond1975}
\APACinsertmetastar {%
Richmond1975}%
\begin{APACrefauthors}%
Richmond, A\BPBI D.%
\BCBT {}\ \BBA {} Matsushita, S.%
\end{APACrefauthors}%
\unskip\
\newblock
\APACrefYearMonthDay{1975}{}{}.
\newblock
{\BBOQ}\APACrefatitle {Thermospheric response to a magnetic substorm}
  {Thermospheric response to a magnetic substorm}.{\BBCQ}
\newblock
\APACjournalVolNumPages{Journal\ of\ Geophysical\ Research}{80}{19}{2839-2850}.
\newblock
\begin{APACrefDOI} \doi{10.1029/JA080i019p02839} \end{APACrefDOI}
\PrintBackRefs{\CurrentBib}

\bibitem [\protect \citeauthoryear {%
Rudd%
, Oliveira%
, Bhaskar%
\BCBL {}\ \BBA {} Halford%
}{%
Rudd%
\ \protect \BOthers {.}}{%
{\protect \APACyear {2019}}%
}]{%
Rudd2019}
\APACinsertmetastar {%
Rudd2019}%
\begin{APACrefauthors}%
Rudd, J\BPBI T.%
, Oliveira, D\BPBI M.%
, Bhaskar, A.%
\BCBL {}\ \BBA {} Halford, A\BPBI J.%
\end{APACrefauthors}%
\unskip\
\newblock
\APACrefYearMonthDay{2019}{}{}.
\newblock
{\BBOQ}\APACrefatitle {How do interplanetary shock impact angles control the
  size of the geoeffective magnetosphere?} {How do interplanetary shock impact
  angles control the size of the geoeffective magnetosphere?}{\BBCQ}
\newblock
\APACjournalVolNumPages{Advances in Space\ {Research}}{63}{1}{317-326}.
\newblock
\begin{APACrefDOI} \doi{10.1016/j.asr.2018.09.013} \end{APACrefDOI}
\PrintBackRefs{\CurrentBib}

\bibitem [\protect \citeauthoryear {%
Sentman%
}{%
Sentman%
}{%
{\protect \APACyear {1961}}%
}]{%
Sentman1961}
\APACinsertmetastar {%
Sentman1961}%
\begin{APACrefauthors}%
Sentman, L\BPBI H.%
\end{APACrefauthors}%
\unskip\
\newblock
\APACrefYearMonthDay{1961}{}{}.
\newblock
\APACrefbtitle {Free Molecule Flow Theory and its Application to the
  Determination of Aerodynamic Forces} {Free molecule flow theory and its
  application to the determination of aerodynamic forces}\
  \APACbVolEdTR{}{\BTR{}}.
\newblock
\APACaddressInstitution{Arlington, VA}{Armed Services Technical Information
  Agency}.
\PrintBackRefs{\CurrentBib}

\bibitem [\protect \citeauthoryear {%
Shi%
\ \protect \BOthers {.}}{%
Shi%
\ \protect \BOthers {.}}{%
{\protect \APACyear {2017}}%
}]{%
Shi2017}
\APACinsertmetastar {%
Shi2017}%
\begin{APACrefauthors}%
Shi, Y.%
, Zesta, E.%
, Connor, H\BPBI K.%
, Su, Y\BHBI J.%
, Sutton, E\BPBI K.%
, Huang, C\BPBI Y.%
\BDBL {}Oliveira, D\BPBI M.%
\end{APACrefauthors}%
\unskip\
\newblock
\APACrefYearMonthDay{2017}{}{}.
\newblock
{\BBOQ}\APACrefatitle {High-Latitude thermosphere neutral density response to
  solar wind dynamic pressure enhancement} {High-latitude thermosphere neutral
  density response to solar wind dynamic pressure enhancement}.{\BBCQ}
\newblock
\APACjournalVolNumPages{Journal\ of\ Geophysical\ Research:\ Space\
  Physics}{122}{11}{11,559-11,578}.
\newblock
\begin{APACrefDOI} \doi{10.1002/2017JA023889} \end{APACrefDOI}
\PrintBackRefs{\CurrentBib}

\bibitem [\protect \citeauthoryear {%
Storz%
, Bowman%
\BCBL {}\ \BBA {} Branson%
}{%
Storz%
\ \protect \BOthers {.}}{%
{\protect \APACyear {2002}}%
}]{%
Storz2002}
\APACinsertmetastar {%
Storz2002}%
\begin{APACrefauthors}%
Storz, M\BPBI F.%
, Bowman, B\BPBI R.%
\BCBL {}\ \BBA {} Branson, J\BPBI L.%
\end{APACrefauthors}%
\unskip\
\newblock
\APACrefYearMonthDay{2002}{}{}.
\newblock
{\BBOQ}\APACrefatitle {{High Accuracy Satellite Drag Model (HASDM)}} {{High
  Accuracy Satellite Drag Model (HASDM)}}.{\BBCQ}
\newblock
\BIn{} \APACrefbtitle {{AIAA/AAS Astrodynamics Specialist Conference, AIAA
  2002--4886}.} {{AIAA/AAS Astrodynamics Specialist Conference, AIAA
  2002--4886}.}
\newblock
\APACaddressPublisher{Monterey, CA}{}.
\PrintBackRefs{\CurrentBib}

\bibitem [\protect \citeauthoryear {%
Sutton%
}{%
Sutton%
}{%
{\protect \APACyear {2008}}%
}]{%
Sutton2008}
\APACinsertmetastar {%
Sutton2008}%
\begin{APACrefauthors}%
Sutton, E\BPBI K.%
\end{APACrefauthors}%
\unskip\
\newblock
\APACrefYear{2008}.
\unskip\
\newblock
\APACrefbtitle {Effects of Solar Disturbances on the Thermosphere Densities and
  Winds from {CHAMP} and {GRACE} Satellite Accelerometer Data} {Effects of
  solar disturbances on the thermosphere densities and winds from {CHAMP} and
  {GRACE} satellite accelerometer data}\ \APACtypeAddressSchool {Ph.{D}
  thesis}{}{}.
\unskip\
\newblock
\APACaddressSchool {Boulder, Colorado}{University\ of\ Colorado}.
\PrintBackRefs{\CurrentBib}

\bibitem [\protect \citeauthoryear {%
Sutton%
}{%
Sutton%
}{%
{\protect \APACyear {2009}}%
}]{%
Sutton2009b}
\APACinsertmetastar {%
Sutton2009b}%
\begin{APACrefauthors}%
Sutton, E\BPBI K.%
\end{APACrefauthors}%
\unskip\
\newblock
\APACrefYearMonthDay{2009}{}{}.
\newblock
{\BBOQ}\APACrefatitle {Normalized force coefficients for satellites with
  elongated shapes} {Normalized force coefficients for satellites with
  elongated shapes}.{\BBCQ}
\newblock
\APACjournalVolNumPages{Journal\ of\ Spacecraft\ and\
  Rockets}{46}{1}{112--116}.
\newblock
\begin{APACrefDOI} \doi{10.2514/1.40940} \end{APACrefDOI}
\PrintBackRefs{\CurrentBib}

\bibitem [\protect \citeauthoryear {%
Sutton%
}{%
Sutton%
}{%
{\protect \APACyear {2018}}%
}]{%
Sutton2018}
\APACinsertmetastar {%
Sutton2018}%
\begin{APACrefauthors}%
Sutton, E\BPBI K.%
\end{APACrefauthors}%
\unskip\
\newblock
\APACrefYearMonthDay{2018}{}{}.
\newblock
{\BBOQ}\APACrefatitle {A New Method of Physics-Based Data Assimilation for the
  Quiet and Disturbed Thermosphere} {A new method of physics-based data
  assimilation for the quiet and disturbed thermosphere}.{\BBCQ}
\newblock
\APACjournalVolNumPages{Space\ Weather}{16}{6}{736-753}.
\newblock
\begin{APACrefDOI} \doi{10.1002/2017SW001785} \end{APACrefDOI}
\PrintBackRefs{\CurrentBib}

\bibitem [\protect \citeauthoryear {%
Sutton%
, Forbes%
\BCBL {}\ \BBA {} Knipp%
}{%
Sutton%
\ \protect \BOthers {.}}{%
{\protect \APACyear {2009}}%
}]{%
Sutton2009a}
\APACinsertmetastar {%
Sutton2009a}%
\begin{APACrefauthors}%
Sutton, E\BPBI K.%
, Forbes, J\BPBI M.%
\BCBL {}\ \BBA {} Knipp, D\BPBI J.%
\end{APACrefauthors}%
\unskip\
\newblock
\APACrefYearMonthDay{2009}{}{}.
\newblock
{\BBOQ}\APACrefatitle {Rapid response of the thermosphere to variations in
  {J}oule heating} {Rapid response of the thermosphere to variations in {J}oule
  heating}.{\BBCQ}
\newblock
\APACjournalVolNumPages{Journal\ of\ Geophysical\ Research}{114}{A4}{}.
\newblock
\begin{APACrefDOI} \doi{10.1029/2008JA013667} \end{APACrefDOI}
\PrintBackRefs{\CurrentBib}

\bibitem [\protect \citeauthoryear {%
Tapley%
, Bettadpur%
, Watkins%
\BCBL {}\ \BBA {} Reigber%
}{%
Tapley%
\ \protect \BOthers {.}}{%
{\protect \APACyear {2004}}%
}]{%
Tapley2004a}
\APACinsertmetastar {%
Tapley2004a}%
\begin{APACrefauthors}%
Tapley, B\BPBI D.%
, Bettadpur, S.%
, Watkins, M.%
\BCBL {}\ \BBA {} Reigber, C.%
\end{APACrefauthors}%
\unskip\
\newblock
\APACrefYearMonthDay{2004}{}{}.
\newblock
{\BBOQ}\APACrefatitle {The gravity recovery and climate experiment: Mission
  overview and early results} {The gravity recovery and climate experiment:
  Mission overview and early results}.{\BBCQ}
\newblock
\APACjournalVolNumPages{Geophysical\ Research\ Letters}{31}{9}{}.
\newblock
\begin{APACrefDOI} \doi{10.1029/2004GL019920} \end{APACrefDOI}
\PrintBackRefs{\CurrentBib}

\bibitem [\protect \citeauthoryear {%
Torge%
}{%
Torge%
}{%
{\protect \APACyear {1980}}%
}]{%
Torge1980}
\APACinsertmetastar {%
Torge1980}%
\begin{APACrefauthors}%
Torge, W.%
\end{APACrefauthors}%
\unskip\
\newblock
\APACrefYear{1980}.
\newblock
\APACrefbtitle {Geodesy} {Geodesy}.
\newblock
\APACaddressPublisher{Berlin, Germany}{de Gruyter}.
\PrintBackRefs{\CurrentBib}

\bibitem [\protect \citeauthoryear {%
Tsurutani%
\ \protect \BOthers {.}}{%
Tsurutani%
\ \protect \BOthers {.}}{%
{\protect \APACyear {2007}}%
}]{%
Tsurutani2007}
\APACinsertmetastar {%
Tsurutani2007}%
\begin{APACrefauthors}%
Tsurutani, B\BPBI T.%
, Verkhoglyadova, O\BPBI P.%
, Mannucci, A\BPBI J.%
, Araki, T.%
, Sato, A.%
, Tsuda, T.%
\BCBL {}\ \BBA {} Yumoto, K.%
\end{APACrefauthors}%
\unskip\
\newblock
\APACrefYearMonthDay{2007}{}{}.
\newblock
{\BBOQ}\APACrefatitle {{Oxygen ion up-lift and satellite drag effects during
  the 30 October 2003 daytime superfountain event}} {{Oxygen ion up-lift and
  satellite drag effects during the 30 October 2003 daytime superfountain
  event}}.{\BBCQ}
\newblock
\APACjournalVolNumPages{Annales\ Geophysicae}{25}{}{569–574}.
\newblock
\begin{APACrefDOI} \doi{10.5194/angeo-25-569-2007} \end{APACrefDOI}
\PrintBackRefs{\CurrentBib}

\bibitem [\protect \citeauthoryear {%
Vallado%
\ \BBA {} Finkleman%
}{%
Vallado%
\ \BBA {} Finkleman%
}{%
{\protect \APACyear {2014}}%
}]{%
Vallado2014}
\APACinsertmetastar {%
Vallado2014}%
\begin{APACrefauthors}%
Vallado, D\BPBI A.%
\BCBT {}\ \BBA {} Finkleman, D.%
\end{APACrefauthors}%
\unskip\
\newblock
\APACrefYearMonthDay{2014}{}{}.
\newblock
{\BBOQ}\APACrefatitle {A critical assessment of satellite drag and atmospheric
  density modeling} {A critical assessment of satellite drag and atmospheric
  density modeling}.{\BBCQ}
\newblock
\APACjournalVolNumPages{Acta\ Astronautica}{95}{}{141-165}.
\newblock
\begin{APACrefDOI} \doi{10.1016/j.actaastro.2013.10.005} \end{APACrefDOI}
\PrintBackRefs{\CurrentBib}

\bibitem [\protect \citeauthoryear {%
Wang%
}{%
Wang%
}{%
{\protect \APACyear {2010}}%
}]{%
Wang2010c}
\APACinsertmetastar {%
Wang2010c}%
\begin{APACrefauthors}%
Wang, T.%
\end{APACrefauthors}%
\unskip\
\newblock
\APACrefYearMonthDay{2010}{}{}.
\newblock
{\BBOQ}\APACrefatitle {{Analysis of Debris from the Collision of the Cosmos
  2251 and the Iridium 33 Satellites}} {{Analysis of Debris from the Collision
  of the Cosmos 2251 and the Iridium 33 Satellites}}.{\BBCQ}
\newblock
\APACjournalVolNumPages{Science\ $\&$\ Global\ Security}{18}{2}{87-118}.
\newblock
\begin{APACrefDOI} \doi{10.1080/08929882.2010.493078} \end{APACrefDOI}
\PrintBackRefs{\CurrentBib}

\bibitem [\protect \citeauthoryear {%
Weimer%
, Bowman%
, Sutton%
\BCBL {}\ \BBA {} Tobiska%
}{%
Weimer%
\ \protect \BOthers {.}}{%
{\protect \APACyear {2011}}%
}]{%
Weimer2010}
\APACinsertmetastar {%
Weimer2010}%
\begin{APACrefauthors}%
Weimer, D\BPBI R.%
, Bowman, B\BPBI R.%
, Sutton, E\BPBI K.%
\BCBL {}\ \BBA {} Tobiska, W\BPBI K.%
\end{APACrefauthors}%
\unskip\
\newblock
\APACrefYearMonthDay{2011}{}{}.
\newblock
{\BBOQ}\APACrefatitle {Predicting global average thermospheric temperature
  changes resulting from auroral heating} {Predicting global average
  thermospheric temperature changes resulting from auroral heating}.{\BBCQ}
\newblock
\APACjournalVolNumPages{Journal\ of\ Geophysical\ Research}{116}{A1}{}.
\newblock
\begin{APACrefDOI} \doi{10.1029/2010JA015685} \end{APACrefDOI}
\PrintBackRefs{\CurrentBib}

\bibitem [\protect \citeauthoryear {%
Weng%
, Lei%
, Sutton%
, Dou%
\BCBL {}\ \BBA {} Fang%
}{%
Weng%
\ \protect \BOthers {.}}{%
{\protect \APACyear {2017}}%
}]{%
Weng2017}
\APACinsertmetastar {%
Weng2017}%
\begin{APACrefauthors}%
Weng, L.%
, Lei, J.%
, Sutton, E.%
, Dou, X.%
\BCBL {}\ \BBA {} Fang, H.%
\end{APACrefauthors}%
\unskip\
\newblock
\APACrefYearMonthDay{2017}{}{}.
\newblock
{\BBOQ}\APACrefatitle {An exospheric temperature model from {CHAMP}
  thermospheric density} {An exospheric temperature model from {CHAMP}
  thermospheric density}.{\BBCQ}
\newblock
\APACjournalVolNumPages{Space\ Weather}{15}{2}{343--351}.
\newblock
\begin{APACrefDOI} \doi{10.1002/2016SW001577} \end{APACrefDOI}
\PrintBackRefs{\CurrentBib}

\bibitem [\protect \citeauthoryear {%
Yamazaki%
, Kosch%
\BCBL {}\ \BBA {} Sutton%
}{%
Yamazaki%
\ \protect \BOthers {.}}{%
{\protect \APACyear {2015}}%
}]{%
Yamazaki2015b}
\APACinsertmetastar {%
Yamazaki2015b}%
\begin{APACrefauthors}%
Yamazaki, Y.%
, Kosch, M\BPBI J.%
\BCBL {}\ \BBA {} Sutton, E\BPBI K.%
\end{APACrefauthors}%
\unskip\
\newblock
\APACrefYearMonthDay{2015}{}{}.
\newblock
{\BBOQ}\APACrefatitle {A model of high-latitude thermospheric density} {A model
  of high-latitude thermospheric density}.{\BBCQ}
\newblock
\APACjournalVolNumPages{Journal\ of\ Geophysical\ Research:\ Space\
  Physics}{120}{9}{7903--7917}.
\newblock
\begin{APACrefDOI} \doi{10.1002/2015JA021371} \end{APACrefDOI}
\PrintBackRefs{\CurrentBib}

\bibitem [\protect \citeauthoryear {%
Zesta%
\ \BBA {} Huang%
}{%
Zesta%
\ \BBA {} Huang%
}{%
{\protect \APACyear {2016}}%
}]{%
Zesta2016b}
\APACinsertmetastar {%
Zesta2016b}%
\begin{APACrefauthors}%
Zesta, E.%
\BCBT {}\ \BBA {} Huang, C\BPBI Y.%
\end{APACrefauthors}%
\unskip\
\newblock
\APACrefYearMonthDay{2016}{}{}.
\newblock
{\BBOQ}\APACrefatitle {Satellite orbital drag} {Satellite orbital drag}.{\BBCQ}
\newblock
\BIn{} G\BPBI V.~Khazanov\ (\BED), \APACrefbtitle {{Space Weather
  Fundamentals}} {{Space Weather Fundamentals}}\ (\BPGS\ 329--351).
\newblock
\APACaddressPublisher{Boca Raton, FL}{CRC Press}.
\PrintBackRefs{\CurrentBib}

\bibitem [\protect \citeauthoryear {%
Zesta%
, Oliveira%
, Schuck%
\BCBL {}\ \BBA {} Wilson%
}{%
Zesta%
\ \protect \BOthers {.}}{%
{\protect \APACyear {2018}}%
}]{%
Zesta2018b}
\APACinsertmetastar {%
Zesta2018b}%
\begin{APACrefauthors}%
Zesta, E.%
, Oliveira, D\BPBI M.%
, Schuck, P\BPBI W.%
\BCBL {}\ \BBA {} Wilson, G.%
\end{APACrefauthors}%
\unskip\
\newblock
\APACrefYearMonthDay{2018}{}{}.
\newblock
{\BBOQ}\APACrefatitle {Ionosphere-thermosphere system response to extreme
  geomagnetic storms} {Ionosphere-thermosphere system response to extreme
  geomagnetic storms}.{\BBCQ}
\newblock
\BIn{} \APACrefbtitle {Final paper abstract number: {SM51A-08}.} {Final paper
  abstract number: {SM51A-08}.}
\newblock
\APACaddressPublisher{Presented at 2018 AGU Fall Meeting, Washington, D.C.,
  10-14 Dec.}{}.
\PrintBackRefs{\CurrentBib}

\end{thebibliography}

\end{document}